\documentclass[useAMS,usenatbib]{mnras}
\usepackage{newtxtext,newtxmath}
\usepackage[T1]{fontenc}
\usepackage{ae,aecompl}
\usepackage{amsmath}	
\usepackage{amssymb}	
\usepackage{epsfig}

{\newif\ifnotend
\notendtrue
\def\veclist{ABCDEFGHIJKLMNOPQRSTUVWXYZabcdefghijklmnopqrstuvwxyz.}
\def\top#1#2.{#1}
\def\tail#1#2.{#2.}
\loop\expandafter\xdef\csname v\expandafter\top\veclist\endcsname%
{{\noexpand\bf\expandafter\top\veclist}}
\edef\veclist{\expandafter\tail\veclist}
\if\veclist.\notendfalse\fi\ifnotend\repeat}

\newcommand {\Vc}{V_{\rm c}}
\newcommand {\Vcd}{V_{\rm c,d}}
\newcommand {\Vch}{V_{\rm c,h}}

\newcommand {\kms}{\,{\rm km}\,{\rm s}^{-1}}

\newcommand {\Myr}{\,{\rm Myr}}

\newcommand {\Gyr}{\,{\rm Gyr}}
 
\newcommand {\pc}{\,{\rm pc}}
\newcommand {\kpc}{\,{\rm kpc}}

\def\d{{\rm d}}
\newcommand {\Msun}{\,{\rm M}_\odot}
\newcommand {\Zsun}{\,{\rm Z}_\odot}

\newcommand {\feh}{\hbox{[Fe/H]}}

\newcommand {\mgh}{\hbox{[Mg/H]}}
\def\feh{\hbox{[Fe/H]}}

\newcommand {\mgfe}{[\hbox{Mg}/\hbox{Fe}]}
\newcommand {\ofe}{[\hbox{O}/\hbox{Fe}]}

\newcommand {\afe}{[\alpha/\hbox{Fe}]}
\newcommand {\dex}{\,{\rm dex}}
\def\figref#1{Fig.~\ref{#1}}

\newcommand {\Rsun}{{R_{0}}}

\newcommand {\Vphi}{{V_{\phi}}}

\newcommand {\Lz}{L_{\rm z}}
\newcommand {\Lzo}{L_{\rm z, 0}}
\newcommand {\SFR}{\rm G}

\title[Inverse gradients in galactic discs]{Understanding inverse metallicity gradients in galactic discs as a consequence of inside-out formation}
\author[R. Sch\"onrich \& P. J. McMillan]
       {Ralph Sch\"onrich$^{1}$\thanks{E-mail: ralph.schoenrich@physics.ox.ac.uk},
        and Paul J. McMillan$^{2,1}$\\
	$^{1}$ Rudolf-Peierls Centre for Theoretical Physics, University of Oxford, 1 Keble Road, OX1 3NP, Oxford, United Kingdom \\
        $^{2}$ Lund Observatory, Department of Astronomy and Theoretical Physics, Lund University, Box 43, 22100, Lund, Sweden}
\date{Draft, \today}
\pagerange{\pageref{firstpage}--\pageref{lastpage}}
\pubyear{2016}

\begin{document}
\maketitle
\label{firstpage}

\begin{abstract}
The early stages of a galaxy's evolution leave an imprint on its metallicity distribution. We discuss the origins and evolution of radial metallicity gradients in discs of spiral galaxies using an analytical chemical evolution model. We explain how radial metallicity gradients in stellar populations are determined by three factors: the radial metallicity profile of the star-forming ISM, radial changes in the star-formation history (in particular inside-out formation), and radial mixing of stars. Under reasonable assumptions, inside-out formation steepens the negative ISM metallicity gradient, but contributes positively to the stellar metallicity gradient, up to inverting the metallicity profile to a positive $\d \feh / \d R$. This reconciles steep negative $\d \feh / \d R$ in some high redshift galaxies to generally flatter gradients in local observations. 

We discuss the evidence for inverse radial metallicity gradients (positive d[X/H]/d$R$) at high redshifts and the inverse relationship between azimuthal velocity and the metallicity (positive $\d \Vphi/\d \feh$) of stars for the Milky Way's thick disc. The former can be achieved by high central gas-loss rates and re-distribution processes, e.g. re-accretion of enriched material in conjunction with inside-out formation, and near-disc galactic fountaining.
For the Milky Way thick disc, we show that the positive $\d \Vphi / \d \feh$ correlation points to comparable timescales for inside-out formation, initial metal enrichment and SNIa enrichment. We argue that the original ISM metallicity gradient could be inferred with better data from the high-metallicity tail of the alpha enhanced population.
Including inside-out formation in our models changes the local vertical metallicity gradient by about $-0.2 \dex / \kpc$, in line with local measurements. 
\end{abstract}


\section{Introduction}
This paper discusses the origins and temporal development of radial metallicity gradients in galactic discs as well as their imprint on local correlations between azimuthal velocities and metallicities.
It is motivated by recent findings of inverse gradients in galactic discs, be it inverse radial metallicity gradients (i.e. positive d[X/H]/dR) found in external galaxies \citep[][]{Cresci10}, as well as inverse relationships between azimuthal speed and metallicity (i.e. positive $\d \Vphi/ \d \feh$)  for local thick disc stars \citep[][]{Spagna10, Lee11}. These gradients are ``inverse" in the sense that they have opposite sign to those found in the present-day star-forming gas in disc galaxies, where metallicity decreases with radius (or equivalently, declines with angular momentum).

In contrast to young, thin disc populations, where the current metallicity gradient imprints a slower local mean rotation with increasing metallicity \citep[see e.g.][]{Casagrande11}, it was discovered by \cite{Spagna10} and \cite{Lee11}, that the metal-poor, alpha-rich stellar population shows an inverse relationship between mean azimuthal speed and metallicity; i.e. in this subgroup,  stars with lower metallicity show slower mean rotation than stars with higher metallicity. The increase of azimuthal velocity with metallicity has been confirmed for thick disc stars in RAVE by \cite{Kordopatis13}. \cite{Curir12} linked this to inverse metallicity gradients (i.e. larger metallicities in the outer regions of a disc) at early times, identified in the chemical evolution models of \cite{Spitoni11} and \cite{Chiappini01}, as well as occasionally observed in high-redshift galaxies, such as those studied by \cite{Cresci10}. 

The finding of \cite{Cresci10} of a positive/inverted $\d \feh/ \d R$ gradient at high redshift can also be linked to the finding of \cite{Queyrel12}, who found positive gradients using nitrogen and oxygen emission lines in two isolated and apparently quiescent galaxies, which cannot relate to the strong gradient changes expected in mergers \citep[][]{Rupke10}. \cite{Stott14} find a similar picture, where metallicity gradients tend to be inverted or very small in galaxies with a high specific star formation rate. At the same time, e.g. \cite{Jones10, Jones13} have observed very steep radial metallicity gradients $\d [X/H]/\d R < -0.2 \dex/\kpc$ in a few galaxies at redshift $z \gtrsim 2$, which by far exceed values measured in the Milky Way or other local galaxies. On the other hand, an analysis of the KMOS sample at redshifts between $z \sim 1-2.5$ \citep[][]{Wuyts16} has yielded mostly flat radial abundance gradients in the gas phase (although with large scatter), as well as no significant correlation with structural parameters. We note that it is also very difficult to interpret these studies at current stage: most investigations rely on nebular oxygen and nitrogen lines, which are produced by various sources, ranging from AGN (usually well controlled), through HII regions (associated with star formation) and planetary nebulae (linked to older stellar populations) to the warm/hot ISM itself.

Most work on the theoretical side has so far been performed with N-body simulations. In their first analysis of a cosmological disc simulation, \cite{Rahimi11} reported stellar radial metallicity gradients of $\d \feh / \d R \sim -0.05 \dex/\kpc$ in their intermediate age stars and $-0.066 \dex/\kpc$ in their young stars. \cite{Gibson13} compare gradients in the ISM and stars of their simulations to the relatively steep negative gradients in cosmological observations from \cite{Jones10} and \cite{Yuan11}. In the spirit of classical chemical evolution models \citep[see e.g.][]{Prantzos00}, which predicted a steepening in the ISM gradient due to the faster decline in star formation rates in the central regions (which increases the ratio between yields and fresh accretion), \cite{Gibson13} link the more negative gradient to inside-out growth. However, we will show that while inside-out growth does tend to steepen the negative ISM gradient, it gives a positive contribution to the gradient in stars.

Both SEGUE \citep[][]{Cheng12} and RAVE \cite[][]{Boeche14} data have also pointed to a weakening of radial abundance gradients with Galactic altitude $z$, up to a mild gradient inversion, a behaviour which is also found in N-body models like that of \cite{Gibson13}.
The inside-out growth of discs has been linked to the issue of gradients by several papers, e.g. \cite{Gibson13, Minchev14, Miranda15}. In particular \cite{Minchev14} link to inside-out growth the tendency of radial metallicity gradients to weaken and then invert with increasing altitude. However, they do not provide a detailed explanation of the mechanisms and causes, while \cite{Rahimi14} show a clear connection between flaring of younger populations towards the outer disc and the radial metallicity trend getting more positive with altitude. We will show that in our simulations, where we can separate the factors, this trend with altitude is linked mainly to radial migration and not to inside-out formation.

To enter our discussion, it is important to note that the findings of inverse radial metallicity gradients \citep[][]{Cresci10} are not directly equivalent to the inverse/positive relation between azimuthal speed and metallicity, discovered by \cite{Spagna10}. As a first, obvious, reason, older stars have been more affected by disk heating than younger ones, and this has opposite effects on the $R-\feh$ and $\Vphi-\feh$ gradients. If we hold the angular momentum distribution constant, heating the most metal-poor populations increases their mean Galactocentric radius (making the radial metallicity gradient more negative), while it increases asymmetric drift, giving lower metallicity stars a comparably lower azimuthal velocity at a given radius (making the azimuthal velocity gradient more positive).

In this paper we show that a clear discrimination between the origins of metallicity gradients in stellar populations is needed, and we identify and discuss the three major sources:
\begin{enumerate}
\item The time-dependent gradient in the star-forming ISM.
\item Inside-out formation and radial variations in the star-formation rates.
\item Radial mixing of stellar populations. 
\end{enumerate}
We demonstrate that variations in the age distribution of the stars born at a given radius is just as important as the metallicity gradient of the star-forming gas at any time in the life of the galaxy in determining the final metallicity gradient of the stellar populations. Inside-out galaxy formation produces these variations in age and typically drives $\d \feh / \d R$ towards positive values, bringing very steep gradients at high redshift in line with a lack of such steep gradients in local disc galaxies. Radial migration by bars \citep[][]{Friedli94} or spiral patterns \citep[][]{Sellwood02} also has a strong impact on disc structure and composition. However, local stellar data show that the mixing of all stellar populations in the disc of our Milky Way by this process is far from complete \citep[][henceforth SB09]{SB09a}. We will show that the main effect of radial migration is to radially average and blur, but not completely destroy, these gradients. 

In addition we will discuss the likely sources of gradient flattening and inversions in the star-forming ISM, offering a slightly different perspective from the usual explanation that this is produced by low-metallicity accretion into the central regions.

This paper provides a more analytic discussion to disentangle the different observations and the processes that shape them. While our standard model is constructed to meet the main observables of the Milky Way, we do not attempt to produce a fiducial model for the Galaxy, and explicitly do not attempt specific fits to observations, which would be questionable given the uncertain errors and interpretation of measurements in the different surveys and more importantly would cloud the main incentive of this work. In Section 2 we quickly introduce two toy models/thought experiments to explain the problem of gradient formation. In Section 3 we quickly discuss the applied chemo-dynamic models, in Section 4 we provide a short analytic framework to understand metallicity gradients in stars better, and discuss changes in radial abundance gradients, including a discussion of mechanisms to invert radial abundance gradients in the star-forming ISM. In Section 5 we discuss relationships between chemistry and local azimuthal velocities, followed by a discussion of gradients at higher altitudes in Section 6. We conclude in Section 7.

\section{Two toy models}
The key concepts explained in this paper can be readily understood by
considering two toy models.

Consider the case of a galaxy which has a metallicity constant in radius at all times, but rising with time. If this galaxy forms inside out, i.e. if the ratio of star
formation rates between the inner disc (small $\Lz$) and the outer disc
(large $\Lz$) is larger at small $t$, then today its lower metallicity stars will still be concentrated at smaller
angular momenta. Or in other words: While at any time the metallicity gradient of the star forming ISM was zero, the
galaxy in our thought experiment exhibits an inverse/positive radial metallicity
gradient in its stars.

Secondly, consider the case where there is no radial gradient in the
metallicity of the star-forming gas at any time during the evolution of a galaxy, and no inside-out growth. Again the older stars
are more metal-poor, and have a higher velocity dispersion (because
there is an age-velocity dispersion relationship). The more metal-poor
stars therefore have higher velocity dispersions, and thus a larger
asymmetric drift. This produces an inverse, i.e. positive gradient in the $v_\phi$-$\feh$
relation while the associated radial expansion gives a small normal, i.e. negative contribution to the radial $\d \feh/\d R$ gradient.

In the following, we will discuss and quantify how these two effects alter
gradients in the disc. In the next Section, we will lay out the analytic chemodynamic models that we will use to compare the size of these effects compares to the impact of other processes which shape
these gradients, and will return to a more formal approach to this problem in Section 4.

\section{Method outline}
A quick definition of our use of "radial migration" terms is necessary. Throughout this work we define "radial mixing" as the general redistribution of stars born at some fixed angular momentum $\Lzo$ or corresponding radius to a position $R$, at which they are observed today. This radial mixing is conceptually split up into "churning" (corresponds to the narrower meaning of radial migration), which redistributes stars in angular momentum $\Lz$ and "blurring" which describes the re-distribution over different positions at fixed $\Lz$ caused by the excursions of stars along their orbits. 

We use an improved version of the analytic chemodynamical galaxy models SMAUG (the models used by SB09). 
These models combine an analytical chemical evolution code with modelling of the galaxy's dynamics, in particular radial migration and the current kinematics and spatial distributions of stars. The model has two gas phases, warm/hot and cold. This allows us to account for the delays in metal enrichment: most of the processed material from dying stars is initially warm/hot, and needs to cool down and condense back into the star-forming cold ISM. The model machinery comprises full population synthesis in order to facilitate the replication of selection effects and to fully account for the time delay in enrichment. SMAUG also accounts for radial flows of the gas, including mixing from radial migration, inflow through the disc connected to accretion and a possibility for additional mixing. We have updated the approach to radial inflow with the angular momentum prescription from \cite{BS12}. In addition, we have updated the kinematic modelling: we now assume both vertical and horizontal action conservation, implemented using the Torus Mapper \citep[TM, ][]{BM16}.

\begin{figure}
\epsfig{file=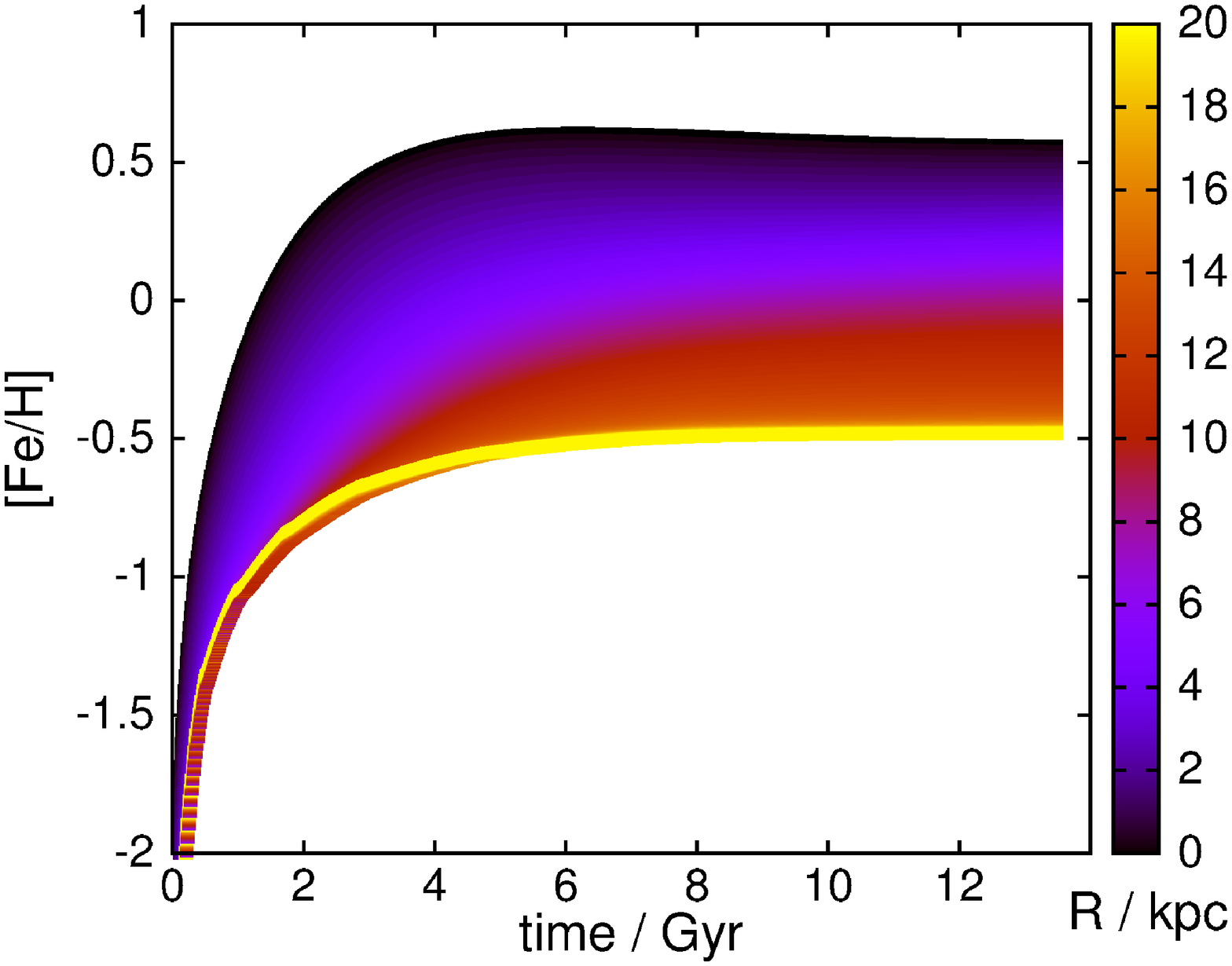,angle=-90,width=\hsize}
\epsfig{file=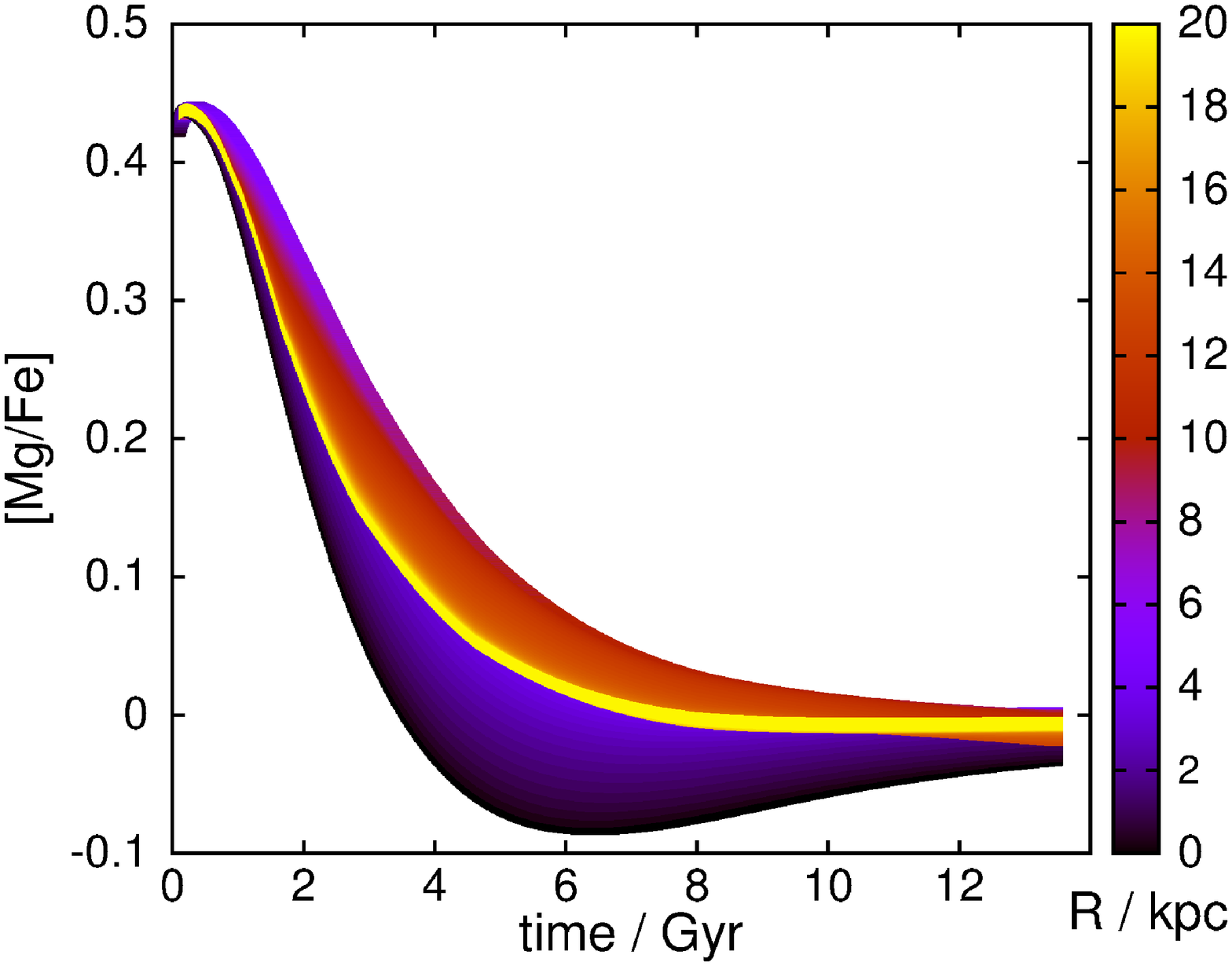, angle=-90,width=\hsize}
\caption{$\feh$ and $\afe$ of the star-forming gas as a function of time for the standard model with inside-out formation. Different colours correspond to different radii.}\label{fig:xtime}
\end{figure}

Otherwise, our assumptions here very much resemble those in SB09, resulting in a similar enrichment history (see \figref{fig:xtime}). We use a radial resolution of roughly $0.25 \kpc$, and an increased time resolution of $15 \Myr$. The accretion history of the galaxy assumes the two-exponential inflow pattern applied in SB09: the disc is set up with an initial $10^8 \Msun$ of primordial gas, and accretes $5\times10^{10} \Msun$ with exponential decay time of $1 \Gyr$ and further $1.0 \times 10^{11} \Msun$ with exponential decay time of $9 \Gyr$. This shifts the accretion to slightly earlier times than in SB09, in concordance with cosmological expectations \citep[see e.g.][]{Wang11}, and other estimates for the star formation history, e.g. \cite{AB09}. A large early accretion rate is also necessary to sustain high star formation rates while the inside-out forming gas disc builds up its mass. As a result of the accretion, overall star formation rates peak early and then slowly subside over the history of the system. The resulting scale lengths and local mass density near the Sun are comparable to estimates from observations \citep[e.g.][]{Holmberg04, Flynn06, Juric08, Read14}. In particular the standard inside-out model has at $t = 12 \Gyr$ local surface mass densities of $36 \Msun \pc^{-2}$ in stars and remnants, matching the derivation of \cite{McKee15}, as well as $\sim 10 \Msun \pc^{-2}$ in gas, slightly lower than the estimate of \cite{McKee15}, with a total baryonic mass of $5.0 \times 10^{10} \Msun$. The total mass is significantly smaller than the accreted mass due to the mass loss over time (see subsection \ref{sec:chemevo}).

\subsection{Inside-out growth and star formation}

Inside-out growth in our simulations is applied by varying the scale-length of the exponential gas disc, $R_g$. When the galaxy accretes new mass at each timestep, setting the specific angular momentum of the infalling material leaves a degree of freedom, i.e. one can either specify the radial density profile of the infalling material or adapt the infall to approximate a shape of the targeted disc. We choose the latter option, keeping the gas disc exponential at all times while varying its scale-length.
While it was shown in \cite{BS12} that reasonable inside-out growth does not significantly affect today's metallicity gradients, its functional shape affects how the enrichment history populates the $\afe$--$\feh$ abundance plane. We found that the simplest possible assumption, i.e. linear growth of the scale radius in time, up to a maximum value, is not viable. Any discontinuities in the growth rate createe artefacts in the metallicity plane and age-metallicity relation when the growth stops. We hence choose the fully differentiable function:
\begin{equation}\label{eq:iogrowth}
R_g (t) = R_{g,0} + (R_{g, e} - R_{g,0}) \cdot \left[\arctan\left(\frac{t-t_0}{t_g}\right) - \arctan\left(-\frac{t_0}{t_g}\right)\right] \cdot N ,
\end{equation}
where $R_{g,0}$ is the initial scale-length of the gas disc and $R_{g,e}$ is that after $12\Gyr$. The normalization $N$ is chosen to ensure that this condition holds at $t=12 \Gyr$. As standard value, we use a growth timescale $t_g = 2 \Gyr$ with an offset time $t_0 = 1 \Gyr$, which implies near-linear growth of the disc during the first $2 \Gyr$ and then a fast transition to a very slow growth of the scale length. The standard inside-out model uses $R_{g,e} = 3.75 \kpc$ and $R_{g,0} = 0.75 \kpc$. We also note that the model's recent growth rates are in line with estimates from \cite{Pezzulli15}.

For star formation we use the Schmidt-Kennicutt law, i.e. the star formation efficiency is given by
\begin{equation}
\eta_{SFR} = 0.15 \left(\frac{\Sigma_g(t)}{\Msun \pc^{-2}}\right)^{0.4} \Gyr^{-1} .
\end{equation}
The prefactor $0.15$ is chosen to meet surface mass density and star formation rate constraints for the Milky Way, see also SB09. Note that the exponent of $0.4$ in the star formation efficiency implies a shorter scale-length of the young stellar disc by a factor $1.4$ compared to the gas disc.

In this work, we assume a cut-off for the Schmidt-Kennicutt law: for a surface density below $4 \Msun \kpc^{-2}$ in cold gas, we assume that the star formation efficiency drops with the third power of the gas surface density. We assume that stars inherit the elemental abundances of the cold ISM phase from which they are born without additional abundance scatter, motivated by observations of very homogeneous abundances in young disc stars \citep[see e.g.][]{Przybilla08}. 

\begin{figure}
\epsfig{file=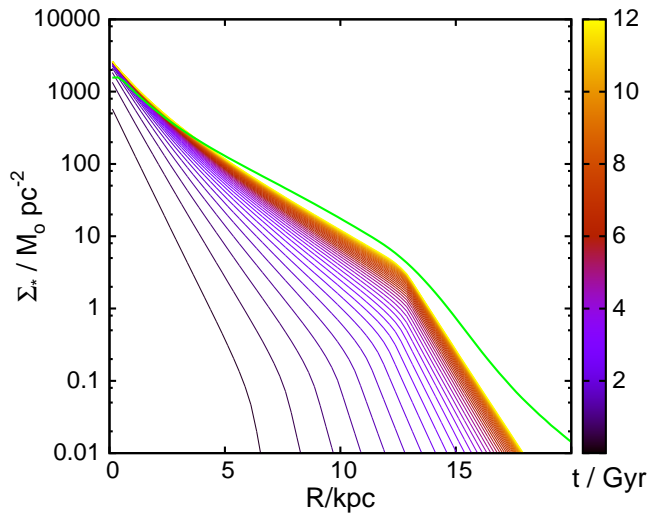,angle=-90,width=\hsize}
\caption{Radial stellar surface density profiles for the inside-out growing simulation when stars are kept fixed in angular momentum at their guiding centre radius, colour-coded in time. This is equivalent to the integrated total star formation as a function of radius. The green line shows today's density profile taking into account full radial mixing (churning and blurring).}\label{fig:hordens}
\end{figure}

\figref{fig:hordens} shows the resulting development of the surface density of the stellar populations in time (colour coding), when neglecting churning and blurring, i.e. keeping stars at their radius of birth. This is equivalent to the integrated total star formation (minus life-time effects) as a function of galactocentric radius. The cut-off in the density profiles results from the applied cut-off in the Schmidt-Kennicutt star formation law (see above). The growth of the scale-length is evident, as well as initial accretion and scale-length growth of the disc move the outer star formation cut-off rapidly outwards at early times. At later stages the growth of the gas-disc's scale-length is partly balanced by the declining total gas mass of the disc, so that the cut-off radius for the star formation remains near $R \sim 14 \kpc$. The thick green line displays today's stellar surface density when fully accounting for radial migration (churning) and the random motions of stars (blurring). In particular the latter term expands the disc, strongly increasing the surface densities in the outer disc and softening the stellar profile around the cut-off radius. 

\subsection{Model stellar dynamics}
While SB09 approximated the vertical structure of the disc by isothermal distribution functions, with conserved vertical energy, the $N$-body simulations of \cite*{Solway12} showed that it is the vertical \emph{action} which is (on average) conserved by stellar migration \citep[The simulations of][suggest this may not be completely true if there is very violent spiral activity]{VC14}. We therefore employ the quasi-isothermal distribution functions of \cite{BM11} in the potential by \cite{Piffl14}, which replace the old isothermal distribution in energy $\exp(-E/\sigma^2)$ with the action distribution $f(J_i) \propto \exp(-(\nu_i J_i)/\sigma_i)$.

We use the same age-velocity dispersion relations as in SB09, which are closely related to the usual power law dependence on time found in the solar neighbourhood e.g. by \cite{DB98, Holmberg09, AB09}:
\begin{equation}\label{eq:disp}
\sigma_i = \max{\left\{\sigma_{i, {\rm min}}, \sigma_{i, 10} \left(t/10 \Gyr\right)^{\beta}\right\}} ,
\end{equation}
where $\beta = 0.33$ is the usual time-exponent.
Strictly speaking, the parameters $\sigma_i$ are not the local
dispersions, but are closely linked to them,\footnote{We will use the
term ``kinematic heat" to refer to the spread in radial and vertical
actions. This is closely related to the velocity dispersion, but not
identical.} Since this equation uses the epicycle frequencies $\nu_i$ at
the guiding centre radius, and since actions diverge when velocities
approach the escape speed, cutting the wings in the velocity
distribution short compared to a Gaussian, the velocity dispersions are of order $\sim 10 \%$
smaller than the parameters $\sigma_i$ in the relevant areas of
parameter space. We use the epicycle frequencies of each population's
birth radius, and keep their action distributions for each population
fixed, no matter how migration has changed their guiding centre radii.

Accounting for this effect, we use $\sigma_{z, 10} = 28 \kms$ and $\sigma_{R, 10} = 43 \kms$ for stars born in the solar neighbourhood. Following the traditional approximation of analytic kinematic models \citep[which among other benefits approximates a constant scale height of coeval local populations, see also SB09a and][for references]{Piffl14}, we assume an exponential behaviour of $\sigma_{i, 10}(R)$, i.e. $\sigma_{i, 10}(R) = \sigma_{i,10}(\Rsun)\exp(-R/R_{\sigma_i})$ with galactocentric radius, and let the dispersion parameters decrease exponentially towards larger radii with scale-lengths $R_{\sigma_i} = 5$ and $7.5 \kpc$ respectively (i.e. $2$ and $3$ disc scale-lengths), setting a minimum dispersion of $\sigma_{\rm min} = 7 \kms$ for the youngest stars.

For each group of coeval stars born at some radius/angular momentum ($R_0/\Lzo$), we hold their action distributions constant regardless of their radial migration. To avoid eventual problems with the epicycle frequencies and kinematics of innermost stars, we use the action distributions of $R = 1.5 \kpc$ for all stars with $R < 1.5 \kpc$, however, we checked that this cut-off does not significantly affect our results.

\begin{figure}
\epsfig{file=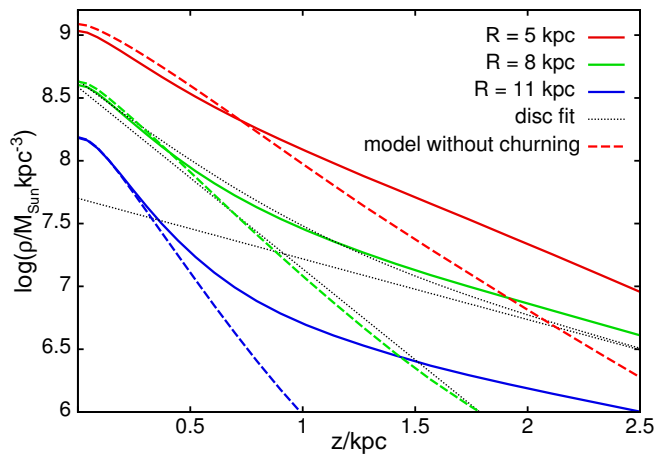,angle=-90,width=\hsize}
\caption{Vertical density profiles at different radii. For comparison we show a common two-exponential decomposition into $\exp{-z/(0.3\kpc)} + 0.13 \exp{-z/(0.9\kpc)}$ (grey lines). The dashed lines show the vertical density profiles we would have if there was no radial migration (i.e. populations kept their initial angular momentum).}\label{fig:vertdens}
\end{figure}

\figref{fig:vertdens} shows that the model disc (with solid lines) has scale heights comparable to those of the Milky Way and displays a thin/thick disc structure. While we did not attempt to fit the Milky Way profile, the model at $R \sim 8 \kpc$ fits quite closely the two-exponential fits from \cite{Ivezic08, Juric08}. With black dashed lines we show the decomposition $\exp{(-z/(0.3\kpc))} + 0.13 \exp{(-z/(0.9\kpc))}$. 


We further assume that the extent of the migration (churning) of a star is independent of its current actions, which is approximately justified by the results of the $N$-body simulations of \cite{Solway12}. This is challenged by the results of \cite{VC14} who found that migrated stars make up a population with a lower initial vertical velocity dispersion than those that do not migrate (implying that typically stars with low vertical actions are more likely to migrate). The difference between the two results is probably due to the \cite{Solway12} simulations being dominated by low-$m$ mode spirals, while those of \cite{VC14} are dominated by multi-armed (high-$m$) spiral patterns. Our choice to assume that the migration coefficients are independent of current actions is motivated by our preference not to complicate the picture here, as the effect of migration is mostly to average and weaken the observed gradients that we are interested in. \cite{Minchev12} claimed that the prevalence of average action conservation instead of vertical energy conservation is the reason why they do not see thickening in their simulated disc under migration \citep[the idea of migration thickening outwards migrating populations was first proposed in][]{SB09b}. \cite{Minchev12} unfortunately convolute the question of heating with migration. Migration does not heat the disc, it just transfers stars with potentially wider action distribution from the inner disc. We can quantify analytically that outwards migrating populations under action conservation achieve larger scale heights \citep[see equation 27 in][]{SB12}. Consistently, this has been shown in simulations, implicitly in \cite{Loebman11}, and with an explicit investigation by \cite{Roskar13}, refuting the claims in \cite{Minchev12}. This model shows that under action conservation we obtain a locally even thicker disc than in the MW, while the comparison case without radial migration (dashed lines in \figref{fig:vertdens}) shows no sign for any thick disc. Hence, it is clear that outwards migrating populations can thicken sufficiently under action conservation. The central question here is, however, not the effect of action conservation, but instead: i) How large is the effect of preferential migration of kinematically colder subsets? ii) Can secular processes provide sufficient kinematic heat in the inner disc regions, or do we require mergers. For the latter, it has been shown that molecular clouds with a fixed mass distribution do not provide sufficient heating in the inner disc \citep[][]{Aumer16}. It remains to be tested how much impact a changing mass distribution can have, and also whether the time-dependent heating for each population needs a more sophisticated recipe in future models.

The radial dependence of migration is formulated as in SB09. We assume a maximum exchange fraction between neighbouring rings of 0.2 per half-timestep (i.e. $7.5 \Myr$). Note that this mass exchange between neighbouring rings conserves the detailed angular momentum balance. I.e. churning itself does not expand the disc\footnote{The one exception to this is due to the participation of the gas disc in the same process. Since the gas disc has a longer scale-length, exchange of angular momentum between gas and stars will cause a mild radial expansion of the stellar populations. This is of particular importance near the cut-off.} Most of the expansion of the disc stems from blurring -- as the stellar populations heat up, they move on orbits with larger mean radii than circular orbits with the same angular momentum. 

\subsection{Model chemical evolution}\label{sec:chemevo}
The chemical evolution is treated as in SB09, with the small difference that we are not using a fixed metallicity for the infalling gas. While it is unclear what metallicity freshly accreted gas had in the early Milky Way, a value close to today's value (near $\feh \sim -0.6$) as assumed in SB09 creates an inwards travelling metallicity wave with an inverse metallicity gradient at early times: the freshly arriving gas has a higher metallicity than the currently present gas and the inner galaxy is better shielded against this infall due to its larger mass density and higher mass-to-infall ratio. This can increase metallicities in the outer regions far faster than the native enrichment in the inner regions and hence creates inverse radial metallicity gradients. While we consider this possible scenario that strong outflowing winds carry large amounts of stellar yields above the disc, from where they can be re-accreted also onto the outer parts of the disc \citep[especially in light of suspecting galactic fountaining, see][]{Marinacci11, Brook12}, more natural and likely than the explanation with cold inflows penetrating mainly to the centre of \cite{Cresci10}, it is not our main subject of study here. What we need for the model is to track a reasonable composition of the inflow and a smooth development of its metallicity. To keep things simple, we use the composition of the cold gas at the current stellar half-mass radius of the model, and scale its metallicity as  $Z_{l}\tan^{-1}{Z_{t}/Z_{s}}$, where $Z_{s} = 0.25 \Zsun$ and $Z_{l}$ is $0.4$ times the metallicity at the half-mass radius. The metallicity of the accreted material is of lesser importance to our models, since it mostly sets the metallicity of outer disc stars, which have a minor impact on local samples. Our assumption is an educated guess from reviewing literature on the Milky Way's high velocity clouds \citep[see e.g.][]{Collins07, Shull11}, which have metallicity measurements ranging somewhere between $0.1$ and $0.55 \Zsun$. The highest values are found for the Smith cloud currently merging with the Galactic disc \citep[][]{Fox16}, consistent with a simple expectation of a mild increase of metallicity by interaction with the more metal-rich outflow of the Galaxy. Our main concern is that our prescription ensures a smooth rise of the infall metallicity, ending at slightly above one quarter solar metallicity.

For the galactic gas we keep the two-phase model of SB09. We now use as our standard initial mass function (IMF) the \cite{Chabrier03} IMF instead of the \cite{Salpeter55} IMF. This increases the ratio between high-mass stellar yields and the lock-up in low-mass stars by a large factor, which needs to be balanced by a higher loss rate to the IGM in concordance with results from cosmological observations, as well as theoretical expectations and simulations \citep[e.g.][]{Meyer87, Aguirre01, Dave01, Nath97}. In particular, \cite{Tumlinson11} find that the amount of metals in the circumgalactic gas likely surpasses the amount held in galactic discs, i.e. those galaxies have to have lost more than half the stellar yields. 

In the model, we have two ways of implementing a metal-loss. Either by a direct loss of stellar yields, or by a mass load parameter $\eta_{\rm l}$, i.e. for every unit mass of star formation, $\eta_{\rm l}$ unit masses of gas are lost from the gas disc. At the same equilibrium metallicity, the latter term has a tendency to induce a quicker early enrichment, since it evaporates larger amounts of the present metal-poor gas and loses less early stellar yields. Both parameters are degenerate in fitting today's metallicities, but we can fix the total loss using today's local ISM metallicity. To allow for a somewhat enriched outflow, we use a direct loss fraction of stellar yields of $0.45$, coupled to a mass load parameter of $\eta_{\rm l} = 0.45$. This parameter set is equivalent to a direct loss of $0.66$ of stellar yields \citep[a Salpeter IMF would require $0.1$; for an extended discussion of IMF impacts, see][]{Andrews16}.

There will be a difference in the redistribution of yields from core collapse (SNII) supernovae and SNIa. Most core collapse SNe will happen within the dense star-forming ISM, while the time-delay of SNIa allows their progenitors to leave the dense regions. This raises the expectation that core collapse SNe give back more yields directly to the cold gas phase, while most of the SNIa yields will go to the warm/hot ISM. We also know that the solar system was enriched with radioactive elements (partly r-process), e.g. Curium 247 \citep[see e.g.][]{Tissot16} with lifetimes of order $10 \Myr$, which demands a fast entrance of yields into the star-forming ISM. In rough concordance with simulations of \cite{Walch15} we split the retained yields from massive stars equally between the cold and the hot ISM, while for SNIa we only add $1.5\%$ directly to the cold phase. With the same argumentation, we set the time constant for the freeze-out of the warm/hot gas to the cold phase to $1 \Gyr$, yielding a total mass in the warm/hot phase of $10^9 \Msun$ today. The precise freeze-out timescale is of minor importance to our results. Its main effect here is to delay the initial enrichment of the cold gas, and hence degenerate with uncertainties in the early loss rates. We discuss its influence on the stellar metallicity gradient in the inner disc in Section 5. 

\subsection{Gas inflow and mixing}

The re-distribution of gas within the galaxy is of great importance. For the inflow through the disc we use the prescription of \cite{BS12}, which is in concordance with a different formulation used by \cite{Pezzulli16}. This prescription assumes that gas accretion onto the disc dilutes the angular momentum of the disc \citep[see][]{Lacey85} and drives disc gas inwards, while conserving total angular momentum. The parameters are constrained by fitting the resulting radial metallicity gradient in the star-forming ISM to the observed gradient in Cepheids \citep[][]{Luck11}. For the angular momentum of the accreted gas we use here $0.35$ in the centre, rising linearly to $0.85$ at the outer edge at $25 \kpc$. 

\begin{figure}
\epsfig{file=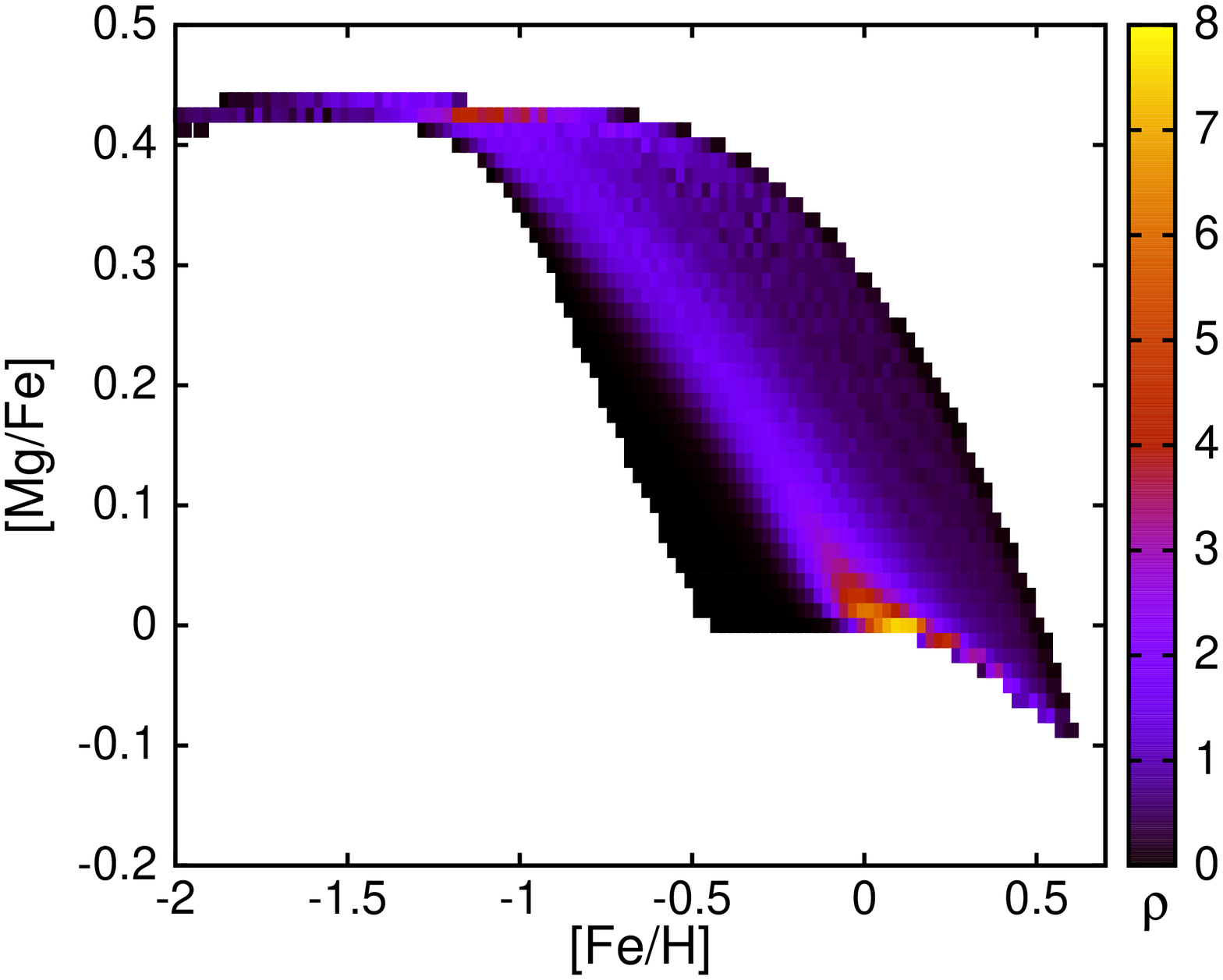,angle=-90,width=\hsize}
\epsfig{file=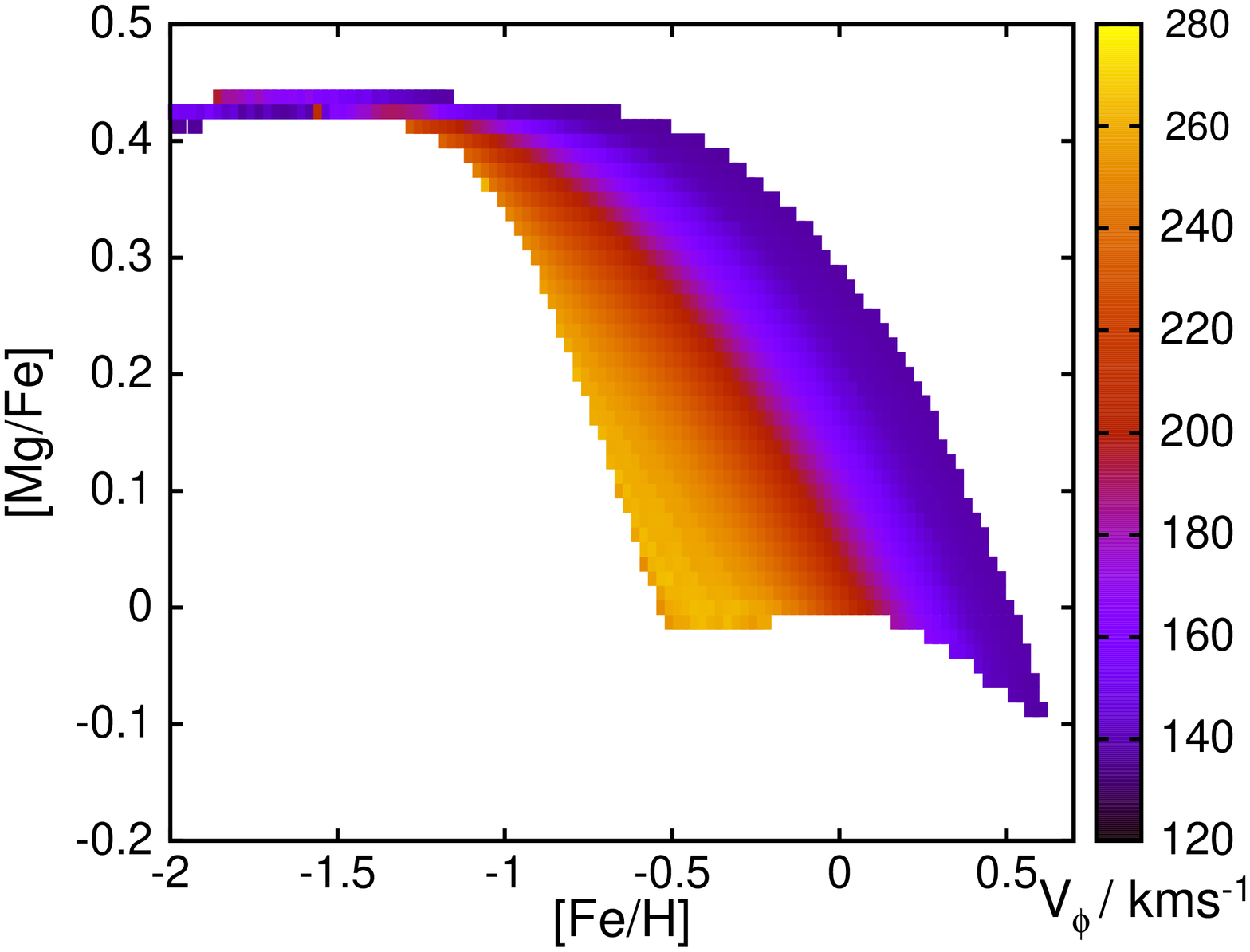, angle=-90,width=\hsize}
\epsfig{file=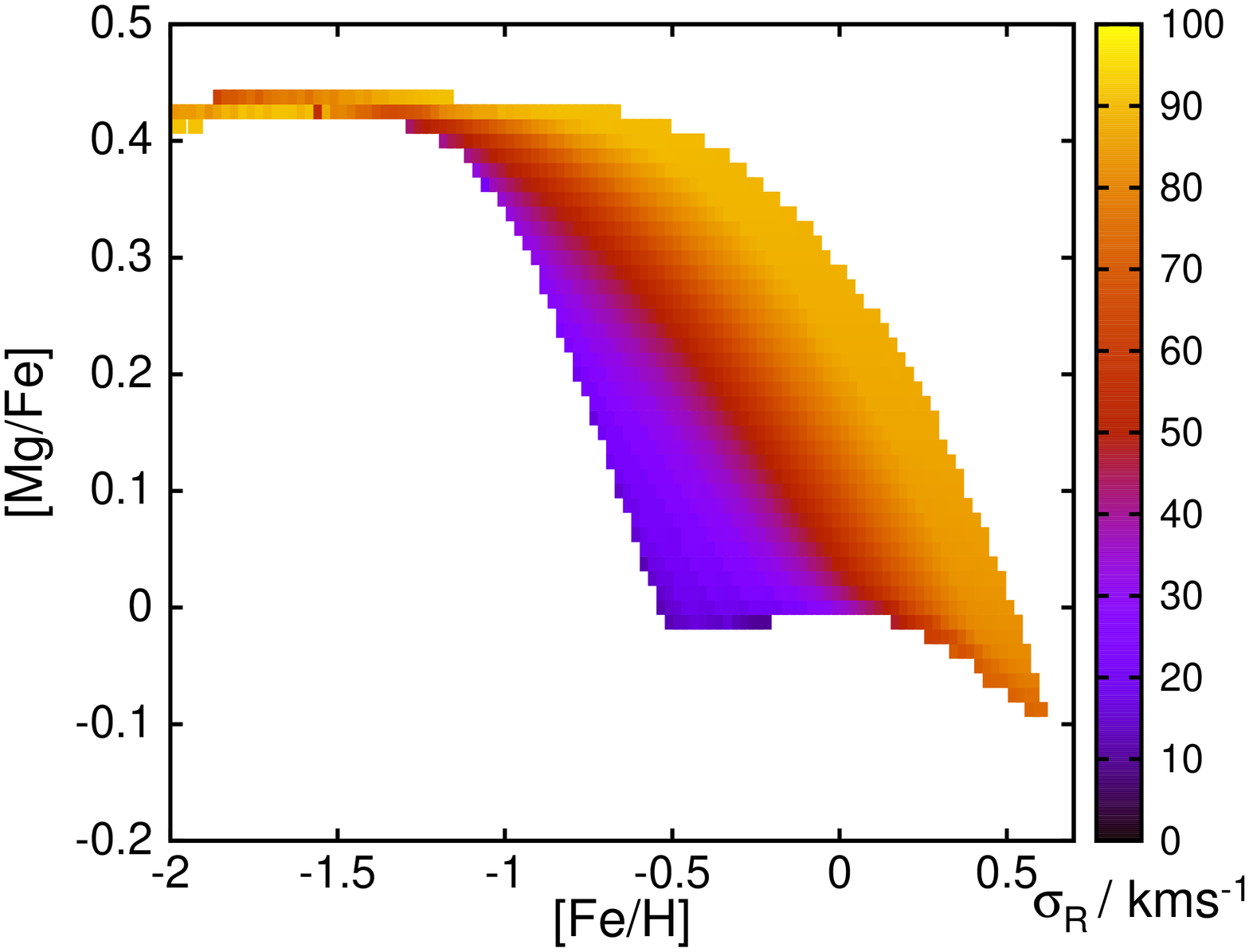, angle=-90,width=\hsize}
\caption{Densities, mean azimuthal velocities and radial velocity dispersions in the metallicity plane at $R = \Rsun, z = 0.4 \kpc$.}\label{fig:metdens}
\end{figure}

As can be seen from \figref{fig:xtime} and \figref{fig:metdens}, the structure of the disc in the age-chemistry-kinematic space still largely resembles the older versions of this model (see SB09a, Fig. 6, SB09b, Fig. 1ff.). The vertical density profile shows the classic two-exponential behaviour, though this is less pronounced in the inner disc, because the model has no explicit bulge nor bar and outwards migrating stars need approximately a scale-length or more to thicken/flare up sufficiently.
\figref{fig:xtime} displays $\feh$ and $\afe$ versus time at each radius (colour coded). We note the relatively quick initial enrichment, followed by the transition into a nearly stable state, comparable with results from \cite{Haywood08}, the GCS survey \citep[][]{Casagrande11}, and the Gaia-ESO survey \citep[][]{Bergemann14}.
\figref{fig:metdens} shows overall density, mean azimuthal velocity and radial velocity dispersions as a function of position in the metallicity plane at $R = \Rsun, z = 0.4 \kpc$.

In addition to the inwards flow, gas is redistributed and mixes throughout the galaxy. As a minimum estimate of this diffusion, we assume the same radial migration for gas as for young stellar populations. However, the mixing may be far larger due to turbulent diffusion. It is also reasonable to believe that yields in the warm/hot gas phase might experience different (stronger) mixing and therefore more re-distribution than the cold gas phase. Some models have played with re-distribution of yields, e.g. launching yields as compact clouds on ballistic trajectories from the disc \citep[see e.g.][]{Spitoni08}, but the observational and theoretical constraints for the properties of the hot phase are unsatisfactory. In principle the disc will also lose some angular momentum to the warm/hot phase, driving a bulk of stellar yields outwards, the balance from stellar yields alone would be of the right order to cancel the metallicity gradient created by the inflow. However, it is not known how this warm/hot medium interacts in turn with the corona and with the infall into the galaxy. 

Our ''standard`` model does not include additional gas mixing, but for this study we will test the simplest and most conservative possible additional mixing model, leaving the radial density distribution of the Galactic gas untouched. This can be achieved by constructing the mixing process out of small, mass conserving interchanges of gas between neighbouring rings:
\begin{equation}\label{eq:redistr}
\Delta m_{i \to i+1} = k \sqrt{m_{i+1}/m_i} \\
\Delta m_{i \to i-1} = k \sqrt{m_{i-1}/m_{i}} \\
\end{equation}
where $m_i$ is the mass residing in ring $i$, and $k$ is a parameter setting the mass transition. For the innermost rings where there are large mass ratios between neighbouring rings (because rings of equal radial width have very different areas) we limit $k$, such that the exchanged mass does not exceed the mass in each ring. Stronger mixing can now be achieved by changing the parameter $k$ and redistributing the mass several times per timestep (the effective mixing length will be proportional to the square-root of the number of steps). In our model variant with additional gas mixing, we use $8$ sub-steps (so an exchange between neighbouring rings every $\sim\Myr$), and let $k = 0.3$ in the first $6 \Gyr$, then decaying exponentially with timescale $6 \Gyr$.

\subsection{Main models used}
Unless otherwise stated, we evaluate the properties of all models after $12 \Gyr$, at $(R,z) = (8.3, 0.7) \kpc$ with a mass weighted selection function.

The three model variants we typically compare, are: The "standard" inside-out model (which we refer to as \textsc{io}), a constant scale-length model, with the sole difference of a constant gas disc scale-length (\textsc{no-io}), and the inside-out model with additional gas mixing (\textsc{io+mix}).

\section{Disentangling the drivers of metallicity gradients}

Most previous discussions of inverse metallicity gradients in disc populations have focused on an inverse gradient in the star forming gas or youngest stellar populations, i.e. considerations of the metallicity profile in galactocentric radius $\feh(R, t)$ or $Z(R,t)$ at a given point in time $t$. However, for an adequate discussion, one has to take into account the full star formation history and re-distribution of stars.
We have to relate the abundances found at the time of our measurement $t_m$ at position $R$ to the past star formation all over the disc. We let stellar populations originate at angular momenta $\Lzo$, which correspond to the radii $R_0$ of a circular orbit; however, using $\Lzo$ is more natural to radial migration and leaves the door open to account for the slight adiabatic shrinking of the galaxy.

Different measurements of metallicity profiles are all related but logically distinct. The basis of the stellar profiles at later times is the original, time-dependent metallicity profile in the cold ISM, $Z_b(\Lzo, t)$, which we will refer to as ''ISM gradient.`` The time-integral of the ISM gradient times the star-formation rate gives then the ''star formation averaged metallicity gradient in the star-forming gas'', which would be a naive predictor for the stellar metallicity profiles at a later time, but is altered by stellar migration, described by a time-dependent mapping (Blurring + Migration) $T = B \cdot M$ between original angular momentum and final position. Throughout this section, we will use abbreviated notation for all differentials, i.e. $\d_R \equiv \d / \d R$ and $\partial_R \equiv \partial / \partial R$

The expectation value for the metallicity of stars found at a certain radius $R$ at time $t_m$ can be described by
\begin{equation}\label{eq:root}
\begin{array}{ll}
\left<Z(R,t_m)\right> = W(R)\iiint & Z_b(\Lzo, t)\; \SFR(\Lzo, t) \;M(\Lzo, \Lz, t)  \\
 & \qquad\quad \times\, B(\Lzo, \Lz, R, t) \;\;dt\,d\Lz\,d\Lzo , \\
\end{array}
\end{equation}
with normalisation $W$, time of birth of each population $t$, original angular momentum $\Lzo$, angular momentum $\Lz$ at time of measurement $t_m$, and current position $R$. The time integral runs from the birth of the disc ($0$) to $t_m$.

The four terms under the integral sign are:
\begin{itemize}
\item $Z_b(\Lzo, t)$, the metallicity of the stars born with angular momentum $\Lzo$ at a time $t$, inherited from the star-forming ISM at the radius $R_0$ of the circular orbit with angular momentum $\Lzo$.
\item $\SFR(\Lzo, t)$, the star formation rate at radius $R_0$ and time $t$. 
\item $M(\Lzo, \Lz,t)$, the transition matrix in angular momentum. Stars born with angular momentum $\Lzo$ have a probability $M(\Lzo,\Lz,t)$ to migrate to the current angular momentum $\Lz$, i.e. 'churning'.
\item $B(\Lzo,\Lz,R,t)$ is the probability of finding a star with angular momentum $\Lz$ at a radius $R$. This describes 'blurring': by the excursions along their orbits, stellar populations with current angular momentum $\Lz$ are distributed over a range of radii and altitudes $(R, z)$. The term also depends on $t$ and $\Lzo$. 
\end{itemize}
We note that for matching observations, all equations would have to carry a selection function, which will depend on the radius of birth (and hence metallicity), the age of each population, and for most surveys on the current position in the galaxy. This will again alter the observed metallicity profiles. For this paper, the accounting for selection, would complicate the picture without providing any further insight.

To simplify equation (\ref{eq:root}), we contract churning and blurring: no other variable depends on the current angular momentum of the stars, so we can execute the integral over $\Lz$ and obtain the new map $T(\Lzo, R, t)$, which is the probability that a star born at $(\Lzo, t)$ is now at position $R$
\begin{equation}\label{eq:Zroot}
\left<Z(R, t_m)\right> = W \iint Z_b(\Lzo, t) \,\SFR(\Lzo, t) \,T(\Lzo, R, t)\;\, \d t \,\d \Lzo
\end{equation}
The normalisation $W(R, t_m)$ is simply the number of stars present at $R$, i.e.
\begin{equation}\label{eq:W}
W(R) = \left(\iint \SFR(\Lzo, t) T(\Lzo, R, t)\;\, \d t \,\d \Lzo \right)^{-1} .
\end{equation}
Note that mass conservation normalises $T$ such that:
\begin{equation}\label{eq:Tnorm}
\int T(\Lzo,R,t) \;\,\d R = 1
\end{equation}
The task is to understand today's metallicity gradient $\d / \d R\left<Z(R, t_m)\right>$. However, we did not find it very instructive to directly differentiate equation (\ref{eq:Zroot}). To gain a better intuition, we introduce the normalised star formation history at each radius/angular momentum
\begin{equation}\label{eq:normalise}
g_{\star}(\Lzo, t) = \SFR(\Lzo, t)(S_g)^{-1} = {\SFR(\Lzo, t)}\left({\int \SFR(\Lzo, t)\;\,\d t}\right)^{-1} .
\end{equation}
where $S_g(\Lzo)$ is the time-integrated star formation density at some original angular momentum $\Lzo$. Now
\begin{equation}
\left<Z\right> = W \iint Z_b g_{\star} S_g T \;\, \d t \,\d\Lzo
\end{equation}
We can further absorb the local normalisation $W(R)$ into $T$, resulting in a new $T'$, so our equation reads:
\begin{equation}\label{eq:Znew}
\left<Z\right> = \iint Z_b g_{\star} S_g T' \;\, \d t \,\d\Lzo
\end{equation}
Note that instead of the old normalisation condition for $W$ and the mass conservation on $T$, we now have the normalisation condition
\begin{equation}
\iint g_{\star} S_g T' \;\, \d t \,\d\Lzo= 1 .
\end{equation}
We can therefore straightforwardly interpret $(g_{\star} S_g T')(\Lzo, R, t)$ as the normalised distribution of birth angular momentum and birth dates of all objects now found at radius $R$.

Differentiating equation (\ref{eq:Znew}) with the help of Leibniz' rule, we obtain 
\begin{equation}\label{eq:drZ}
d_R \left< Z \right> = \iint Z_b g_{\star} S_g \partial_R (T') \;\, \d t \,\d\Lzo .
\end{equation}
This is interesting, because the classic way of thinking about metallicity distributions as the cumulated abundance gradient in the star forming gas does not appear here (no spatial derivative of $Z_b$). The abundance gradient in the present stellar populations is determined by the distribution of origin of stars at a given radius (both in time and initial angular momentum). This depends on the star formation and enrichment history, as well as the radial re-distribution. 

To assess the situation, we need to link this equation to the original gradients, i.e. we need to recover a derivative in the radius of birth, or better original angular momentum $\Lzo$:
we split the radial mixing term $T = T_1 + T_2$ into a ``trivial/well-behaved'' part $T_1$ that does not change the re-distribution function with radius, i.e. $T_1(\Lzo(R_1), R_2, t) = T_1(\Lzo(R_1 + \delta R), R_2 + \delta R, t)$, and a part $T_2$ that does change with $\Lzo$ (and covers the more complicated/higher order aspects of radial mixing). With the simple assumption of a constant rotation speed $\Vc$, the condition for $T_1$ implies:
\begin{equation}\label{eq:partT1}
\partial_R T_1 = - \Vc \partial_{\Lzo} T_1
\end{equation}
We define $T'_1$ just as before by absorbing $W(R)$ into $T_1$. Inserting eq.(\ref{eq:partT1}) into equation (\ref{eq:drZ})
\begin{equation}
\begin{array}{lll}
\d_R \left<Z\right> &=& - \iint Z_b g_{\star} S_g (\Vc \partial_{\Lzo} T'_1 - T_1 \partial_R W) \;\, \d t \,\d\Lzo + \\
&& + \iint Z_b g_{\star} S_g \partial_{R} T'_2 \;\, \d t \,\d\Lzo $,$
\end{array}
\end{equation}
where the second term in the first line comes from the definition of $T'_1$.
Nicely $S_g = 0$ both for $R = 0$\footnote{This is fulfilled by definition. $S_g(R) = \int_0^{2\pi}{\Sigma_{\star}(R,\phi)d\phi}$ where $\Sigma_{\star}$ is the time-integrated star formation surface density, which is by construction finite. If we formulated this in $\Lzo$, the same argument would hold as long as $\lim\limits_{R \to 0} \Vc(R) > 0$.} and $R \to \infty$, so that we can integrate by parts, i.e.
\begin{equation}
\iint Z_b g_{\star} S_g \left(\Vc \partial_{\Lzo} T'_1\right) \, \d t \d\Lzo = -\iint T'_1 \partial_{\Lzo} (Z_b g_{\star} S_g)\, \d t \d\Lzo
\end{equation}

Regrouping our terms from the first line, we obtain
\begin{equation}\label{eq:gradbasis}
\begin{array}{lll}
\d_R \left< Z\right> &=& \iint \Vc S_g T'_1 (g_{\star}\partial_{\Lzo} Z_b + Z_b \partial_{\Lzo} g_{\star}) \;\, \d t \,\d\Lzo\;+\\
&& + \iint Z_b g_{\star} (\Vc T'_1 \partial_{\Lzo} S_g + S_g T_1 \partial_R W) \;\, \d t \,\d\Lzo \;+ \\
&& + \iint Z_b g_{\star} S_g \partial_{R} T'_2 \;\, \d t \,\d\Lzo
\end{array}
\end{equation}

We will see that this has broken down today's stellar metallicity gradient profile into three basic concepts: the star-formation averaged metallicity gradient (first term), and two terms widely neglected in the literature: the impact by SFR profile changes/SFR part of the gradient (second term), and some rather annoyingly convoluted terms caused by migration (second and third line).

The first line links directly to the original star formation history, but it consists of two terms: The inherited star formation weighted metallicity gradient $g_{\star} \partial_{\Lzo} Z_b$ and the metallicity weighted star-formation gradient $Z_b \partial_{\Lzo} g_{\star}$. We can intuitively see that these two terms will be of the same order and in most cases have opposite signs. The first term is typically of order $-0.05 \dex/(\kpc \Vc)$, while the second term links $\sim 1 \dex$ of enrichment with time to the temporal change of the star formation rate. This temporal change (in particular in the inner disc regions) in the fraction of metal-poor stars formed early on, will easily exceed $\gtrsim 0.1$ between the inner and outer radii, hence giving positive radial contribution of order $ \sim 0.1 \dex/(\kpc \Vc)$ to the stellar metallicity gradient. In simple words: if there is strong inside-out formation on a timescale comparable to the enrichment timescale, and if a significant number of stars are formed during that phase, the overall metallicity gradient of the stellar populations in a galaxy can be inverted. $T'_1$ in the first line fulfils the zeroeth order contribution of radial mixing: it makes the gradient term from the first line just the average of all regions contributing to the stellar populations found at $R$.

The second line in eq.(\ref{eq:gradbasis}) balances the density gradients in the disc. If there is no radial mixing, then $T'$ approaches an identity map, i.e. $T'_1 \to \delta(\Lzo - R \Vc)$. In that case $W \to S_g^{-1}$ and the second line vanishes, as well as the third line, which summarizes the irregular part of radial mixing.

\begin{figure}
\epsfig{file=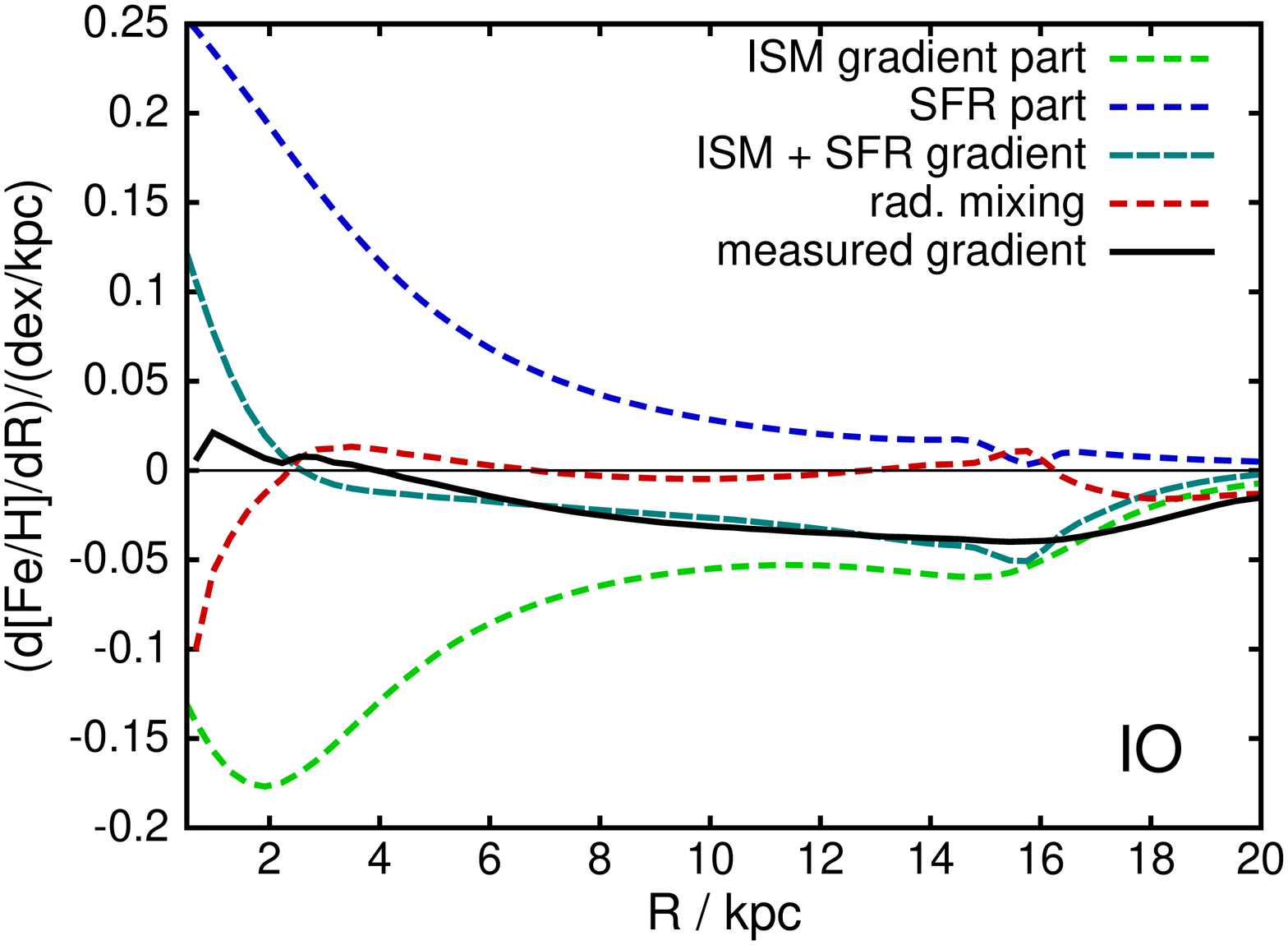,angle=-90,width=\hsize}
\epsfig{file=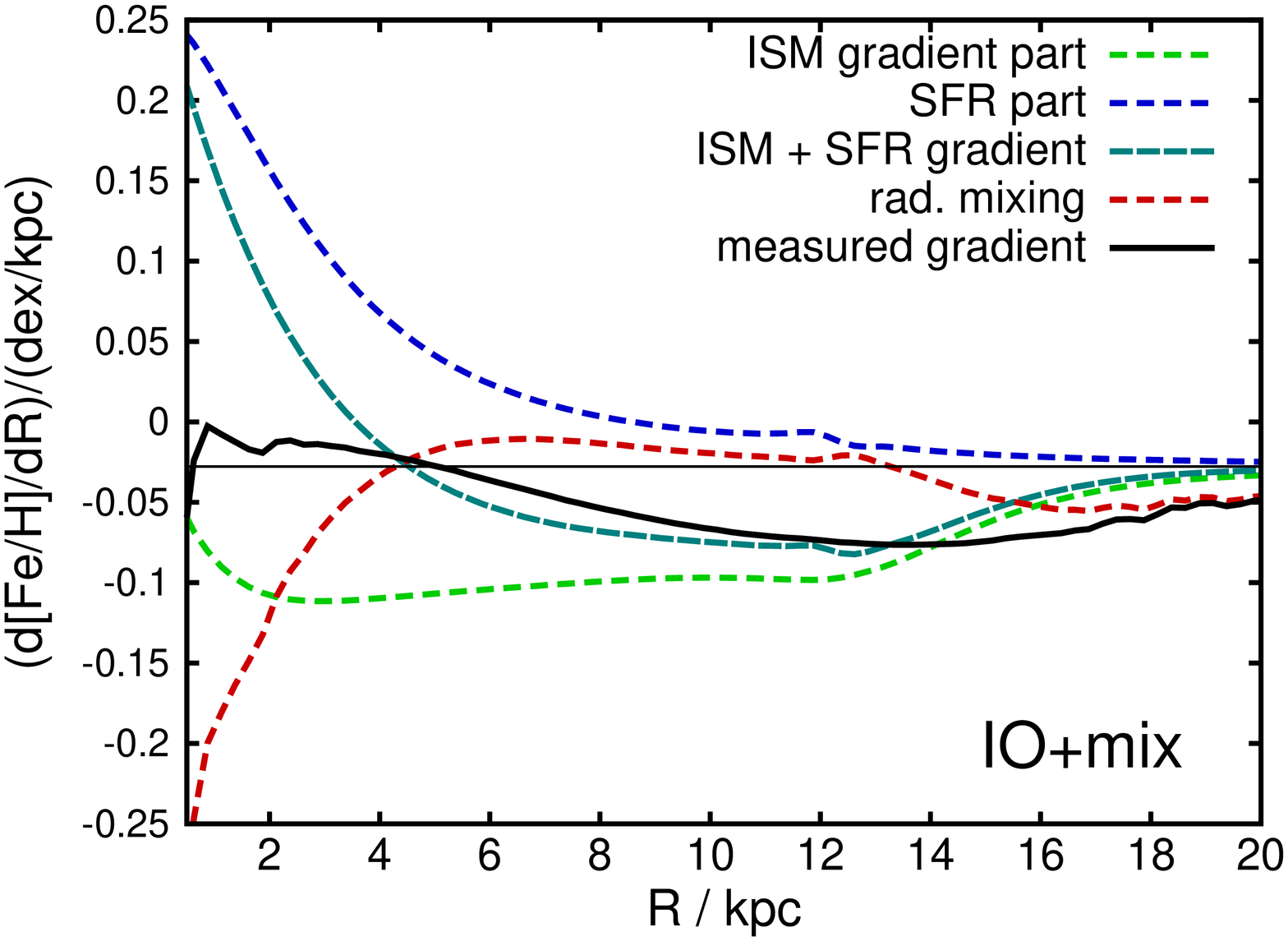,angle=-90,width=\hsize}
\epsfig{file=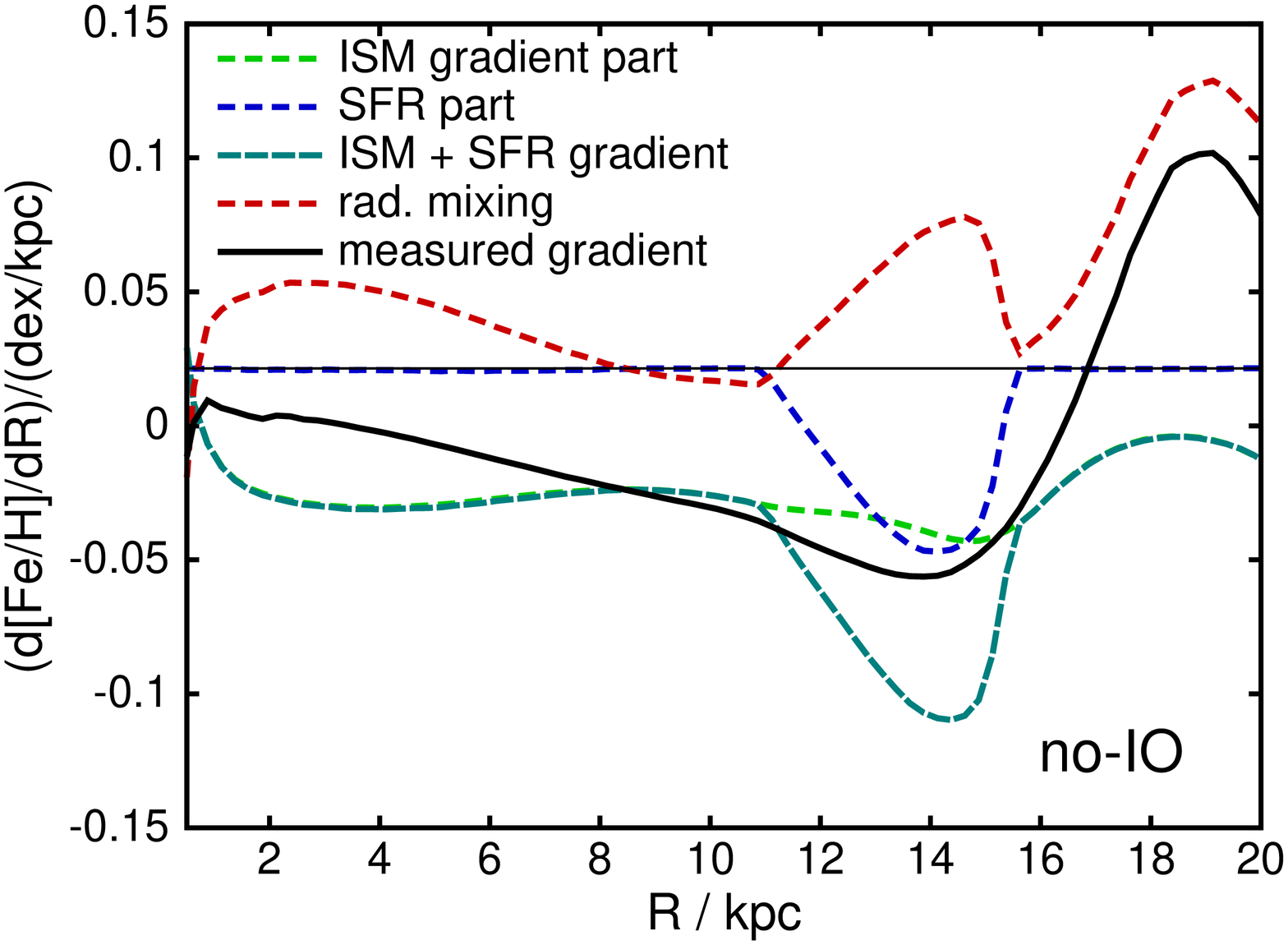,angle=-90,width=\hsize}
\caption{Metallicity gradients and their contributing parts in the standard simulation with inside-out formation (top), in the inside-out model with enhanced gas mixing (middle), and in the model with constant scale-length of the gas disc (bottom). The gradient without radial mixing is shown in dashed turquoise, split according to equation (\ref{eq:Lzt}) into the star-formation averaged past gradient (green) and the contribution from star-formation rate changes (blue). The red dashed line shows the change of gradient by radial mixing, resulting in the final (observable) metallicity gradient (black solid line). Note that the model without inside-out formation still experiences a change in the cut-off radius, the dominant effect being a shrinking of the cut-off radius, when the accretion rates of the disc diminish. This leads to a negative contribution from the SFR term, somewhat compensated by radial mixing of more metal-rich stars into the cut-off region.}\label{fig:gradan}
\end{figure}

Having understood the nature of the three major players for today's abundance gradients (time-dependent ISM gradient, SFR differences, radial mixing), we can now split them conceptually apart. In the absence of radial mixing the integral in $\Lzo$ in the first line of equation (\ref{eq:gradbasis}) disappears, so that we can divide
\begin{equation}\label{eq:Lzt}
\begin{array}{lll}
(\d_R \left<Z\right>) &=& (\d_R \left<Z\right>)_{\rm no mix} + (\d_R \left<Z\right>)_{\rm mix} \\
 &=& \color{green}{\Vc\int g_{\star} \partial_{\Lzo} Z \;\d t}\; + \color{blue}{\Vc \int Z\partial_{\Lzo}g_{\star} \;\d t} +  \color{red}{(\d_R \left<Z\right>)_{\rm mix}}  $.$
\end{array}
\end{equation}

\figref{fig:gradan} summarizes this, showing the three parts: i) the star formation-averaged metallicity gradient in the star-forming gas (green, first term of eq. \ref{eq:gradbasis}), ii) the change in the age-distribution of stars with radius (blue, second term in eq. \ref{eq:gradbasis}), and iii) changes via the radial mixing of stars (red, all other terms). The turquoise lines summarize the gradient without radial mixing, the black lines show the resulting gradient profiles in the observable stellar populations.

As discussed above, the inside-out formation (top panel of \figref{fig:gradan}) gives a strong positive contribution to the abundance gradient throughout the entire disc, which greatly weakens the gradient in the outer disc and even reverses the gradient at small galactocentric radius. 
Radial mixing, which is quite moderate in this model, tends to weaken the gradients, mostly by averaging between regions with large differences. 
The transition matrix in angular momentum, $M$, which describes the churning by giving the probability of having angular momentum $\Lz$ for a star with initial angular momentum $\Lzo$, will have larger off-diagonal terms for older stars, reflecting that the very metal-poor stars born at early times can migrate a bit further out than their metal-rich, younger counterparts. This effect is quite mild, if migration rates do not vary strongly with time, and if the angular momentum distribution is approximately conserved, in which case the short disc scale length limits outwards migration at very early times. The blurring term $B$ has a straight-forward effect on the gradient. We expect inner disc populations to be intrinsically hotter than their outer disc counterparts due to stronger heating, with another contribution from their older age. This means that inner disc populations get blurred towards larger radii. For MW like set-ups this should weaken any gradient inversion, but in particular near cut-offs in the disc it will play a dominant role. Interestingly, in the standard inside-out model two separate effects counteract each other: On the one hand, radial mixing weakens the central metallicity peak, i.e. gives a positive gradient contribution, but it also pushes older and kinematically hotter inner disc stars further out, giving a negative gradient contribution. 
We also note that it is a well-known fact in chemical evolution modelling that inside-out formation strengthens the negative metallicity gradient in the star-forming ISM \citep[][]{Prantzos00}. Steep radial metallicity gradients far in excess of $-0.1 \dex/\kpc$ have been observed in other galaxies \citep[e.g.][]{Jones10, Jones13}.

The middle panel of \figref{fig:gradan} shows the inside-out model with additional mixing in the gas phase. This particularly flattens the gradient in the inner disc and reduces the contribution of the inherited gradient, while the star-formation related term stays nearly the same. As a result, the disc without radial mixing shows a very strong gradient inversion in the central region at $R < 5 \kpc$. This is ameliorated by radial mixing.

For contrast, the bottom panel of \figref{fig:gradan} shows the simulation with constant scale-length of the gas disc. Accordingly the inner parts of the disc show no star-formation contribution to the gradient, so only the averaged gradient and radial mixing play a role. In this case, the radial mixing contribution becomes significantly positive. The big negative excursion in the star-formation rate related part around $12-15 \kpc$ derives from the cut-off in the Schmidt-Kennicutt law, which very quickly expands to $\sim 15 \kpc$ after the start of the simulation, and then slowly travels inwards due to the declining accretion rates. Radial mixing partly compensates this by bringing in more metal-rich stars, and in the outermost zone of this model, where densities vanish, even overcompensates due to the larger kinematic heat of inner disc stars.

\begin{figure}
\epsfig{file=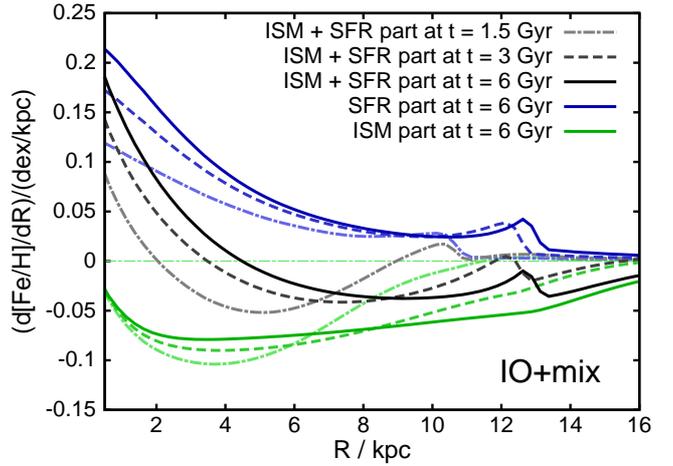,angle=-90,width=\hsize}
\caption{Time evolution of the ISM (green lines) and SFR (blue) contributions (first two terms of equation (\ref{eq:Lzt}) in the model with inside-out formation and enhanced gas mixing to the radial stellar metallicity gradient (black lines) without accounting for stellar radial mixing. Results are shown at $t = 1.5 \Gyr$ (dash-dotted lines), at $3 \Gyr$ (dashed) and at $6 \Gyr$ (solid lines).}\label{fig:gradansold}
\end{figure}

In \figref{fig:gradansold} we show the evolution of the impact of inside-out formation on stellar radial metallicity gradients (different line-types mark the different ages). We show in green the star-formation averaged radial metallicity gradient, while blue lines depict the inside-out contribution to $\d \feh / \d R$, and black lines show the resulting radial metallicity gradient in the stellar populations before accounting for stellar radial mixing. With our parameters, the impact of inside-out formation grows significantly in time up to an age of about $6 \Gyr$, which is mostly caused by the longer baseline of inside-out growth and the fact that the difference in the star-formation histories $g_{\star}$ grows with time. Even at the earliest times or respectively at high redshift, the inside-out growth exerts a strong positive bias of order $0.1 \dex/\kpc$, resulting in a mildly positive $\d \feh/ \d R$. 

\begin{figure*}
\epsfig{file=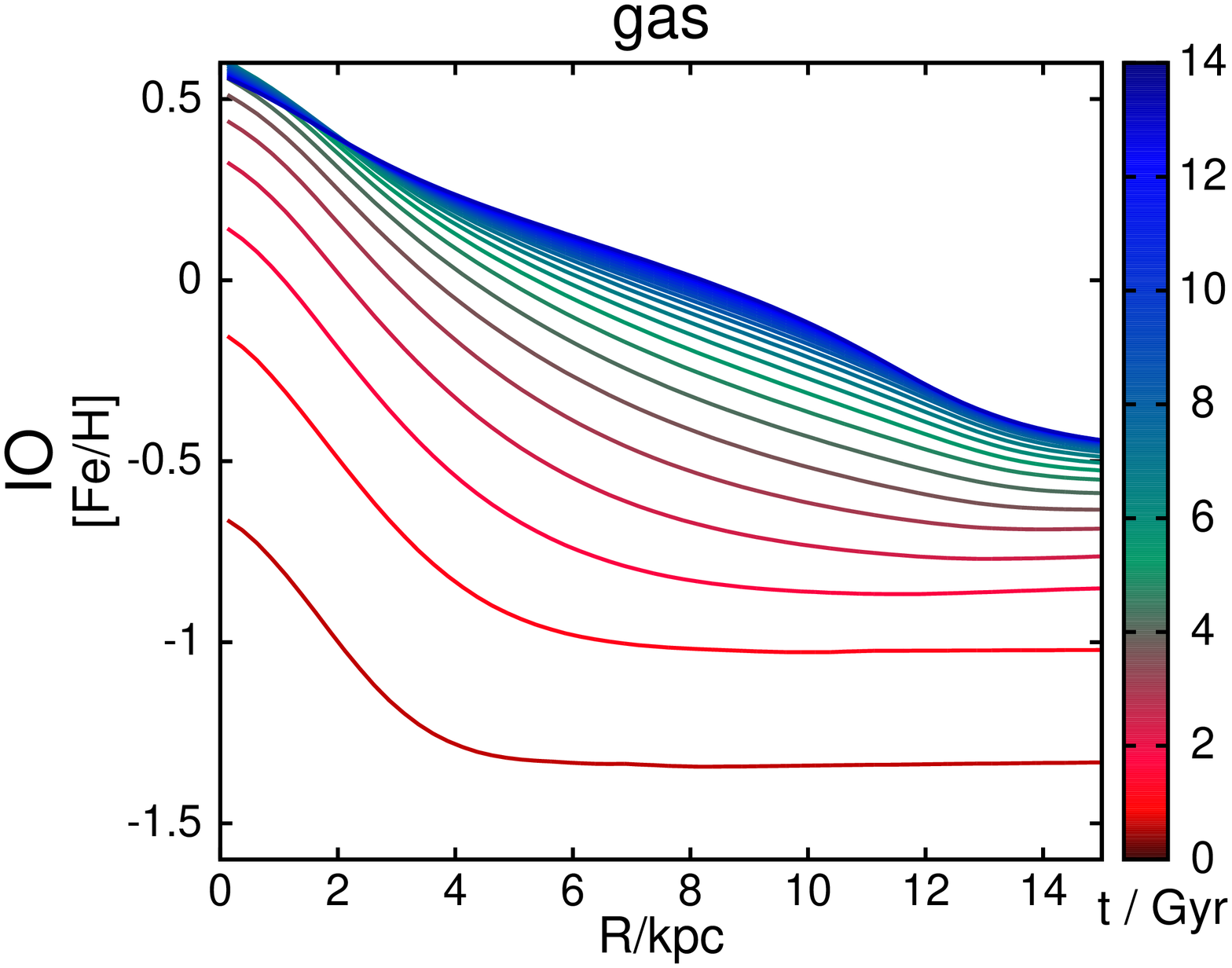,angle=-90,width=0.325\hsize}
\epsfig{file=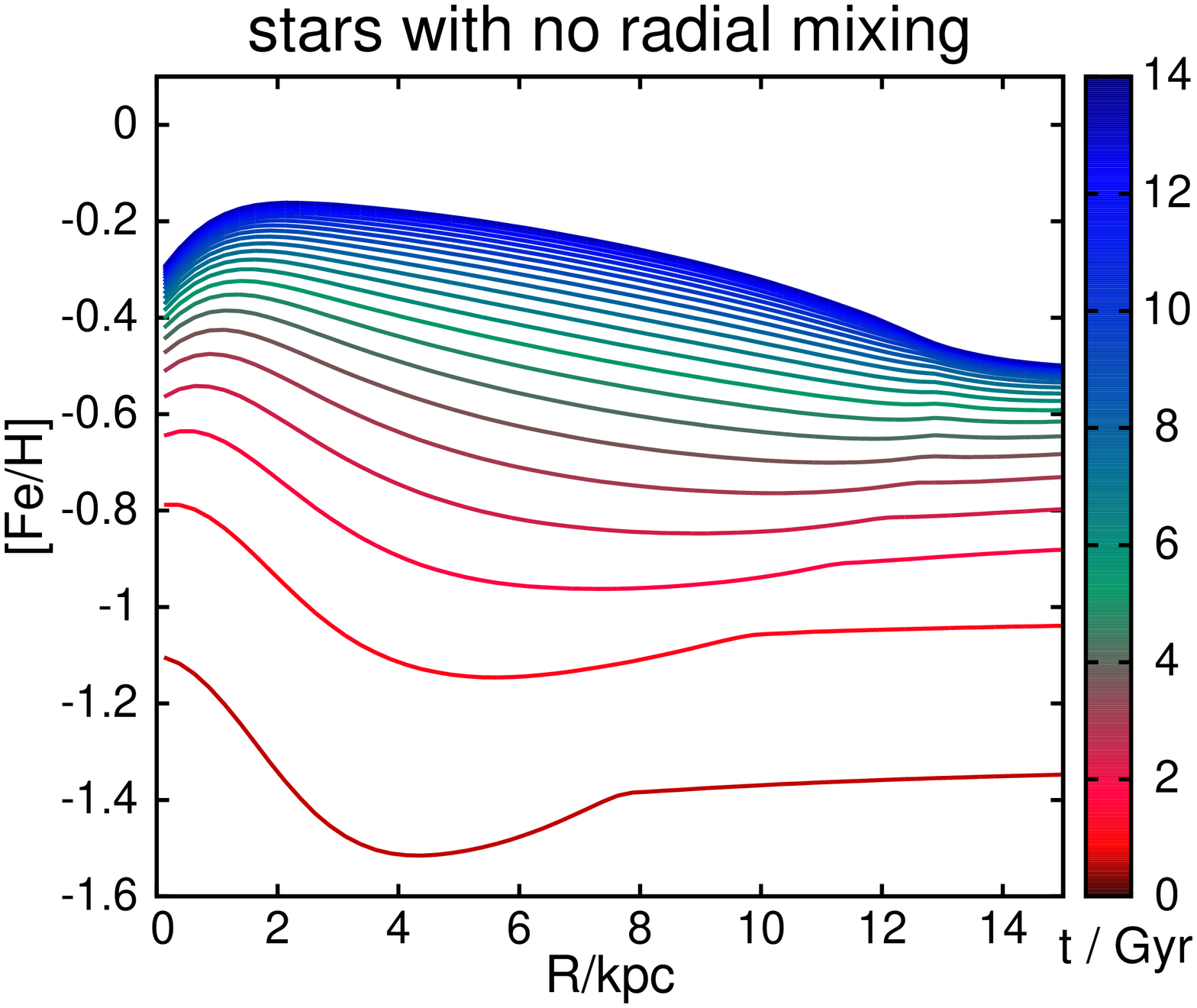,angle=-90,width=0.325\hsize}
\epsfig{file=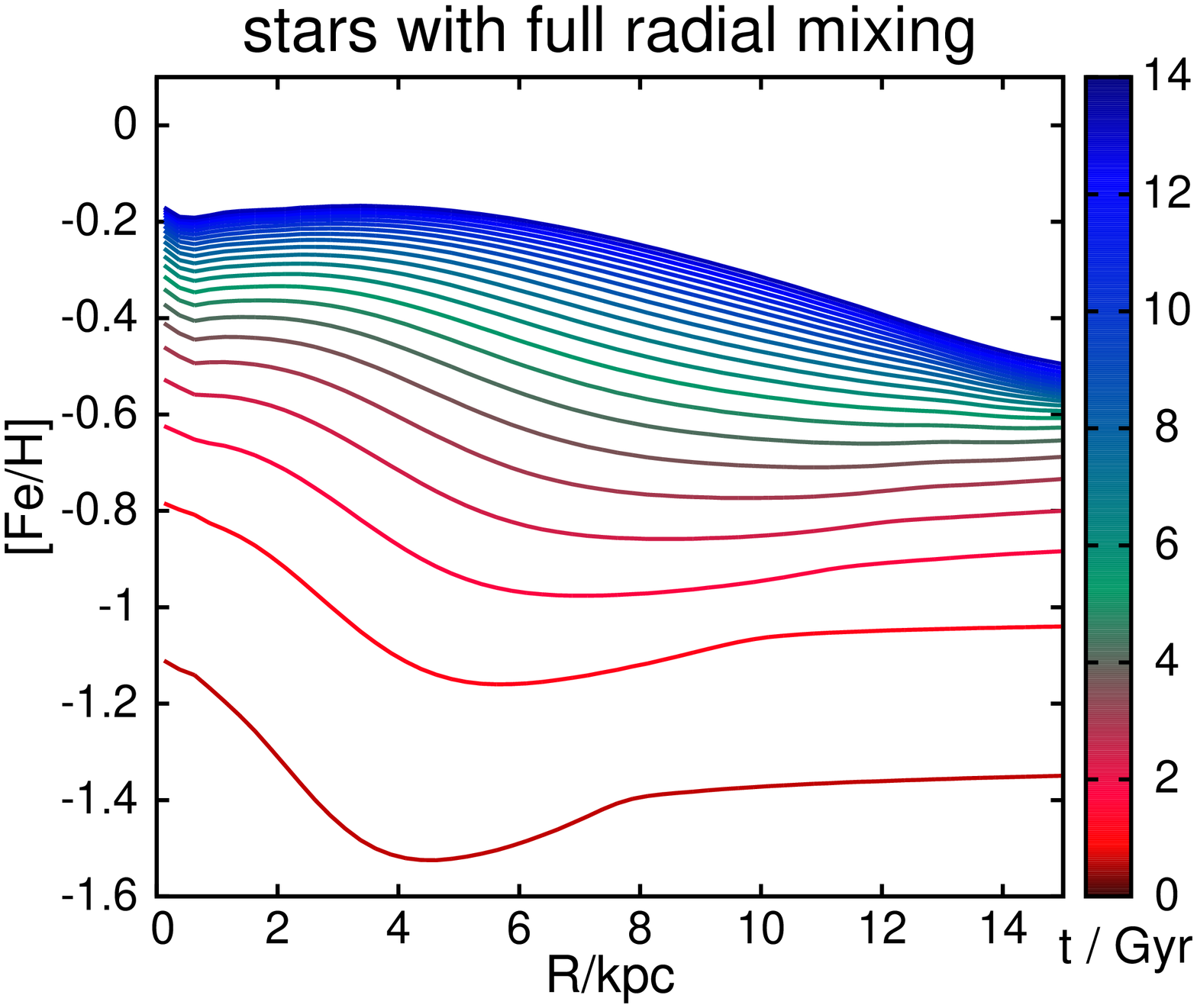,angle=-90,width=0.325\hsize}
\epsfig{file=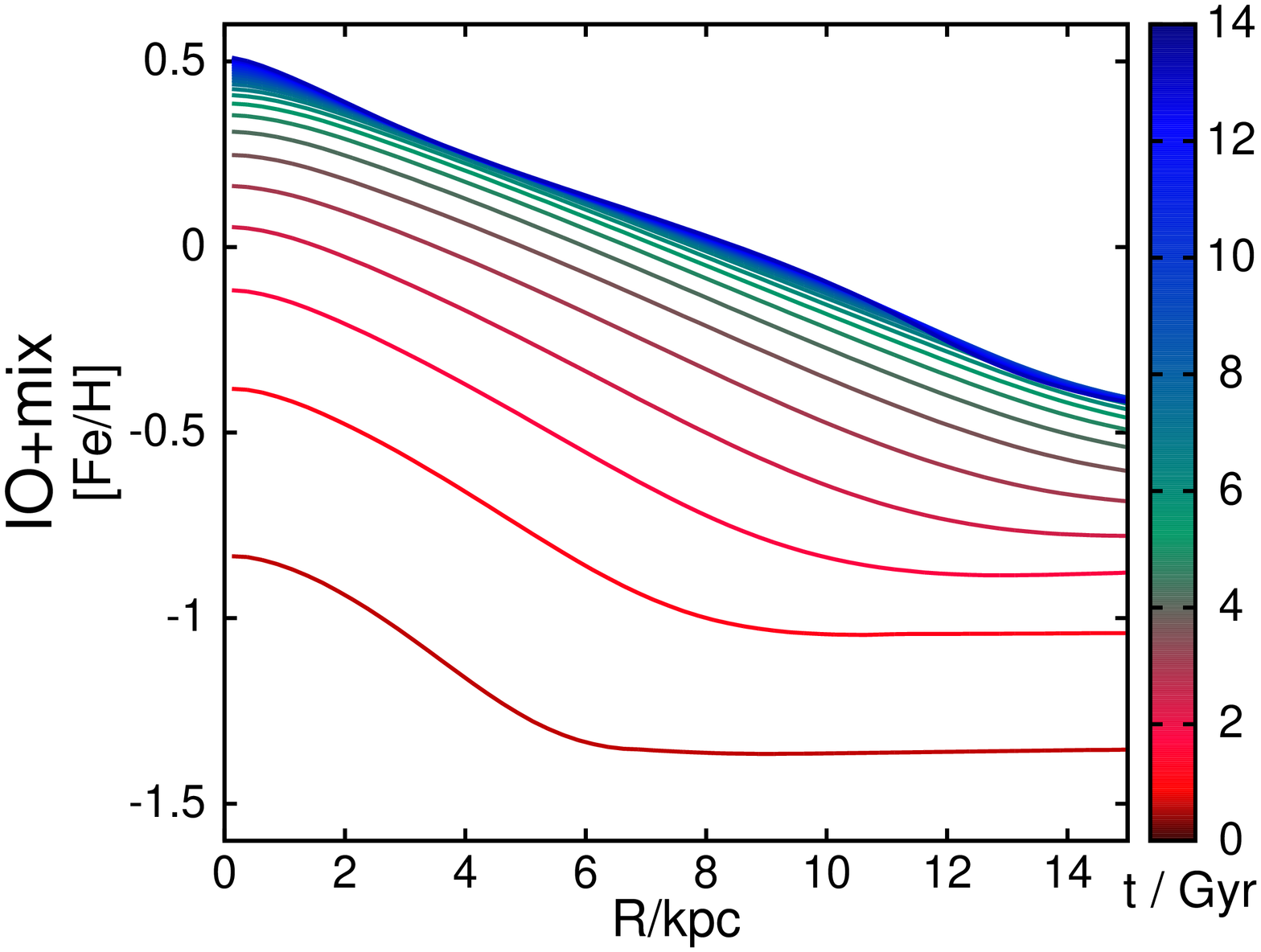,angle=-90,width=0.325\hsize}
\epsfig{file=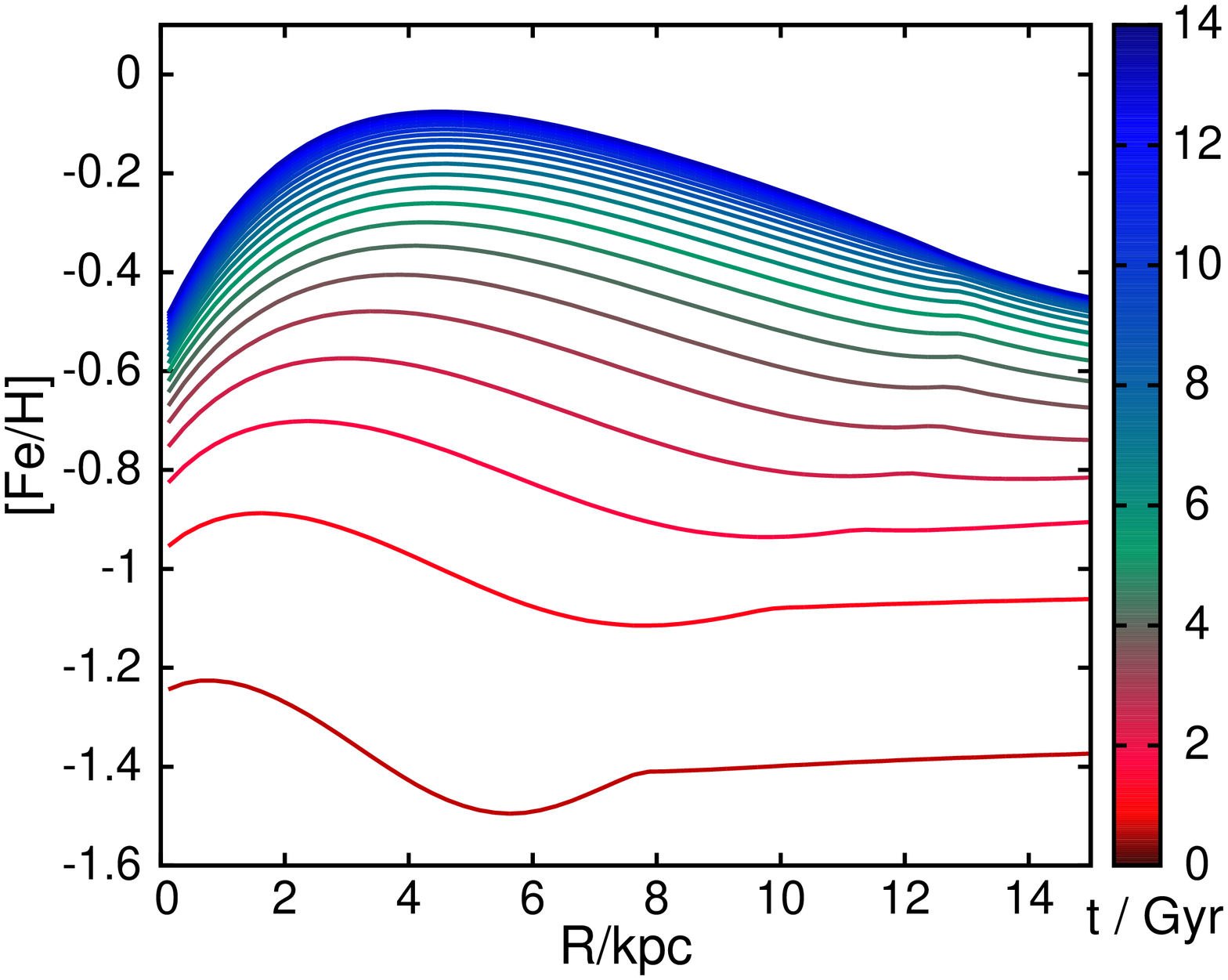,angle=-90,width=0.325\hsize}
\epsfig{file=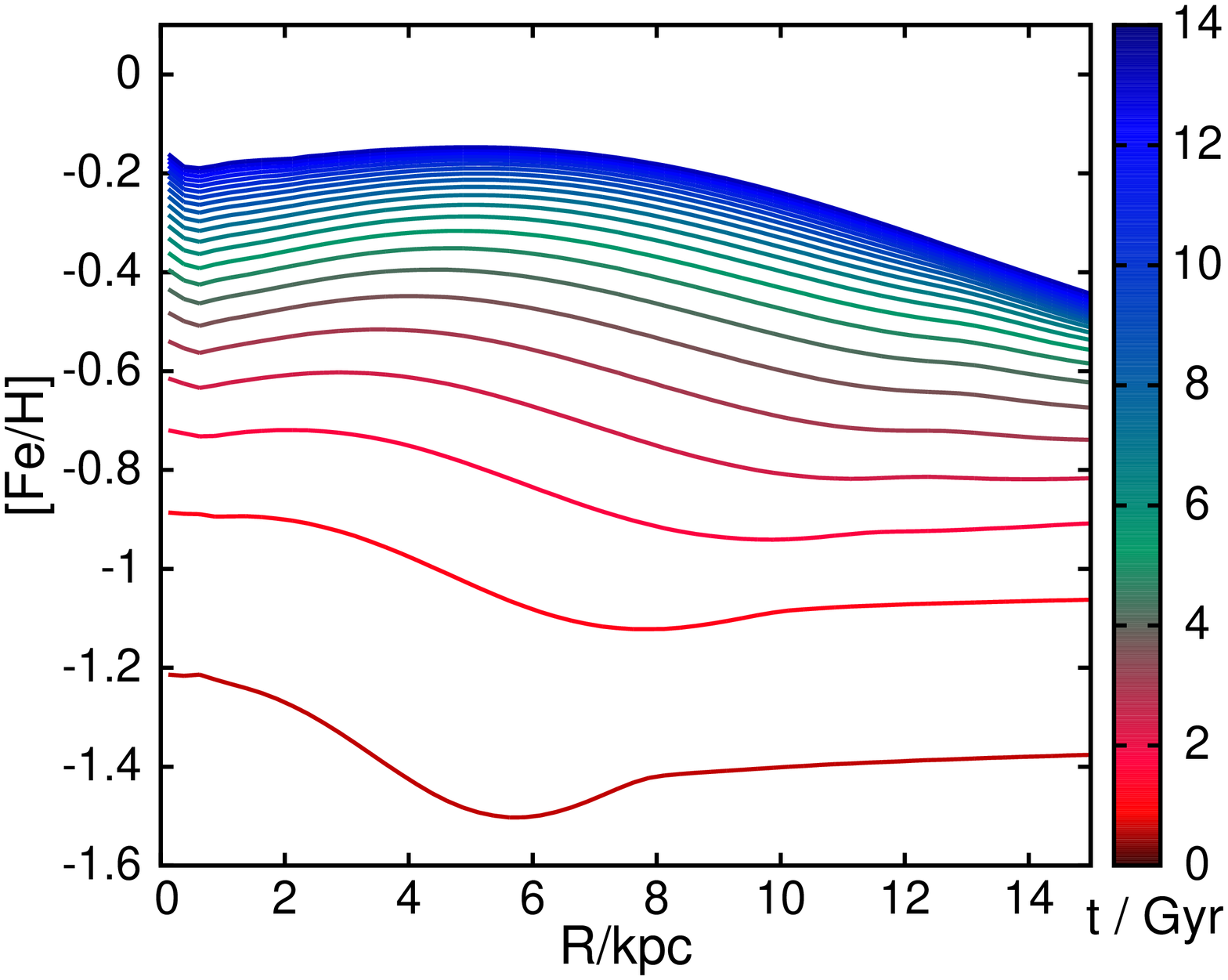,angle=-90,width=0.325\hsize}
\epsfig{file=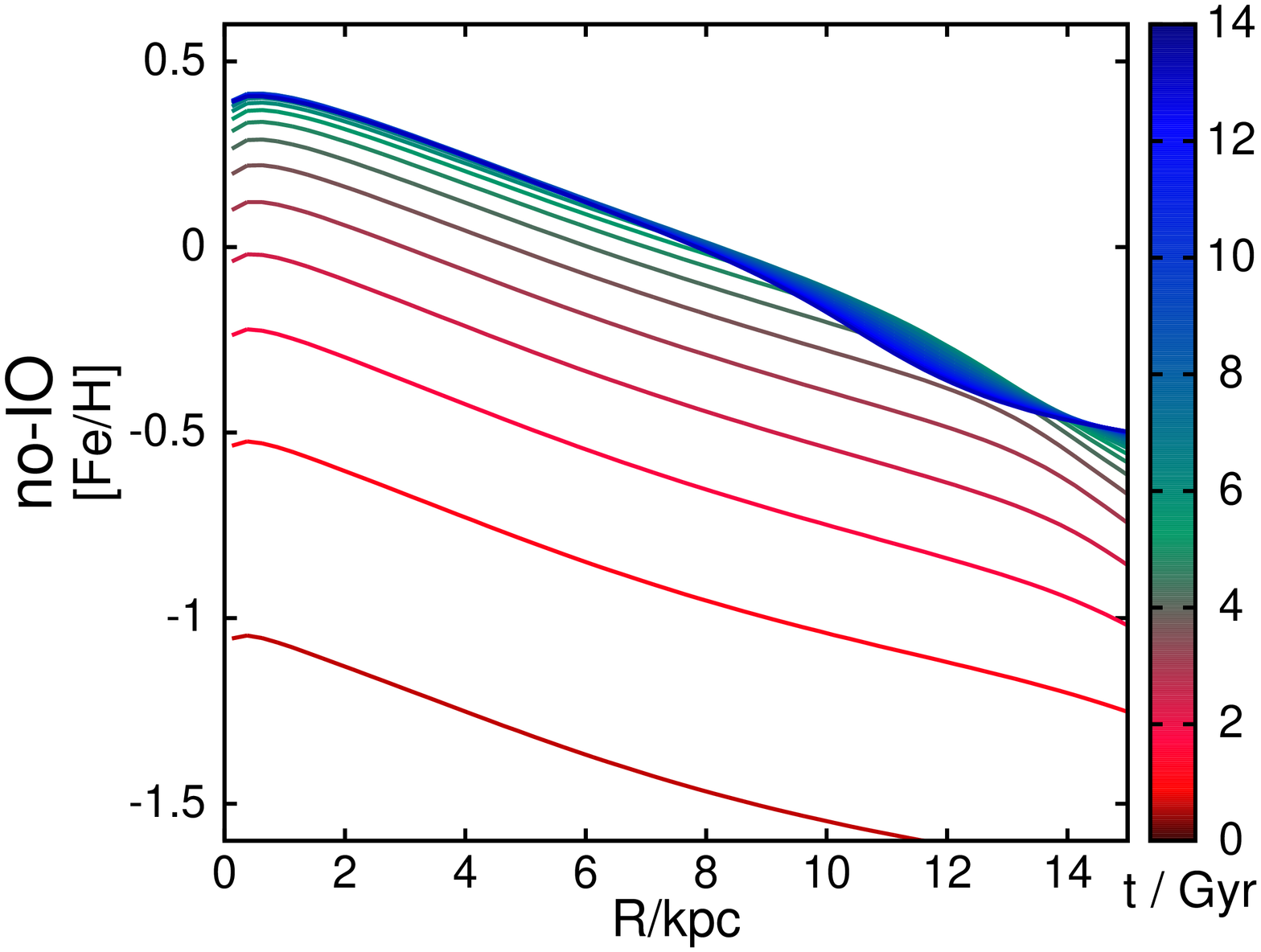,angle=-90,width=0.325\hsize}
\epsfig{file=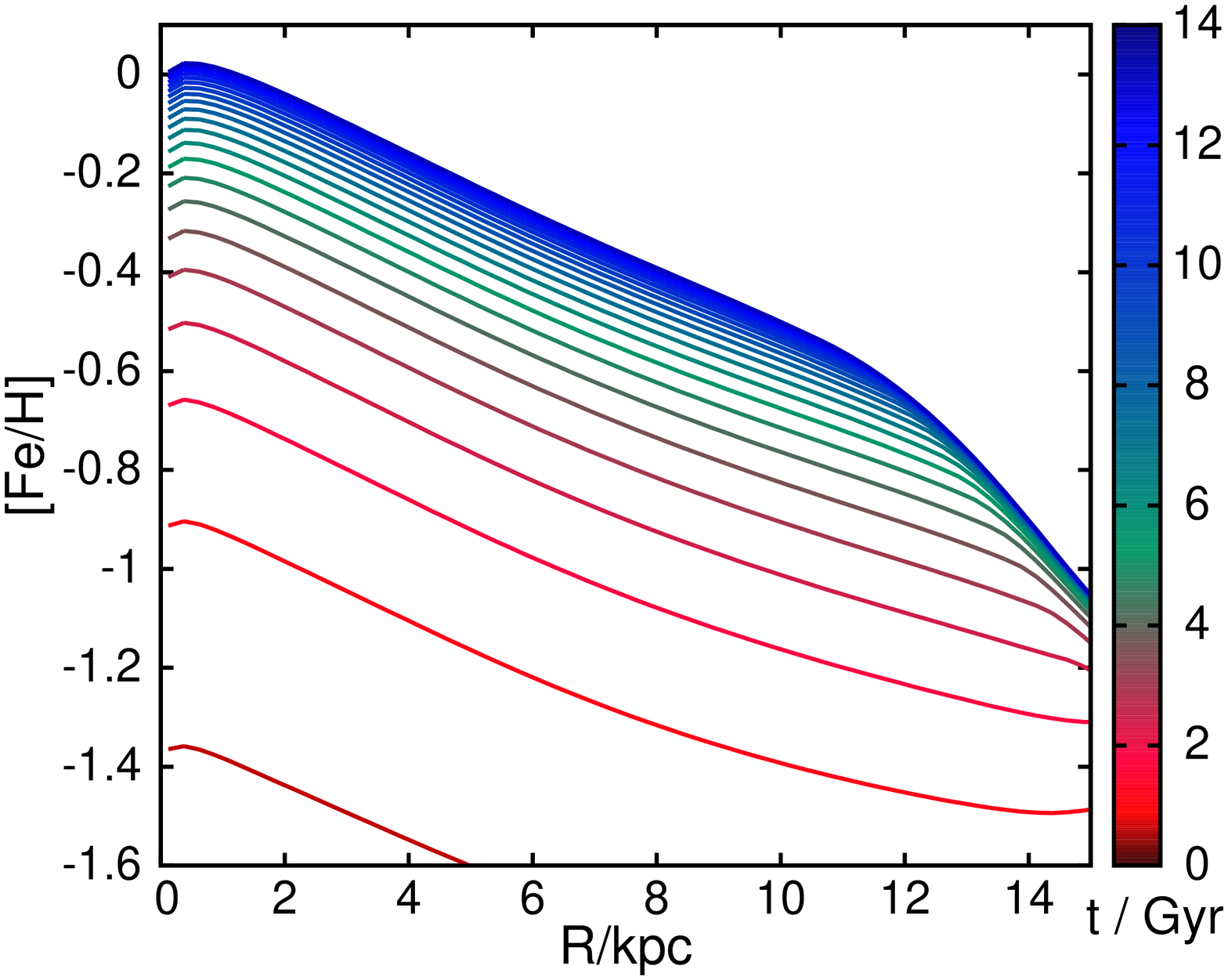,angle=-90,width=0.325\hsize}
\epsfig{file=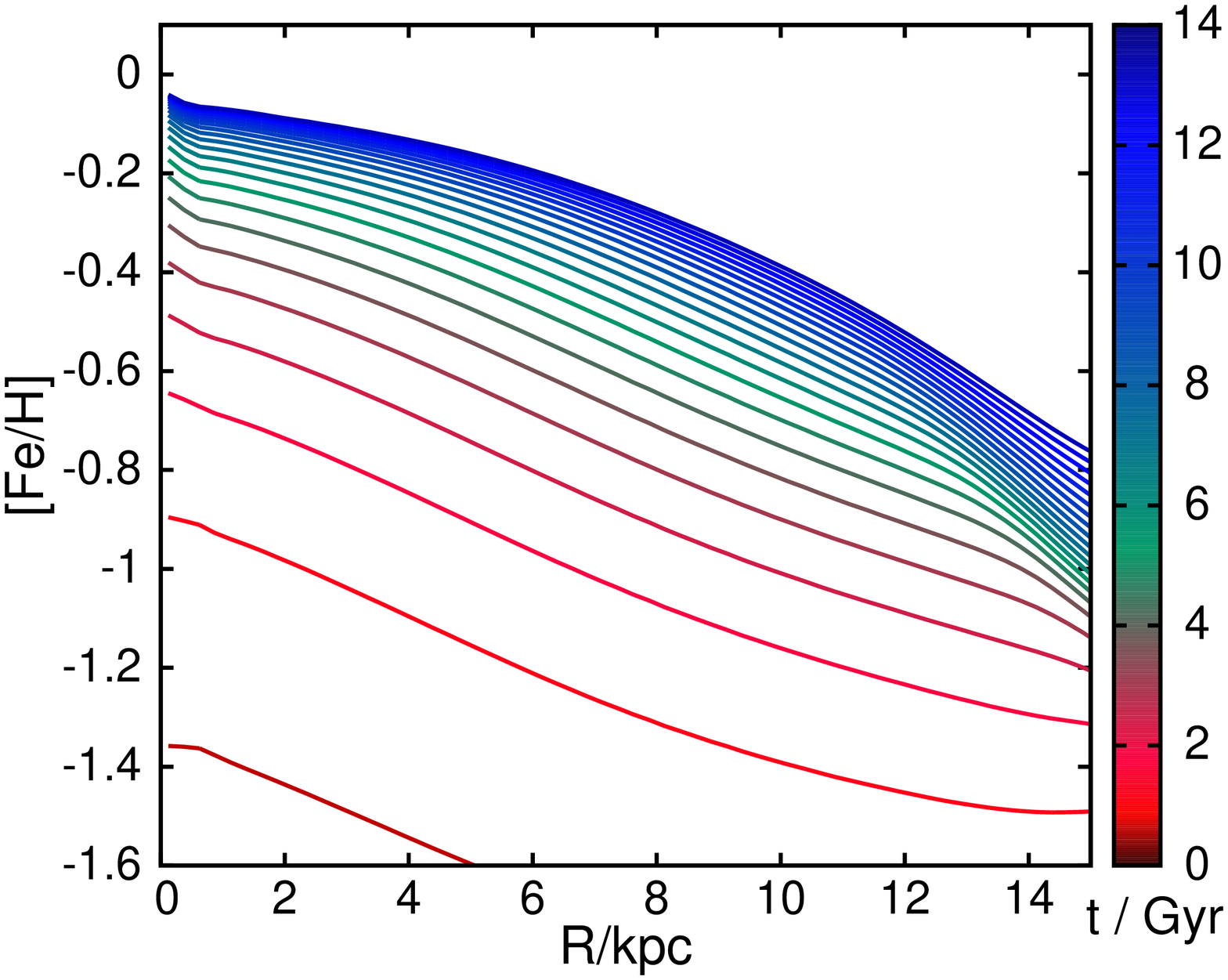,angle=-90,width=0.325\hsize}
\caption{Left column: Metallicity profiles in the star-forming gas at different times in the simulation (colour coded). We show the same models as before, i.e. the "standard" inside-out model on top, the inside-out model with additional gas mixing in the centre, and the model with constant scale-length in the bottom row. The centre column shows each model the average metallicity of stellar populations without accounting for radial mixing, while the right-hand column accounts for radial mixing.}\label{fig:metprofiles}
\end{figure*}

\subsection{Inside-out growth and observed stellar metallicity profiles}

\figref{fig:metprofiles} compares radial metallicity profiles of the cold (star-forming) gas (\emph{left} column) and of the stars (centre and right columns) at various different times in the simulation. The \emph{middle} column shows the stellar metallicity profiles if we ignore any migration effects and the \emph{right} column depicts the profile including the effects of typical blurring and churning over time. The main effects of the blurring and churning are an averaging between the metallicity gradients of neighbouring rings (see the factor $T'_1$ in the first line of eq.\ref{eq:gradbasis}), to exert a negative bias on the radial metallicity gradient by moving the oldest (and more metal-poor) stars to larger radii, and with strong mixing to reduce (though not completely remove) radial gradients in the disc. The top row displays the results for our standard inside-out formation model (\textsc{io}), the middle row shows that from the model with additional gas mixing (\textsc{io+mix}), and the bottom row shows the reference model without inside-out formation (\textsc{no-io}). 

Note, in particular, that in the inside-out models at early times the abundance profile in all components is steeper than in model \textsc{no-io}. This is due to the shorter scale-length of the inner disc, the initially larger star formation efficiencies, and the faster decline of the star formation rate in the central regions, which shifts the balance of yields to freshly accreted gas. However, while the simulation without inside-out formation shows persistently negative $\d \feh/\d R$ at all times, the inside-out formation models show a flattening of the (initially steeper) stellar abundance gradients with time, up to an inverted gradient in the central regions at later times. 

The middle row shows our inside-out model with weakened radial metallicity gradient produced by strong radial mixing of both gas phases. This mixing is still counteracted by the inflow towards the disc centre, so a significant, albeit shallower gradient remains. It is not of interest to discuss if such a mixing is physical or not. Since there are many weakly constrained factors that influence abundance gradients in galaxies, we just use this one to achieve a weaker gradient without risking a gradient inversion within the star forming gas. As a result, we see now that in the inner regions of the galaxy, the present abundance gradient in the stellar populations is dominated by the combined effect of rising metallicity with time and inside-out formation. The fact that more stars are formed at early times and lower metallicities becomes more important than the abundance gradient in the star-forming gas, which remains negative at all times. While radial migration again delivers a bias in the opposite direction, weakening the positive gradient, the inner regions still display the inverse gradient.

\begin{figure}
\epsfig{file=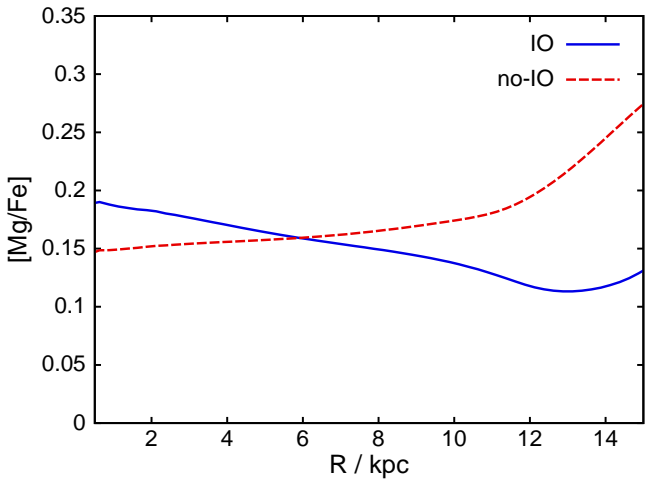,angle=-90,width=\hsize}
\caption{Stellar $\mgfe$ profiles of the standard model with inside-out formation (blue), and the model without inside-out formation (red).}\label{fig:mgfeprofiles}
\end{figure}

On a side-note we remark that inside-out growth very likely infers a negative $\d \afe \d R$ gradient. This has been noticed by \cite{Gibson13}. We would, however, like to discourage oxygen for these discussions: oxygen yields critically depend on the shell structure and hence in particular wind losses of massive stars (simply put, strong wind losses reduce the material available for helium burning), which in turn are related to metallicities \citep[see e.g. early models from][]{Maeder92}. This is less critical for heavier alpha elements like magnesium. As an example, the $\d \ofe / \d R$ gradient in SB09, was positive while any other alpha to iron element ratio in SB09 would be nearly flat in radius. \figref{fig:mgfeprofiles} displays the radial profiles for the $\mgfe$ ratio. Some recent observations find the expected different trends in magnesium and oxygen \citep[][]{McWilliam08, Genovali15}.
The model without inside-out formation (red line) displays a very weak positive $\d \mgfe / \d R$, which is mostly caused by a very weak metallicity dependence of magnesium vs. iron yields. Another contributing factor is the slight outwards expansion by blurring, which is stronger for the oldest, alpha-rich populations. The stronger rise beyond the cut-off is again caused by the retraction of the cut-off with slowing infall. As expected, in the model with inside-out formation (blue line), the inside-out formation dominates over the other effects, leading to a negative $\d \mgfe / \d R$, again exacerbated at the cut-off. There is a couple of other complications preventing easy interpretations of observations: the inside-out formation leads to a central depression of the $\afe$ values at intermediate to late times, which partially (and depending on the specific yield and flow models) compensates the inside-out effect. In addition, the exact behaviour depends also on the different redistribution of SNIa yields compared to SNII yields, loss rates and inflow abundances.

To summarize: We have shown that if inside-out evolution has a comparable timescale to the enrichment timescale in a galaxy, the stellar populations are liable to acquire an inverse (positive) radial gradient, even if the radial metallicity gradient in the star-forming gas is negative at all times.

\subsection{A source of inverse/positive metallicity gradients in the cold gas: Redistribution}\label{sec:redistr}

We have shown in the previous sections that inverse gradients in the stellar populations are a natural consequence of inside-out formation. However it is still important to discuss the possibility of inverse gradients in the star-forming gas itself. So far the two main explanations invoked for this possibility have been a direct accretion of cold metal-poor gas into the central regions \citep[][]{Cresci10} and the rather peculiar chemical evolution argument of \cite{Curir12}. In the first case it is unclear how accretion of metal-poor material would manage to pierce into the centre in a sufficient amount to offset the higher star formation efficiencies and prevent an inward flow through the disc (both of which drive a negative radial metallicity gradient). The second case \citep[][]{Curir12} produces a very small signal, and, while an intriguing idea, also depends on the specific and arbitrary changes in the star formation efficiencies applied in their underlying chemical evolution models. It can be shown analytically, and we have also confirmed in tests, that our chemical evolution model in its standard set-up does not produce any gradient inversion in the star-forming gas  when we remove radial flows and mixing from our model to match the classical evolution models of \cite{Spitoni11} and \cite{Curir12}, unless we impose significantly higher loss rates in the central regions (which is a different argument, see below).

However, the radial redistribution of metals within the galaxy does open a viable path to creating inverse gradients in the star-forming gas. Only a fraction of the stellar yields will be directly recycled to the cold star-forming gas phase (the direct enrichment parameter in our model), while half the yields or more end up in the warm and hot ISM, a part of them being expelled from the galaxy, a part of them lingering near the disc. The unhappy truth is that we know nearly nothing about this redistribution, apart from some information on the rotation of the coronal gas \citep[see e.g.][]{Marasco11, Marinacci11}. The angular momentum difference with the disc might not only drive inflow in the disc, but angular momentum and pressure from stellar yields might push in turn the coronal gas outwards. This can result in the enriched, hot material condensing back onto the disc at larger radii than its production. If this effect is strong enough, it can invert the metallicity gradient in the star-forming gas. 

We can identify three natural processes that would yield an early gradient inversion:
\begin{itemize}
\item Redistribution of the yields near the disc as detailed above. 
\item Loss of material/yields. The intense star formation of the central regions and potentially activity of the central black hole will drive a stronger outflow. Outflow from the central regions of disc galaxies (usually perpendicular to the disc plane) has been observed, for the MW see \cite{BH03}, for M82 see \cite{Shopbell98}. It is natural to assume a higher loss-rate for yields from central regions in particular at early times when the star formation rates are higher and more concentrated.
\item Re-accretion of enriched material. Little is known about the spatial distribution of yields from galaxies near the boundary to the IGM. It can be assumed though that a fraction of the enriched outflow mixes with the infalling material especially at high redshifts. If the timescale of re-accretion is shorter than the star formation timescale of the disc, the (re-accreted) material can quite easily obtain higher metallicities than the gas disc. Two factors help here: Due to the inflow through the disc, the outer disc will have more recently accreted material, i.e. an increasing metallicity of the accreted ISM contributes to an inverted metallicity gradient. More importantly, the inside-out formation of the disc demands that in relation to the present mass more fresh material is accreted onto the outskirts.
\end{itemize}

\begin{figure}
\epsfig{file=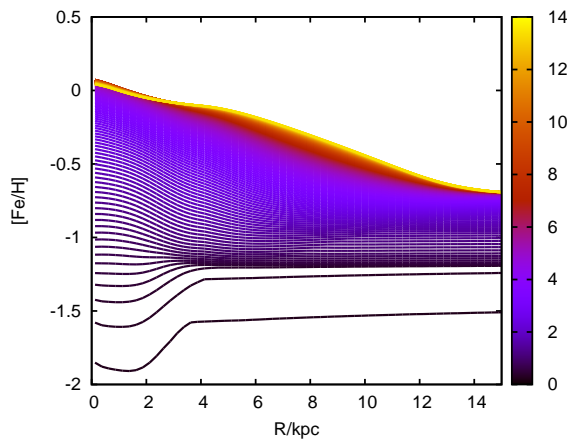,angle=-90,width=\hsize}
\caption{Metallicity profiles in the star forming gas for an alternate model with higher loss rates and re-accretion of one quarter of the lost material with the infall of fresh gas. With the inside-out growth, a lot of material is present in the inner rings, while the outer rings (beyond $\sim 4 \kpc$) experience quite rapid accretion of relatively more enriched material. This results in a gradient inversion of this model in the first $\sim 600 \Myr$. The model is plotted every $75\Myr$ for the first $\Gyr$ and then at multiples of $0.75 \Gyr$.}\label{fig:metprofalt}
\end{figure}

\figref{fig:metprofalt} shows an example of gradient inversion by a galactic fountain. As a parameter change from the standard model, we increase the fraction of lost material to $80\%$ in the outer regions and $90\%$ in the innermost $4 \kpc$. Instead of pre-setting the metallicity of the freshly accreted gas, we assume that one quarter of the lost gas is re-accreted, mixing with primordial gas, up to a maximum metallicity for the infalling gas of $\feh=-0.7$. It can be seen that during the first $\sim 0.6 \Gyr$ the radial metallicity gradient is inverted, due to the re-accretion of material at higher metallicities than the young disc. This inversion could last significantly longer and to higher metallicities with a more metal-rich inflow, as the re-accretion fraction might be higher, and we do not see an a priori restriction on the upper bound to the inflow metallicity.
At later times the gradient normalises. As pointed out above, this is just one of many natural factors that favour inverse metallicity gradients in young discs and that do not require any cold streams piercing into the central galaxy.

\begin{figure*}
\epsfig{file=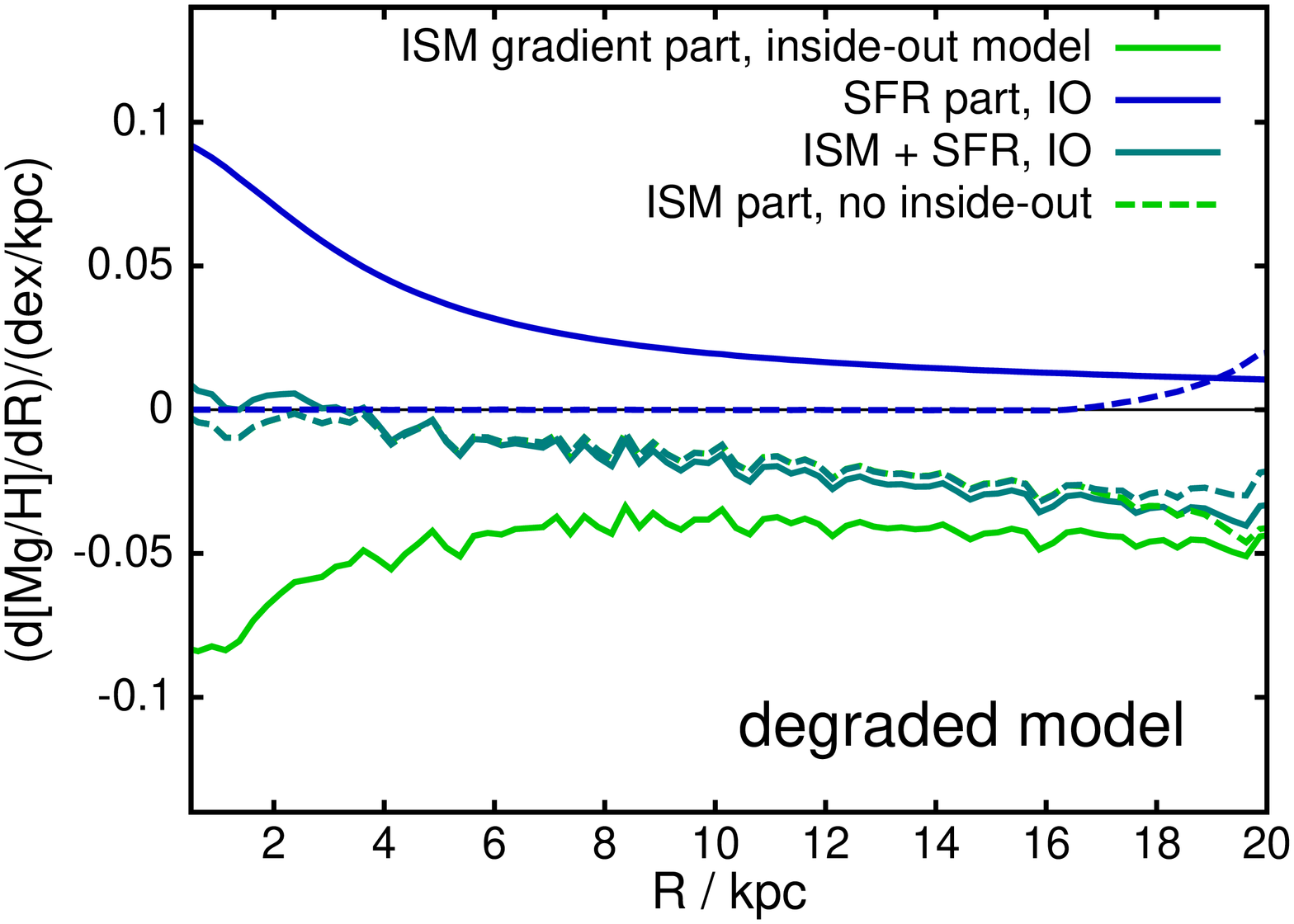,angle=-90,width=0.49\hsize}
\epsfig{file=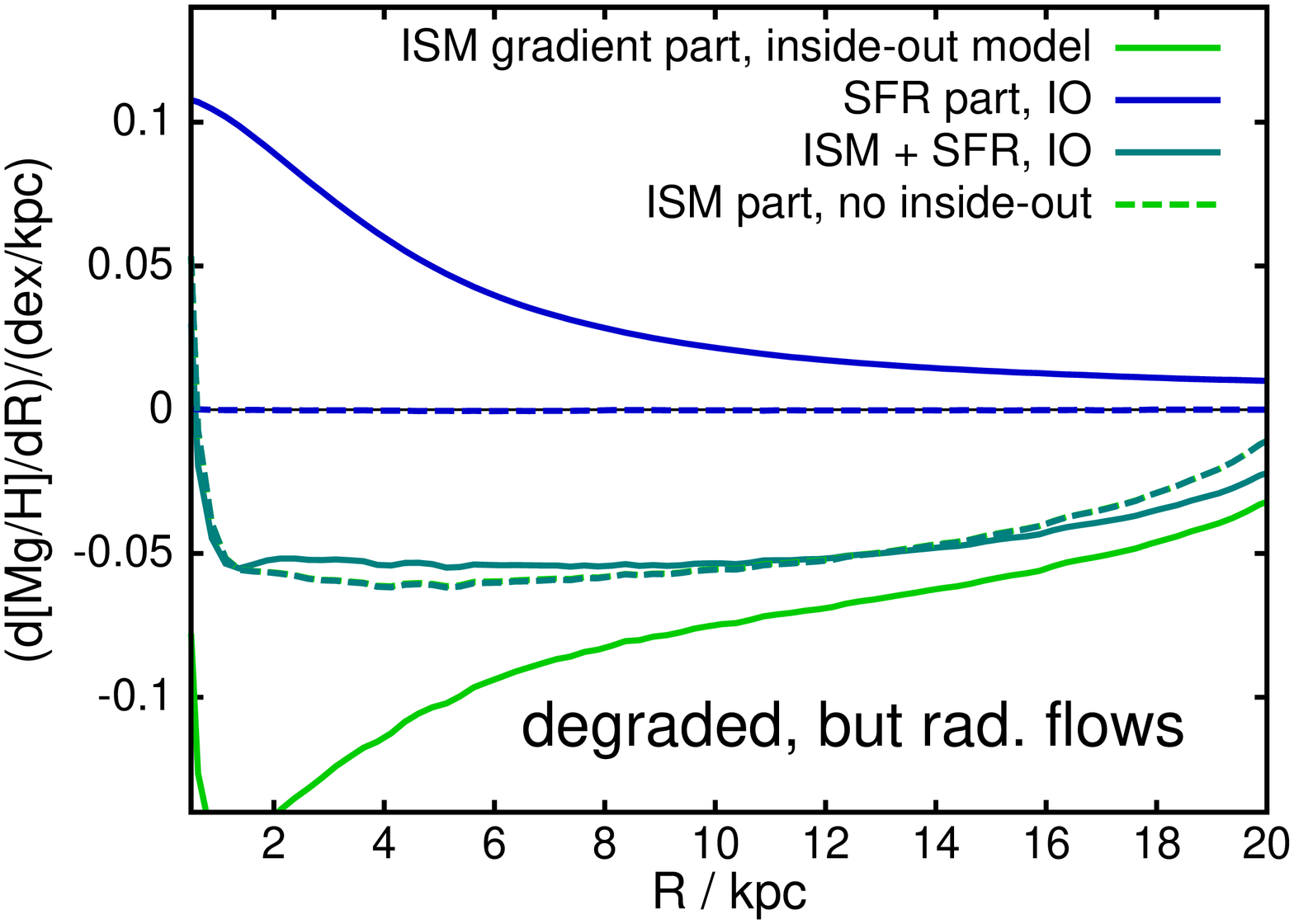,angle=-90,width=0.49\hsize}
\epsfig{file=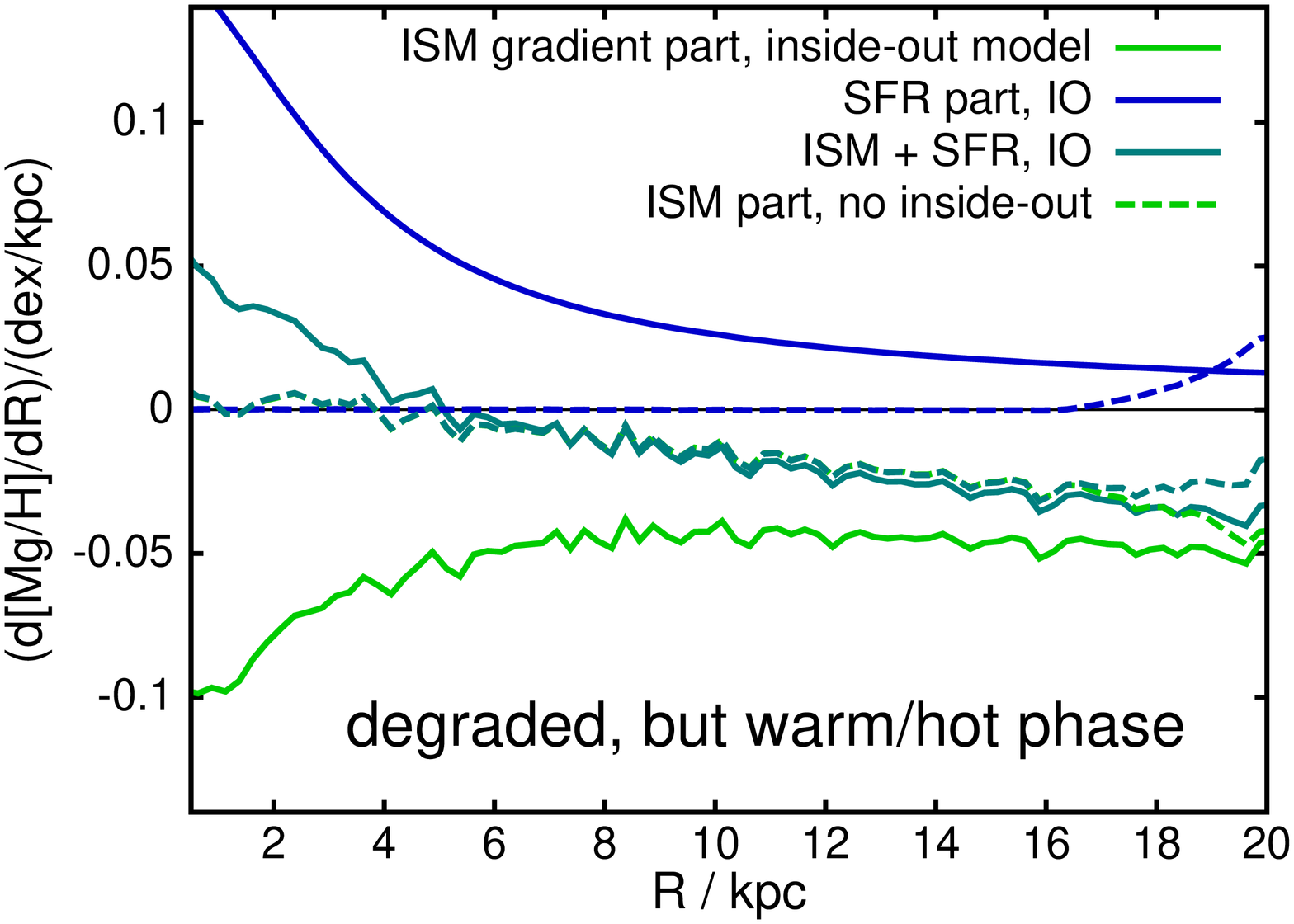,angle=-90,width=0.49\hsize}
\epsfig{file=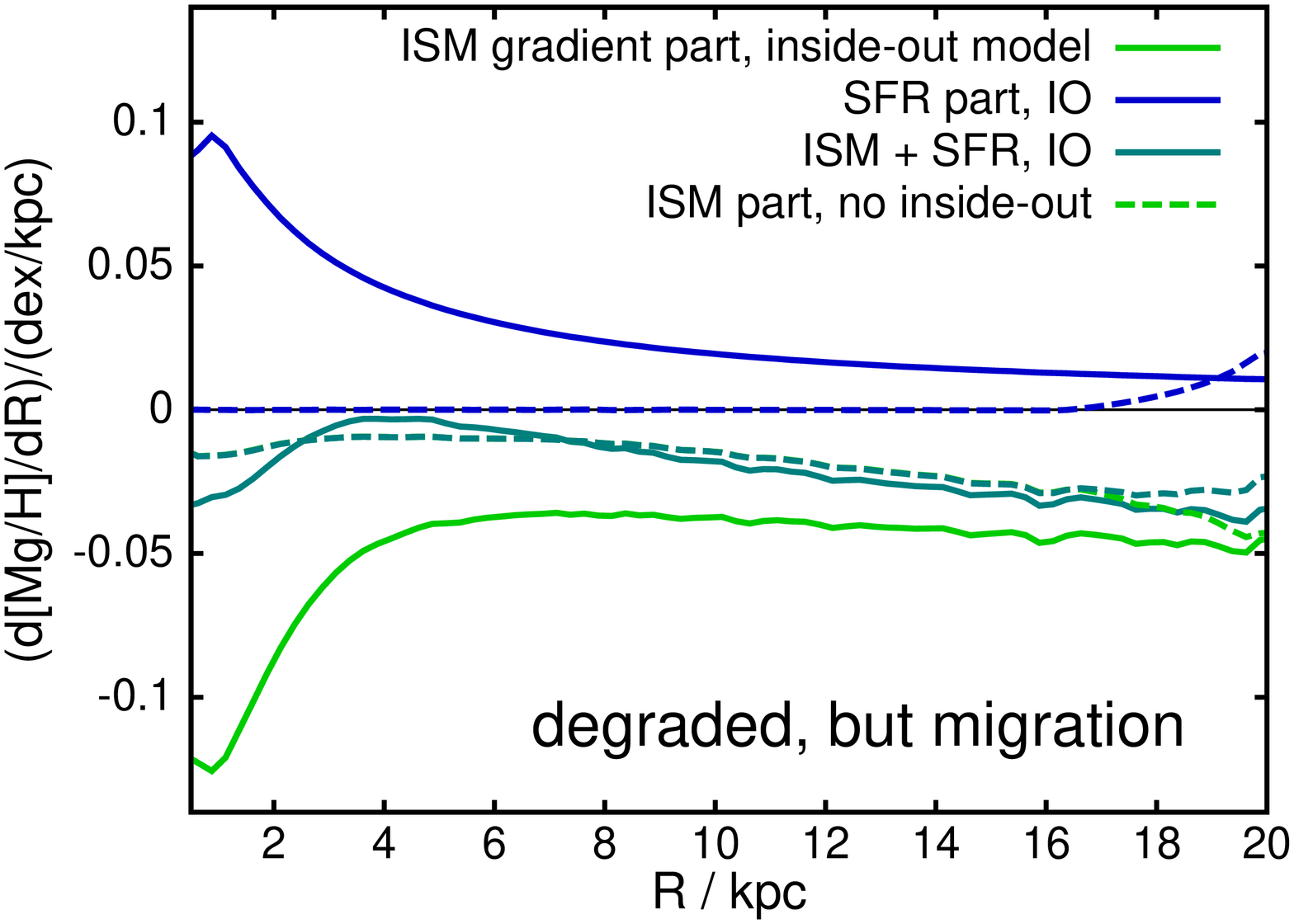,angle=-90,width=0.49\hsize}
\caption{The top left plot shows $\d \mgh/ \d R$ profiles for simplified models with near-instantaneous enrichment, no hot gas phase, and no migration. Dashed lines are the model without inside-out formation, solid lines show its inside-out formation counterpart. As in \figref{fig:gradan} the resulting gradient profile (without migration) in the stellar populations is painted in turquoise. Its contributions from the time-averaged metallicity gradient and from the changing star formation history with radius are painted in green and blue. In the other three panels we selectively re-instate one of our more realistic approximations for either radial flows (top right), a warm/hot phase (bottom left), and radial mixing in the gas and stars (bottom right).}\label{fig:gradansdeg}
\end{figure*}

\subsection{Relation to simple chemical evolution models}\label{sec:simplistic}

Classical chemical evolution models have failed to predict gradient inversions in the stellar populations under inside-out formation, while discussing in great detail the steepening effects of inside-out formation on the star-forming ISM \citep[][]{Prantzos00}. At least a part of the cause is that the classical evolution assumptions lead to an unnaturally strong coupling between star formation and enrichment rates. This is primarily due to the one-phase treatment of the gas and missing radial exchange, and in some cases the assumption of instantaneous recycling. Many chemical evolution models assume that star formation cycles can be broken down into infinitesimally short time bins, in which each infinitesimally small star formation cohort enriches the gas for its successors immediately (an assumption that is still frequently taken for SNII yields). Under those assumptions, the metallicity in the ISM would be mostly a function of the present stellar mass. To give a simple example: The simple accreting box model has a metallicity of:

\begin{equation}
Z_b(t) = p\cdot\left(1-\exp\left({1-\frac{M_{\rm s}(t)}{M_g}}\right)\right)
\end{equation}
where $p$ and $M_{\rm s}(t)$ are stellar yields and the stellar mass at time $t$ while $M_g$ is the (constant) gas mass. 
This model is sufficiently simple that we can write the equivalent of equation (\ref{eq:Zroot}) as 
\begin{equation}
\left<Z\right> = \frac{1}{M_{\rm s,final}}\int_0^{t_{\rm final}}  Z_b(t)\; \frac{\d M_{\rm s}(t)} {\d t} \d t .
\end{equation}
As $Z_b$ does not depend explicitly on $t$, but only on $M_{\rm s}(t)$
\begin{equation}
\left<Z\right> = \frac{1}{M_{\rm s,final}}\int_0^{M_{\rm s,final}}  Z_b(M_{\rm s})\; \d M_{\rm s} .
\end{equation}
Therefore the final mean metallicity at each $L_{z,0}$ is independent of the details of the star formation history, and only depends on the total star formation in the given radial bin. As a result, this simple model (wrongly) predicts that inside-out formation leaves the stellar radial metallicity gradient nearly unaffected.

To test this with our models, several parameters have to be changed to approximate a simplified chemical evolution code. We set the fraction of yields going to the warm/hot ISM to $0.005$, and freeze out $0.99$ of this per timestep. Also, we free all yields from the first $7.5 \Gyr$ in the first timestep, set the SNIa delay time to $15 \Myr$, and evaluate magnesium instead of iron. Further, we set the inflow composition to primordial, eliminate radial flows by setting the specific angular momentum of the infalling gas to that of a circular orbit at each radius, eliminate gas mixing and switch off churning, by setting the migration parameter to $10^{-6}$. With this degraded model, the top left panel of \figref{fig:gradansdeg} analyses the standard inside-out formation history (solid lines) in comparison with a constant scale-length model with gas disc scale-length of $3.25 \kpc$ (dashed lines). The colours are chosen to match \figref{fig:gradan}, however, since radial mixing is switched off, only the first terms of equation (\ref{eq:gradbasis}) are of interest. As expected, the inside-out model has a far steeper star-formation averaged gradient (green solid line) than its constant scale-length counterpart. However, this steeper gradient is almost fully compensated by the gradient contribution from the radially dependent star-formation history (the contribution is near-zero for the model with constant scale-length). As a result, the observable metallicity gradient in today's stellar populations is almost exactly the same in both cases. Minor differences arise mostly from the finite time-resolution of the model, which sets a minimum delay time for the stellar yields at the time resolution of $15 \Myr$ and has some importance at the lowest metallicities, in particular in the intensely star-forming inner regions. Further, the star formation histories and gradients will not perfectly match up due to different degrees of saturation (in the outer rings the ratio between stellar mass and present gas is low, and hence metallicity has yet to approach its equilibrium value). The other three panels of \figref{fig:gradansdeg} show the effects of switching back on single aspects of our full model. As can be expected, the presence of a warm-hot gas phase mostly affects the quick changes in the inner regions, giving a very large positive effect on the gradient in the inner regions. The re-distribution via migration has a tendency to move metals outwards and hence contributes negatively in the innermost regions and positively at inner to intermediate radii. Radial flows create the general abundance gradient, but have only minor effects on behavioural differences between inside-out and constant scale length models. This is reasonable, since the gas flows only a relatively short distance during a typical enrichment time, so that flows do not connect vastly different regimes.

This comparison conveys two important messages: the detailed star formation history has to remain strictly coupled to chemical evolution, otherwise the mismatch between enrichment and star formation rates will alter the metallicity gradients. While useful for other problems, "painted" N-body models with separately calculated, more sophisticated chemical evolution, e.g. \cite{Curir12} should not be used for assessing radial metallicity gradients or relationships between chemistry and kinematics, unless they exactly preserve the same star formation history in both the N-body and chemical evolution part. This is particularly the case for combinations which have large discrepancies between the early star formation in their chemical evolution models vs. the N-body part \citep[see Fig. A.1 in][]{Minchev14}. 

More importantly, this demonstrates the importance of radial flows, mixing, and processes that delay enrichment in young galaxies. These processes greatly alter the observed metallicity gradient: i) radial mixing flattens the ISM gradients, while radial inflow tends to steepen them. ii) The metallicity of the inflowing gas will change with time as the enrichment of the IGM proceeds, iii) metals get locked up in the warm and hot ISM of a galaxy, from which they only freeze out with a significant time delay, and iv) the time between the initial collapse of a molecular cloud and the first yields is of order $10 \Myr$. No matter the assumptions, any star formation spike on a timescale shorter than this will (apart from direct enrichment from neighbouring supernovae) be dominantly influenced just by the preceding metallicity of the ISM. Points i), iii), and iv) are the main factors driving inverted gradients in stellar populations, but most chemical evolution models neglect at least some of them.

\begin{figure}
\epsfig{file=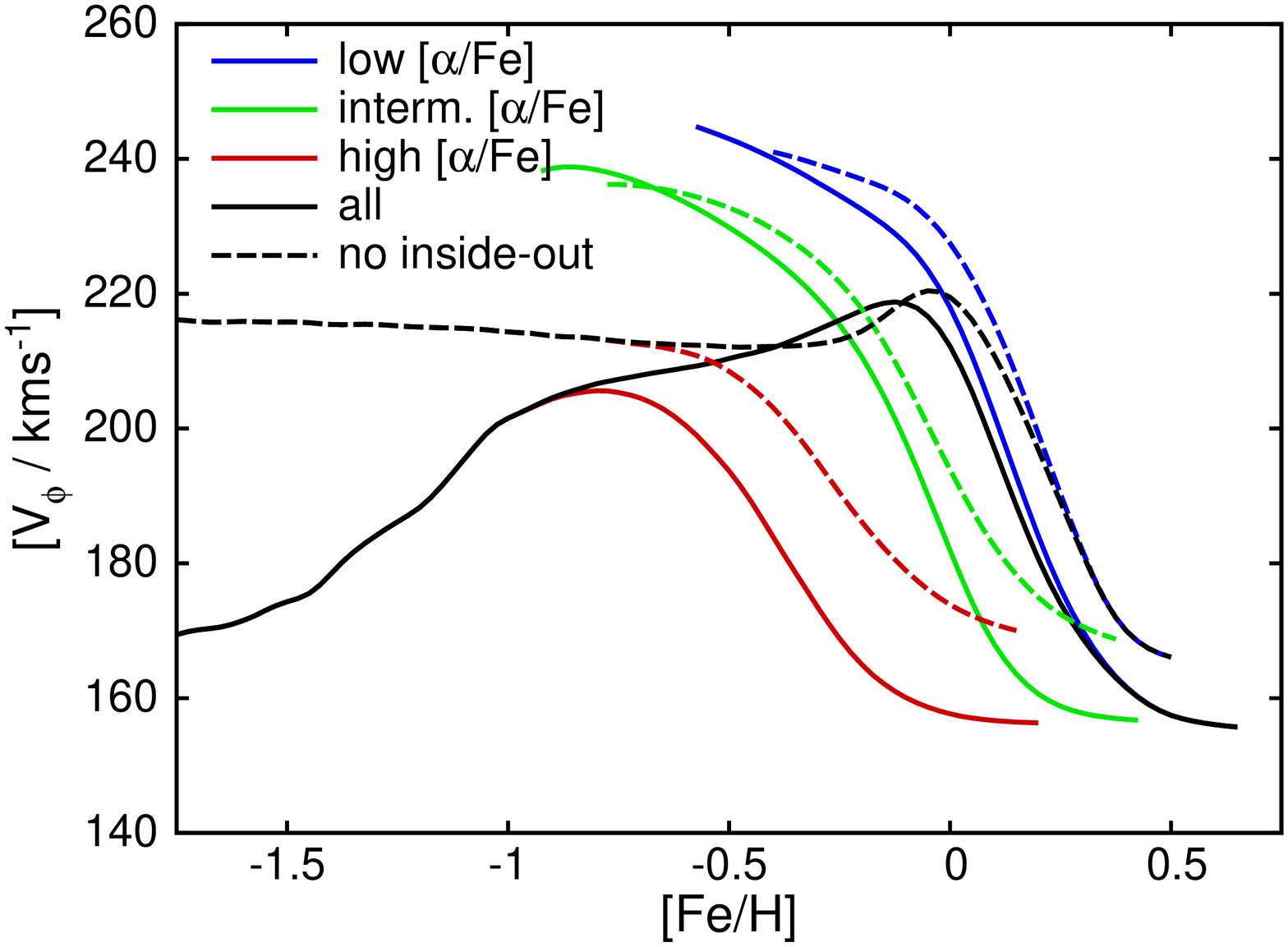,angle=-90,width=\hsize}
\epsfig{file=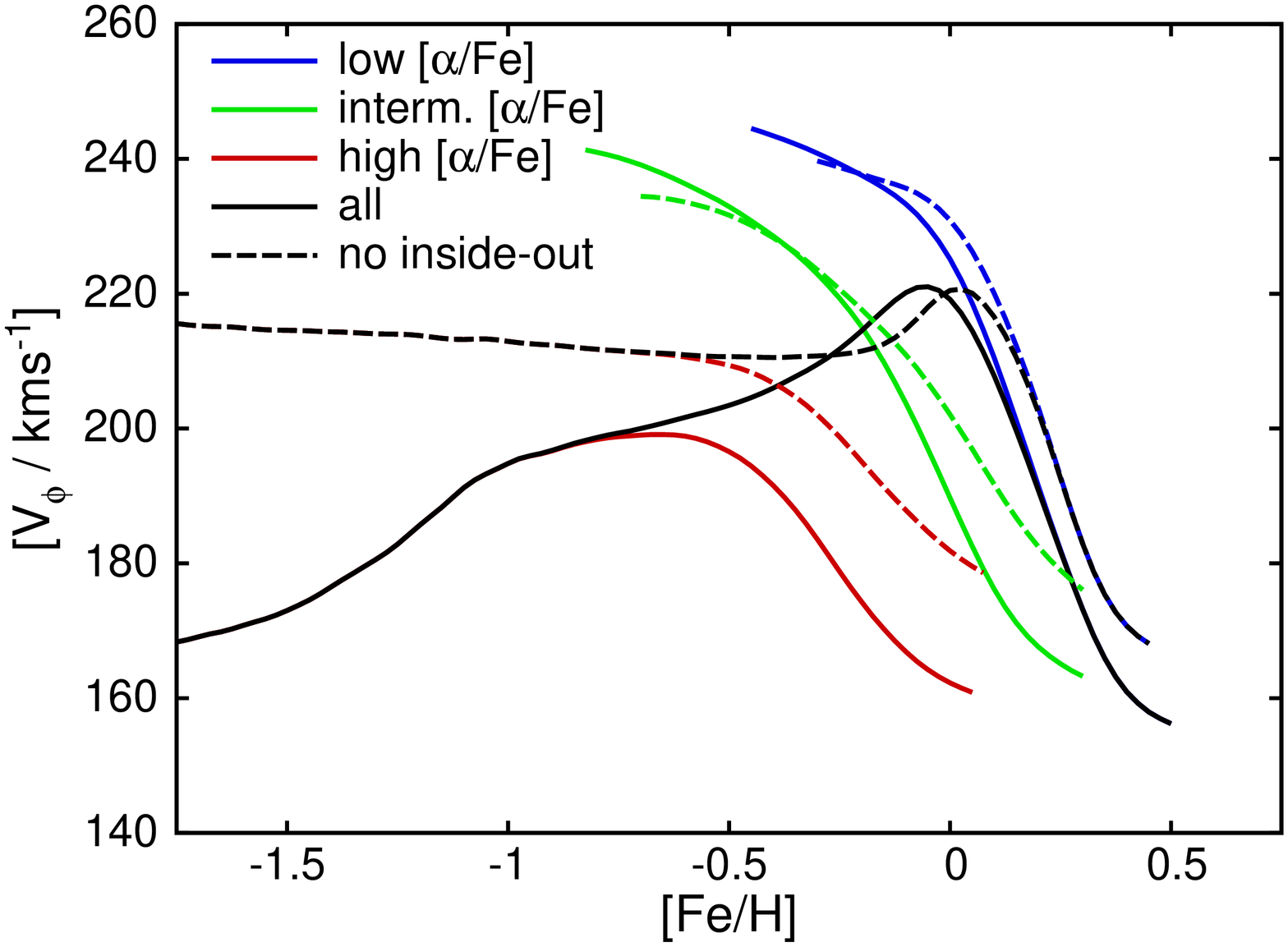,angle=-90,width=\hsize}
\caption{Mean azimuthal velocity at each metallicity bin in the inside-out model (black lines), and the mean $\Vphi$ in the $\afe$ bins with delimiters $-0.05, 0.10, 0.25 \dex$ (colours). To make a more "realistic" case, for this plot and all subsequent $\Vphi$ vs. $\feh$ statistics, we sum all stellar populations (mass weighted) in the model in the altitude range $0.4 \kpc < z < 0.8 \kpc$ and fold the abundance determinations with optimistic Gaussian errors of $\sigma = 0.05 \dex$ both in $\feh$ and in $\afe$. The top panel shows the "standard" inside-out model (solid lines) compared to a model with constant gas disc scale-length (dashed lines) of $3.5 \kpc$ (i.e. no inside-out formation at all). The bottom panel shows the same constant scale-length model, but the inside-out model with strong gas mixing and hence reduced radial metallicity gradient.}\label{fig:velgrad}
\end{figure}

\section{Kinematic gradients}

Gradients in mean azimuthal speed depending on metallicity as first found by \cite{Spagna10} have to be explained by two different factors: asymmetric drift by kinematic heat of each metallicity cohort, and their angular momentum distributions, both factors determining the local asymmetric drift. Following the example of eq.(\ref{eq:root}), we can write
\begin{equation}\label{eq:feLz}
\begin{array}{ll}
\left<\Vphi(Z_x,R)\right> = W_f^{-1} \iiint \frac{\Lz}{R} & \;s_{Z_b = Z_x}\,g_{\star}(\Lz,t) \,M(\Lzo, \Lz, t) \\
& \times\; B(\Lz, R, t) \;\;dt\,d\Lz\,d\Lzo  ,
\end{array}
\end{equation}
where $s_{Z_b \in Z_x}$ selects populations only when $Z_b$ is in the intervall $Z_x$, and $W_f$ is a normalisation:
\begin{equation}\label{eq:weightLz}
\begin{array}{ll}
W_f = \iiint \;s_{Z_b \in Z_x}\,g_{\star}(\Lz,t) \,M(\Lzo, \Lz, t) \; B(\Lz, R, t) \;\;dt\,d\Lz\,d\Lzo .
\end{array}
\end{equation}
It can be directly seen that while structurally similar, this has a different behaviour from eq.(\ref{eq:Lzt}). In contrast to the mean metallicity, the factor $\Lz$ prevents us from merging the churning and blurring terms. If we take the bin $Z_x$ to be infinitesimally narrow, and if we assume that the metallicity rises monotonically, the object $s_{Z_b = Z_x}$ will pick the populations at the point when the metallicity of $Z_x$ is reached with a weight inversely proportional to $\partial_t Z(\Lzo, t)$, i.e. to the rate of change of the metallicity at that time. This raises again the important point about delayed enrichment. In the overly simple model from Section~\ref{sec:simplistic}, a faster star formation rate just speeds up enrichment roughly proportionally, which implies that the changes in $g_{\star}$ and the evolution of $Z$ cancel out in this equation. This does not happen in a model that incorporates delayed enrichment and chemical evolution. 

\figref{fig:velgrad} shows the mean azimuthal velocities at each metallicity bin in the inside-out model (coloured) versus the same quantities in a model with constant scale length ($R_{\rm d, gas} = 3.5 \kpc$) of the gas disc (green points). It is apparent that while the metallicity gradients have similar values in both cases, the scale length growth produces a strong dependence between mean azimuthal velocity and metallicity. To create this plot we calculated the abundance plane at $R = 8 \kpc$ integrating over $0.4 \kpc < z < 0.8 \kpc$, folded it with a Gaussian observational error in $\afe$ and $\feh$, and evaluated the mean azimuthal velocity versus metallicity for the entire sample (yellow points) and for different $\afe$ bins with delimiters at $\afe = -0.05, 0.10, 0.25 \dex$. The behaviour can be quite easily explained: The youngest stars with largest azimuthal velocities dominate the two lower $\afe$ bins. For them, the recent/current abundance gradient of the galactic disc dominates the velocity structure: While radial migration with the \cite{SB09a} coefficients induces strong mixing in angular momentum of stars, enabling us to see large numbers of stars with different metallicities, the migration is not complete, i.e. low metallicity stars are still preferentially located at larger guiding centre radii than their metal-rich counterparts. Together with mildly reduced velocity dispersions for the metal-poor stars (they are outer disc populations), those metal-poor low-alpha stars can even reach mean azimuthal velocities  above the local circular speed \citep[this observation was discussed e.g. by][]{Haywood08}, while the metal-rich stars still concentrated in the inner regions attain low azimuthal velocities. For an in-depth discussions of Str\"omberg's equation governing this behaviour, see \cite{SBD}.

A very important characteristic of these kinematics is the slope of $\d \Vphi/ \d \feh$ at the metal-rich rim of the high $\afe$ population. This is very difficult to assess in current surveys because of the decreasing number of high $\afe$ stars with increasing $\feh$. Hence the fraction of low $\afe$ stars at the same metallicity increases, so that increasing contamination with higher $\Vphi$ populations is a major issue (in this light it is also unwise to use a sample selection sloping in $\afe$ versus $\feh$). However, if the structure of this branch can be determined, it gives direct insight into the situation of the old disc, when significant SNIa enrichment sets in. If there was a significant radial metallicity gradient in the disc at that time, the trajectories of the outer regions of this disc will have their $\afe$ versus $\feh$ knee at a lower $\feh$ than their inner disc counterparts. While all environments run through the same high $\afe$-low $\feh$ points, the metal-rich end of this sequence will hence show some differentiation. In the presence of negative $\d \feh/ \d R$, only the innermost radii can still contribute to the highest metallicity bins, implying negative $\d \Vphi / \d \feh$. If the ISM gradient at this time was indeed inverted/positive, the age-dispersion relation would compete with a positive correlation from the radial gradient, which could be resolved by comparing velocity dispersions and the asymmetric drift.

\begin{figure}
\epsfig{file=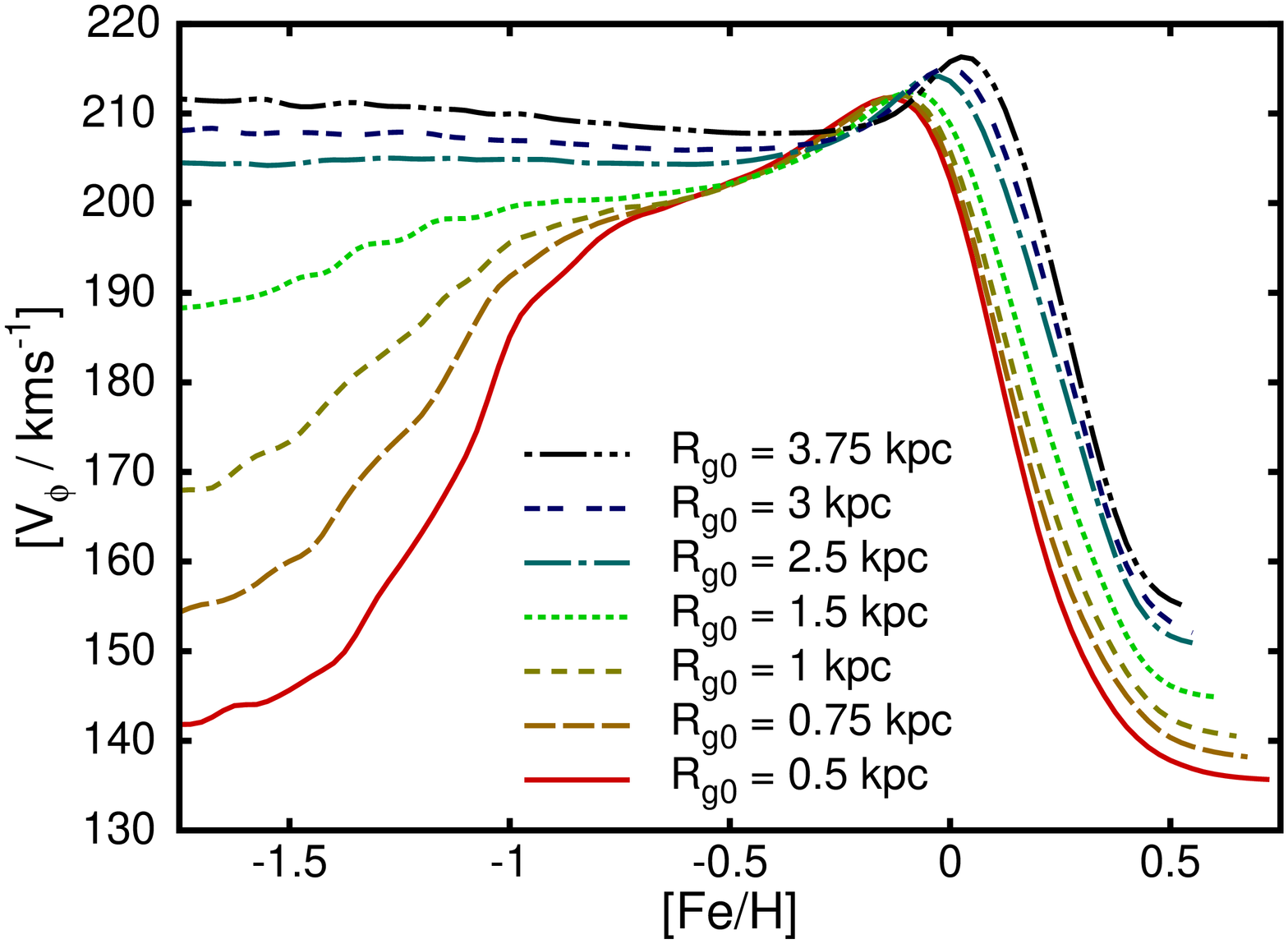,angle=-90,width=\hsize}
\epsfig{file=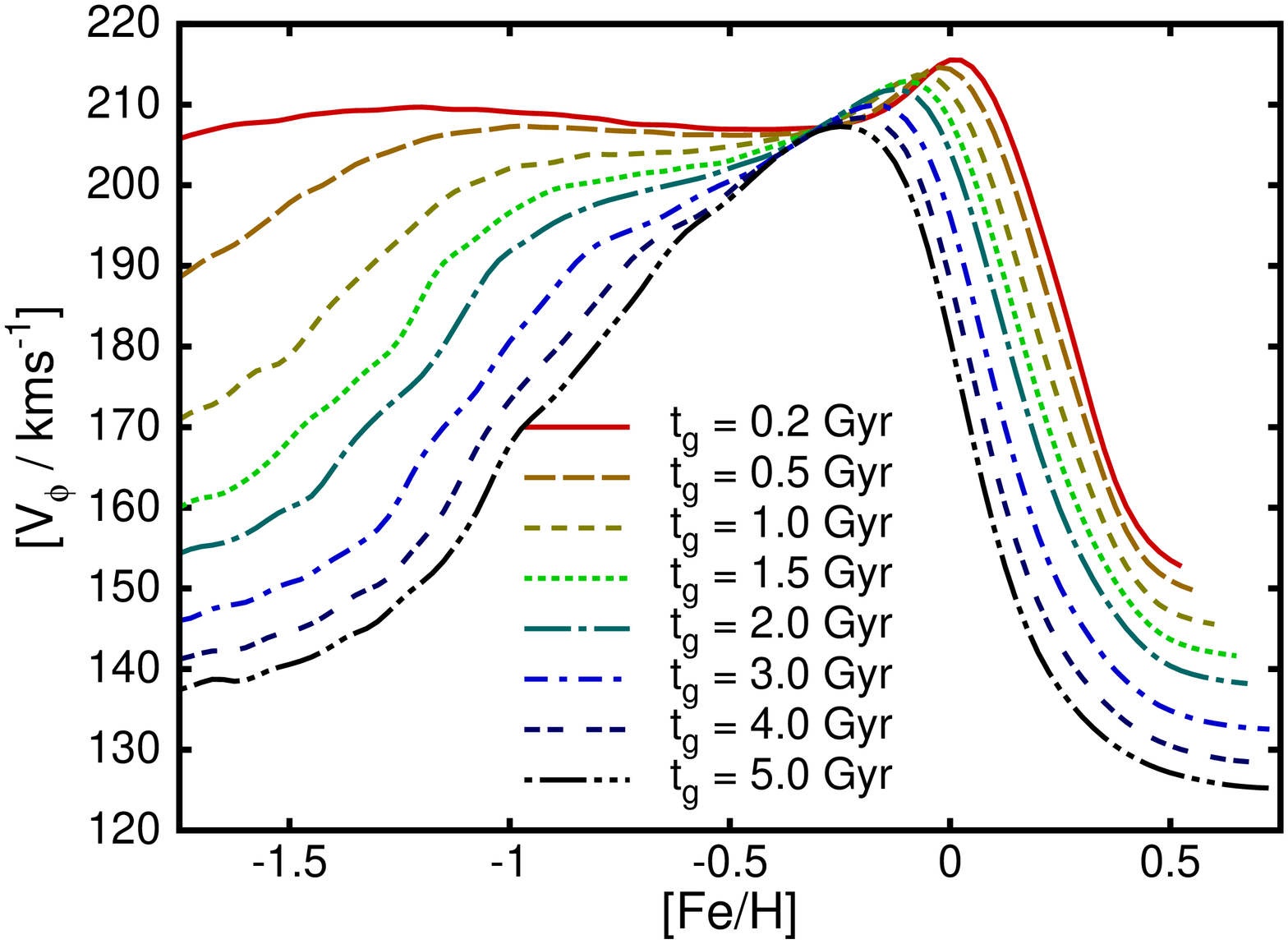,angle=-90,width=\hsize}
\caption{Varying the parameters on the inside-out formation. The top panel shows the mean azimuthal velocity vs. metallicity as in Fig. (\ref{fig:velgrad}) for  different initial scale-lengths of the gas disc at a growth timescale of $t_g = 2 \Gyr$. The bottom panel shows the same relation for an initial scale-length of the gas disc of $R_{g,0} = 0.75 \kpc$, but different growth rates $t_g$.}\label{fig:velgradsyn}
\end{figure}

\figref{fig:velgradsyn} explores the parameter space for the inside-out growth of the disc. The top panel shows the mean azimuthal velocity as in \figref{fig:velgrad} for different initial scale-lengths $R_g$ of the gas disc (cf. eq. \ref{eq:iogrowth}). We need substantial inside-out growth to achieve a reasonable effect on $\d \Vphi/ \d \feh$. The effect is nearly proportional to the amount of disc growth, which is evident, since the current disc scale-length translates directly into asymmetric drift, and churning does little to change the distribution. 
We also note that the mild bump in the velocity-metallicity relations for the entire disc around $\feh \sim -1$ is partly caused by a strong early slowdown in the increase of the infall metallicity, and would be entirely removed, if the cut-off of the early disc would be at smaller radii, or if the infall metallicity had a more linear time-dependence. We tested that it does not affect the general behaviour of our models beyond this bump.

The timescale of disc growth is limited by the bottom panel of \figref{fig:velgradsyn}. If the disc evolves too fast, there will be no effect, if the growth is too slow, the disc will have the major part of its evolution after the high $\afe$ stars have branched off. For example, in the model with the longest inside-out formation timescale of $5 \Gyr$, the branching point for the high $\afe$ stars lies only around $\Vphi \sim 170 \kms$. 

Note that there is a shift in the mean metallicity in particular of the outer regions. This is because we keep all other parameters fixed, i.e. we use the same time-dependent accretion rates for all models, which results in significant differences between models in today's stellar mass profiles and hence enrichment.

\begin{figure}
\epsfig{file=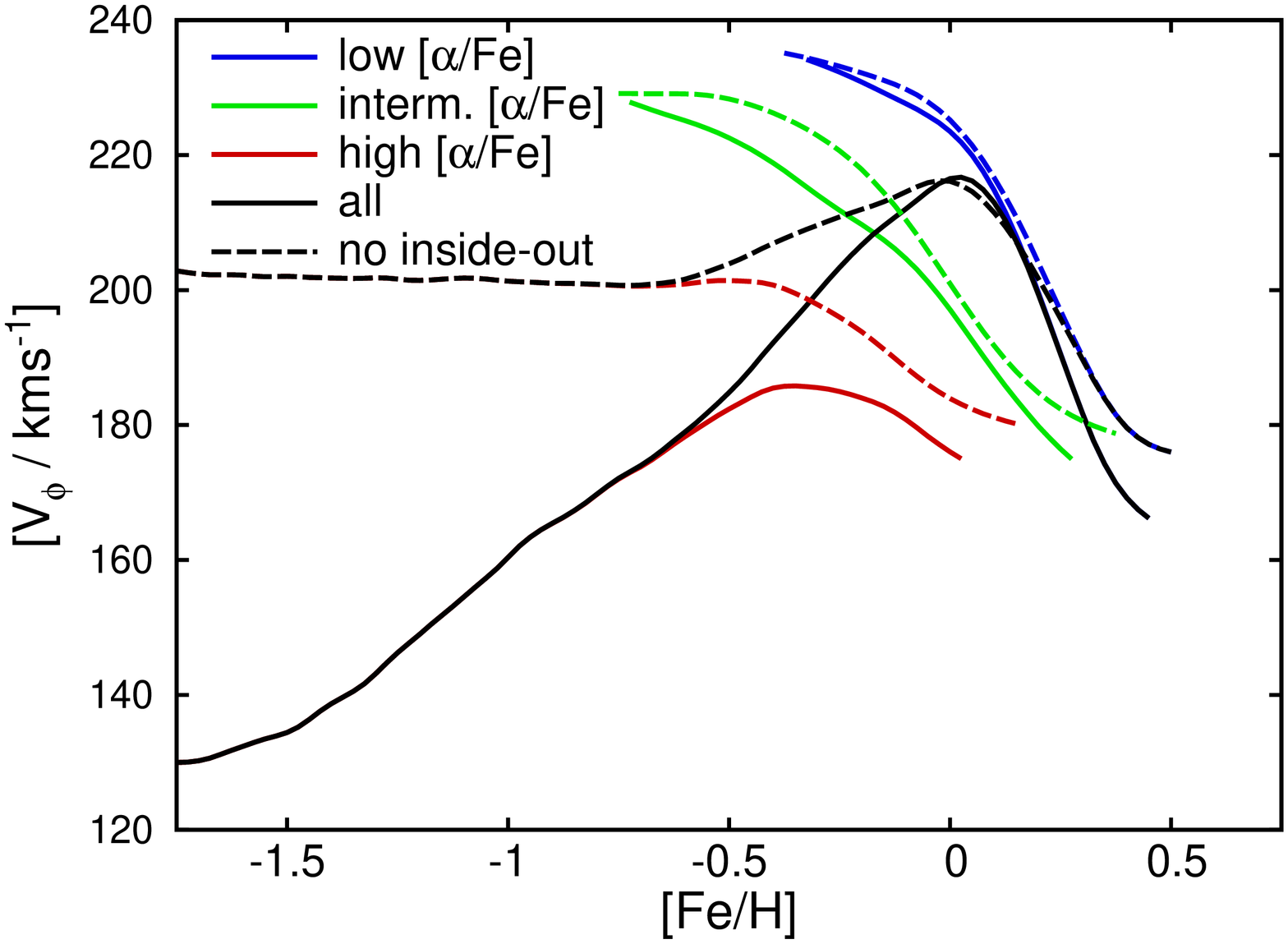,angle=-90,width=\hsize}
\epsfig{file=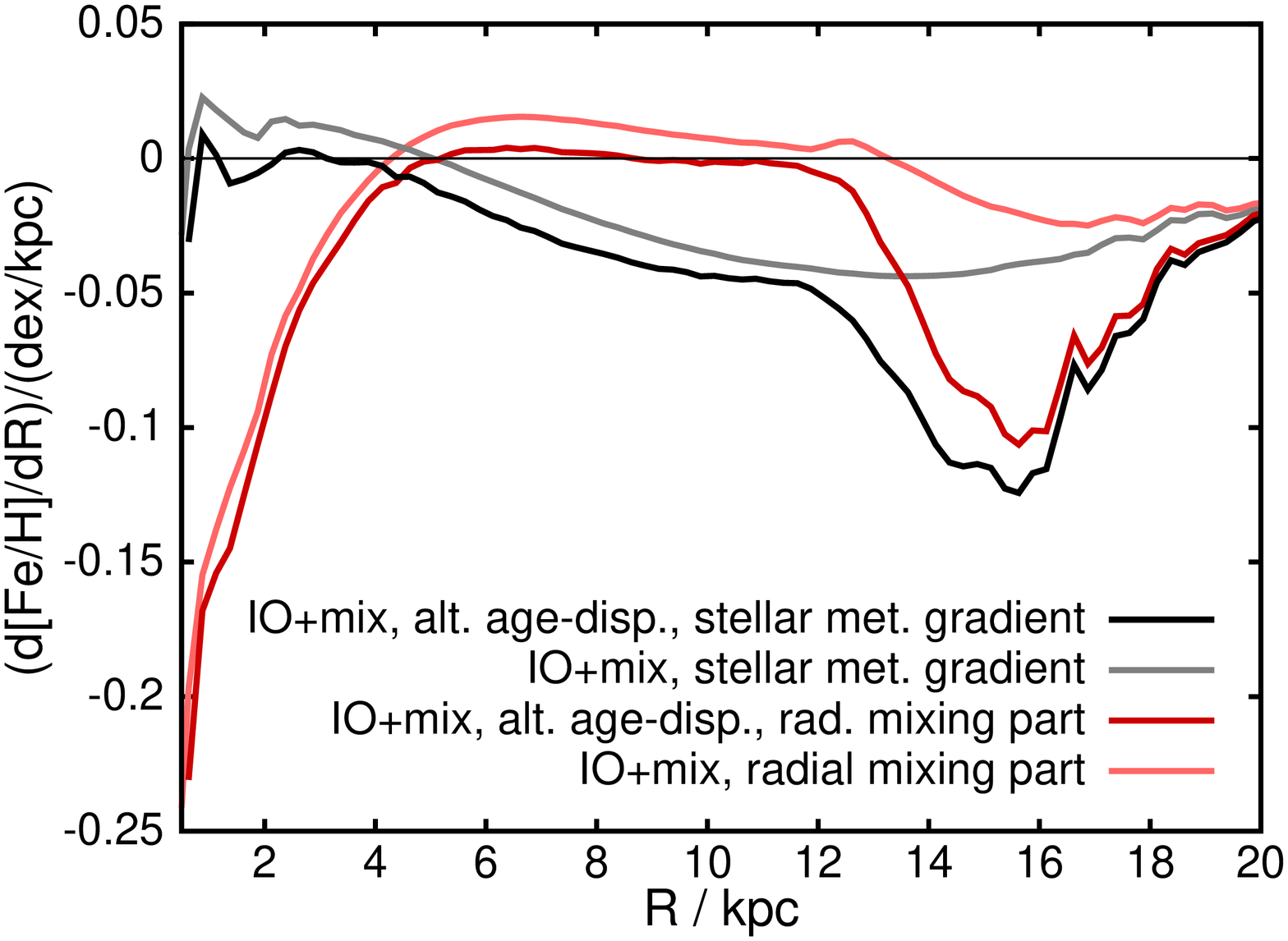,angle=-90,width=\hsize}
\caption{Top panel: $\Vphi$ vs. $\feh$ profiles (black lines) as in Fig. (\ref{fig:velgrad}) for the standard-inside out (IO, solid lines) and constant scale-length models (no-IO, dashed), but with the altered age-dispersion relation. Coloured lines display $\afe$ bins with delimiters $-0.05, 0.10, 0.25 \dex$ as in \figref{fig:velgrad}. Bottom panel: observable radial metallicity gradient profile in the stellar populations and its radial mixing contribution for the model with altered age-velocity relation compared to the inside-out model with strong gas mixing.}\label{fig:altblurr}
\end{figure}

\subsection{Kinematic heat and blurring}

\figref{fig:altblurr} shows the changes experienced by the two gradients when we change the heating law in our galaxy. For this purpose we exchange the standard heating law, which is a space-dependent $\tau^{0.33}$ power law in age, with a "broken" thin-thick disc law. For this purpose we keep the $\tau^{0.33}$ power law up to age $10 \Gyr$, but scale all velocity dispersions down by a factor $1.5$, and then set a linear rise to double the dispersion between $10 \Gyr$ and $11.5 \Gyr$, keeping the dispersion flat afterwards. This distribution resembles somewhat the broken time-dependence in \cite{Sanders15} and looks also similar to ``upside-down'' galaxy simulations \citep[][]{Bird13}. As we discussed before, this impacts the statistics in $\d \Vphi / \d \feh$ and $\d \feh / \d R$ in opposite ways. The spreading of the oldest stars makes the gradient more negative throughout the disc and essentially eliminates the region with inverse/positive gradient. However, this implies a larger asymmetric drift for the most metal-poor stars, which increases the gradient of rotational speed with metallicity for thick disc stars. While the general structure is unchanged, there is a significant decrease in the $\d \Vphi / \d \feh$ slope at the high-metallicity end of the high $\afe$ subgroup, essentially rendering the trajectory flat. While population densities are extremely low down this tail, this could be an interesting diagnostic in large data samples. From the constant scale-length model, we can see that the effect on the azimuthal velocities is of order $20 \kms$, which is already comparable to the effects observed in local samples. The take-away message is that, while the gradient for the thick disc points to an inside-out formation, the exact quantification has to await more precise constraints for the heating law, and metallicity data throughout the disc, since the blurring effect has an inverse sign to the inside-out formation impact on gradients.

\begin{figure*}
\epsfig{file=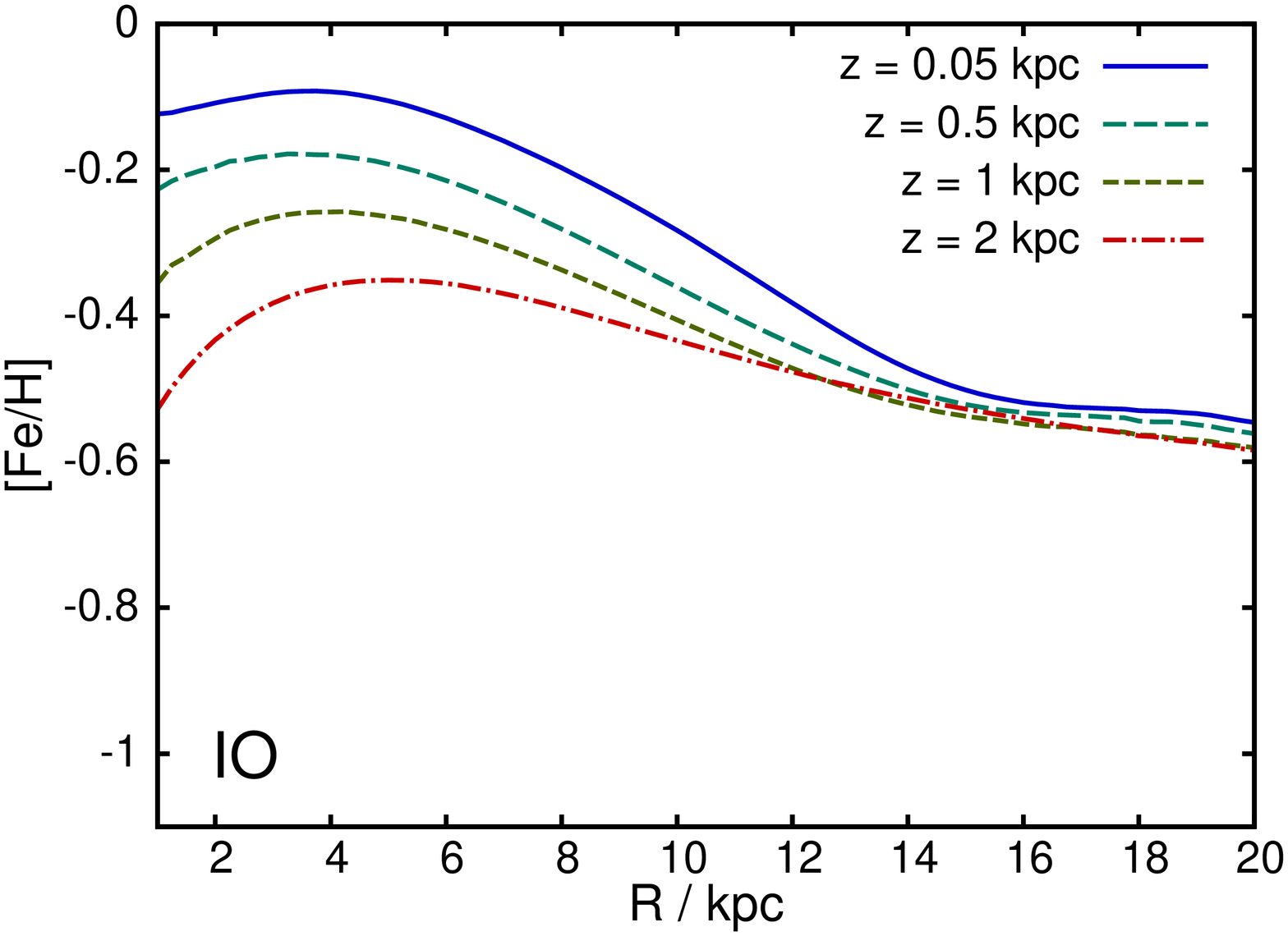,angle=-90,width=0.49\hsize}
\epsfig{file=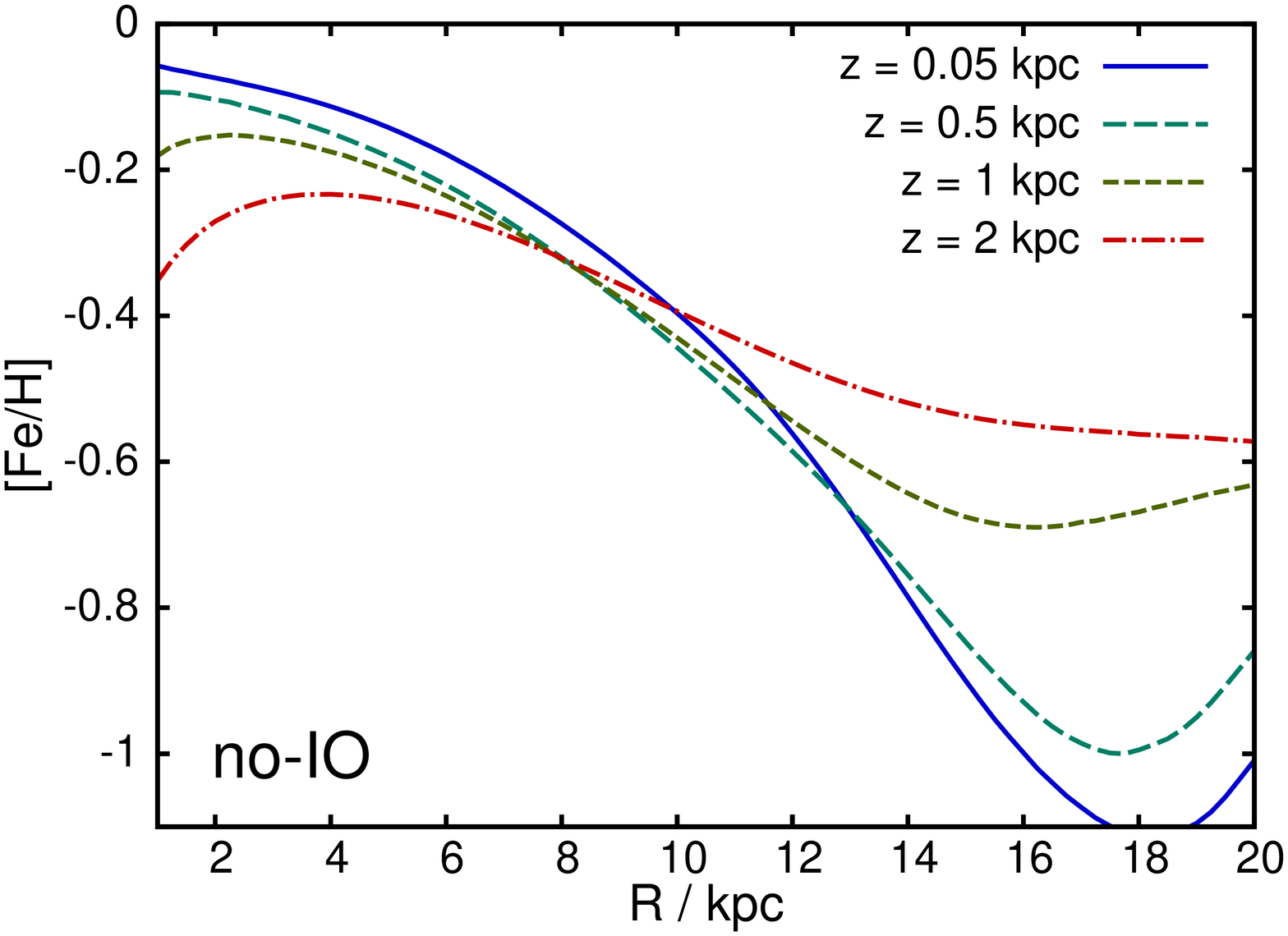,angle=-90,width=0.49\hsize}
\epsfig{file=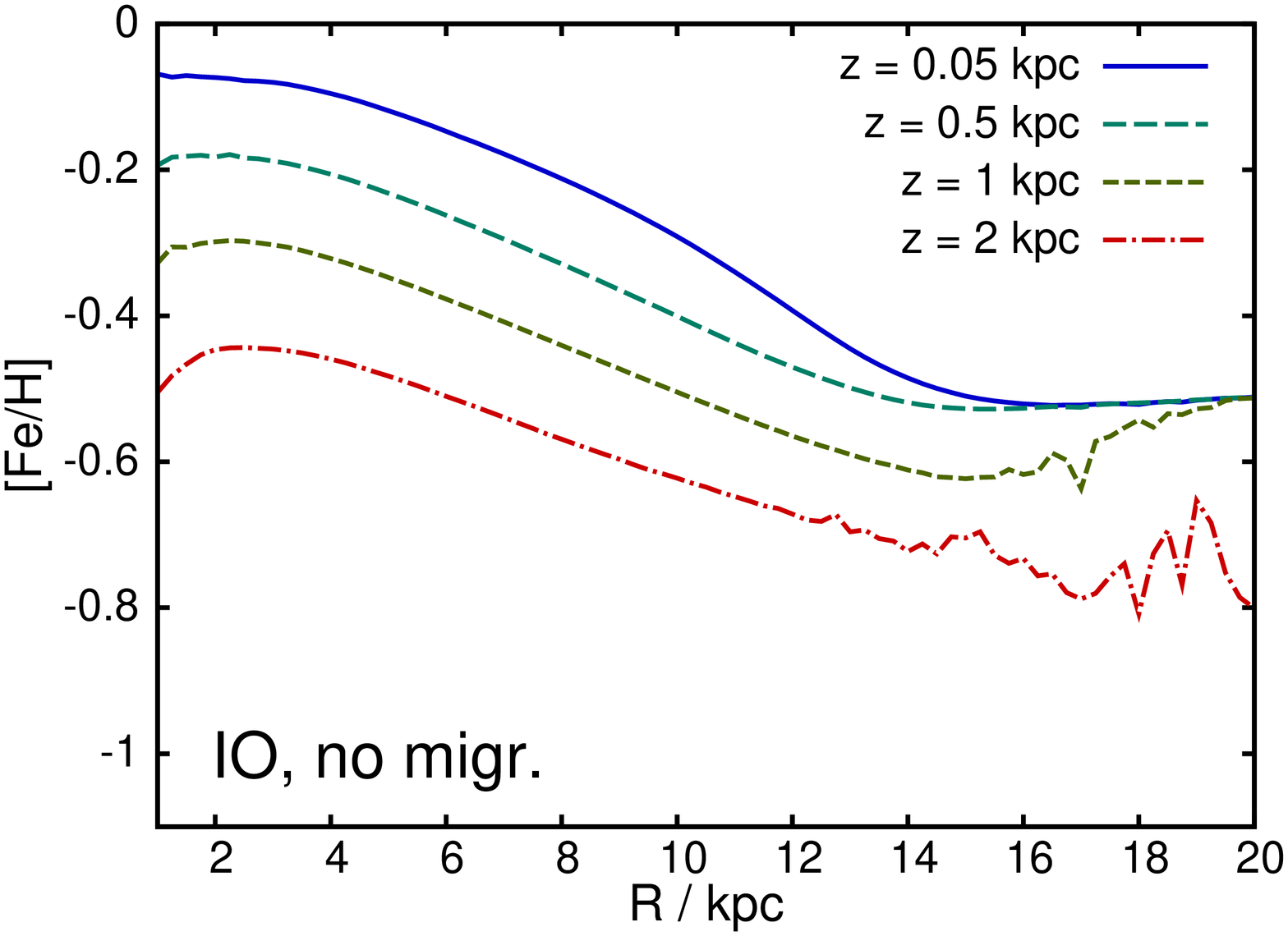,angle=-90,width=0.49\hsize}
\epsfig{file=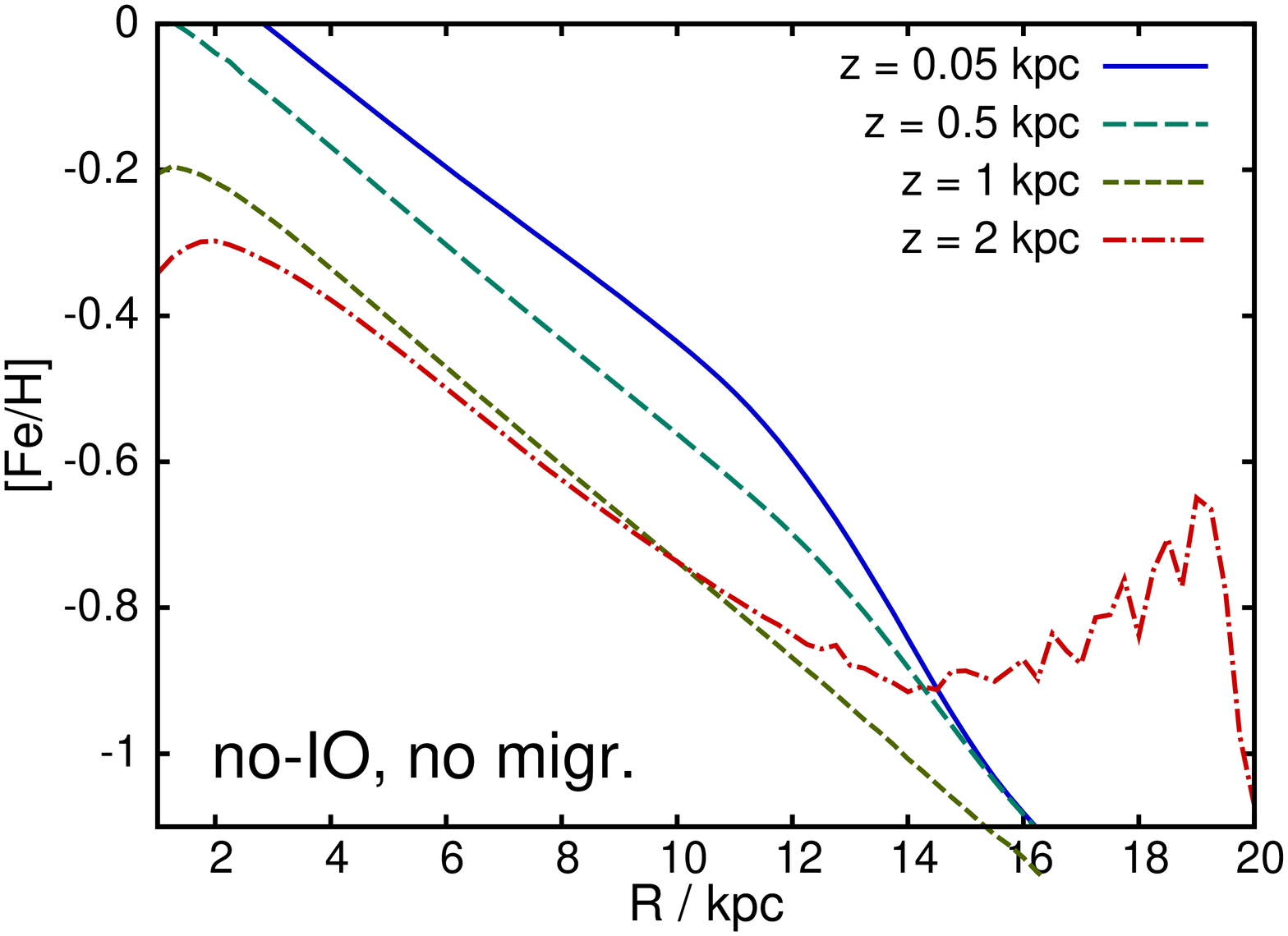,angle=-90,width=0.49\hsize}
\caption{Metallicity profiles at different altitudes versus radius. The flattening and partly inversion of radial metallicity gradients is apparent in all models. The top-left panel shows the standard inside-out model. The four panels are arranged in a matrix with inside-out models on the left versus constant scale-length on the right hand side, while models with radial migration are in the top row, and models without churning are in the bottom row. While inside-out formation generally gives a positive contribution to the radial metallicity gradient, the dominant term behind the trend with altitude above the plane is radial redistribution of the stars (churning).}\label{fig:vertfehprof}
\end{figure*}

\section{Gradients off the plane}

Lastly, we look into the effect on gradients in and above the disc plane. It is a priori clear that, with these assumptions, models with inside-out formation should display a steeper negative vertical metallicity gradient than constant scale-length models. This is because inside-out formation produces more low-metallicity stars with hot kinematics, and in particular a larger number of low-metallicity stars in the innermost regions (which we believe to be kinematically hotter) than in the outer disc regions. This is mediated by the transition term $T$ in equation (\ref{eq:Zroot}). Towards higher altitudes the fraction of older stars generally increases, but at larger radii, the younger populations can still contribute more stars because some are radially migrated outwards, bringing more intermediate age / younger ages to larger altitudes at intermediate to large galactocentric radii, due to the increasing scale-height in outwards migration. I.e. radial migration typically gives a positive trend towards $\d \feh/\d z$. This mixture of different ages, is fully in line with observations. \cite{Casagrande16} find a major contribution from stars at intermediate ages out to $z \sim 1 \kpc$. Their gradient of the mean age of stars with altitude, $\d \tau/ \d z = (4 \pm 2) \Gyr \kpc^{-1}$, is quite exactly in line with our model, which ages from a mean of $5 \Gyr$ at $z \sim 0.05 \kpc$ to a mean age of $9 \Gyr$ at $z = 1 \kpc$. 

\figref{fig:vertfehprof} shows the corresponding metallicity profiles. In the two models with radial mixing (top row), the inversion point of the radial metallicity gradient shifts to successively larger radii towards larger altitudes. In addition, the inside-out formation (left column vs. constant scale-length models on the right) clearly enhances the vertical metallicity gradient at inner and intermediate radii, just because more metal-poor kinematically hot stars are present. This pushes the gradients also into the range of estimates from \cite{Kordopatis11} and \cite{Schlesinger14}, with the same structure as noted in \cite{Schlesinger14} that the decrease in metallicity is largely associated with the dominance of high $\afe$ over low $\afe$ stars.

Comparing the no-churning model to the inside-out model with churning, we see the same effect as noted in \cite{Kawata16}: For a disc with a near-constant initial scale height (as set in our model) stars from the inner disc have larger vertical actions and attain larger scale heights, which can give a positive contribution to $\d \feh / \d z$. However, as noted above, the vertical gradient depends critically on the enrichment history of the disc. Due to large age uncertainties, this is even true when attempting to select a single age group. The age-metallicity relation will generally give a negative contribution to the vertical metallicity gradient, while the weakening of radial gradients by inside-out formation makes the inner disc stellar populations more metal-poor and thus again negatively contributes to $\d \feh / \d z$.  

The effect near the disc cut-off is also very interesting. The cut-off is a dominant factor in inverting the vertical metallicity structure of the disc. Because the immigrated objects have a stronger impact than the local populations, the structure is mostly determined by the rates of immigration from different radii, and we lose the always negative contribution to $\d \feh / \d z$ from the local populations. In the model without inside-out formation, the more metal-rich populations from further inside dominate the effect, while, in contrast, the inside-out models get a strong negative contribution from the large number of metal-poor stars reaching larger $z$. The trend towards very low metallicities at the lower altitudes of the model without inside-out formation stems from the contraction of the cut-off radius with the slowing infall, which in this model makes the vast majority of the kinematically cold, locally formed populations beyond the cut-off old and metal-poor.
We also warn that the heating assumptions here cast a bias, mostly carried by intrinsic outer disc populations migrating inwards. Since those would experience the higher heating rates in the inner regions, our assumption underestimates their dispersions and holds them too close to the plane.

\begin{figure}
\epsfig{file=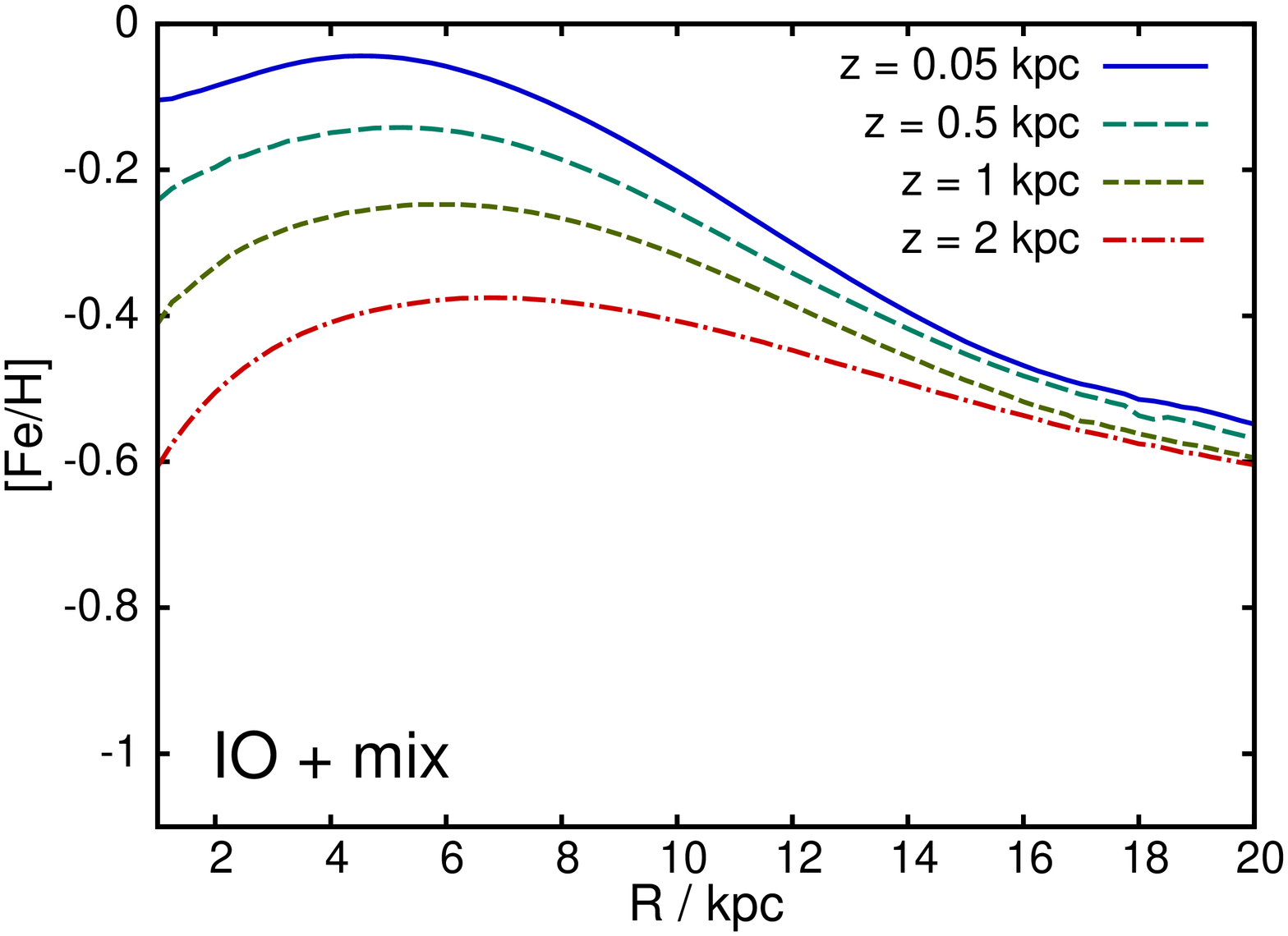,angle=-90,width=\hsize}
\epsfig{file=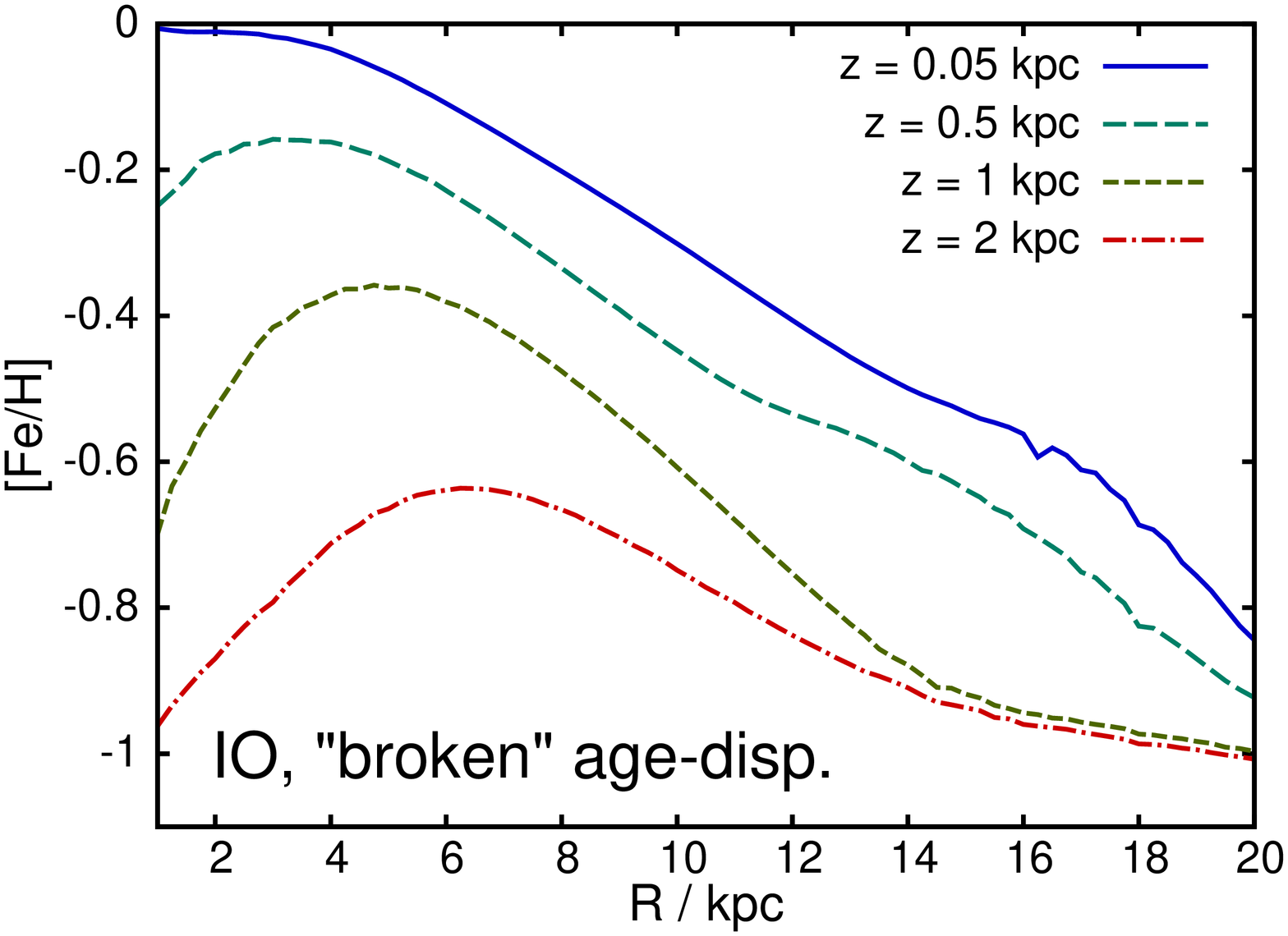,angle=-90,width=\hsize}
\caption{Metallicity profiles at different altitudes versus radius in two alternative versions of the model. The top panel shows the "standard model" with additional gas mixing. The bottom panel shows the standard inside-out model, but with a ``broken'' thin-thick disc heating prescription, reducing the velocity dispersions of thin disc stars, while increasing them strongly for the oldest populations.}\label{fig:vertfehprofb}
\end{figure}

To complete this discussion, \figref{fig:vertfehprofb} shows two alternative versions of the standard model. The top panel shows the inside-out model with additional mixing. As can be expected from our prior discussions on radial metallicity gradients for the entire disc, the main difference to the ``standard'' inside-out model is moving the gradient inversion point to larger radii, while the general behaviour with altitude is very similar. 

The bottom panel of \figref{fig:vertfehprofb} shows the same model with the "broken" thin-thick disc heating law. Due to the disproportionately increased vertical velocity dispersions for the oldest and most populations, this model displays vertical metallicity gradients that are at the upper end of what may still be compatible with observations, indicating that the vertical metallicity gradient alone could serve to rule out the most extreme models in both galactic and extra-galactic data. The drastic decrease in velocity dispersion over short timescales, while the alpha-rich population is forming, is responsible for the particular steepening beyond $z > 1 \kpc$. Also, this enhances the gradient inversion towards higher altitudes in the inner regions, and the steep radial metallicity gradients at large altitudes will be a major constraint for this type of models in the upcoming data sets.

To summarize: Inside-out formation leads to a significantly more negative vertical metallicity gradient throughout the inner and intermediate disc regions. Gradient flattening and gradient inversions towards higher altitudes should not be ascribed to inside-out formation or radial migration alone, but both factors are synergetic. Thin-thick disc like changes in the dispersions are particularly important to make outer disc vertical metallicity gradients more negative, but we caution that a similar effect can be achieved by stronger inside-out formation, and if the inner disc star formation dies out quite quickly (within a timescale of $\sim 2 \Gyr$).

\section{Conclusions}

In this paper we have shown that when interpreting metallicity distributions of stellar populations, the full star formation history of the system has to be taken into account. We have demonstrated that radial metallicity gradients $\d \feh / \d R$ cannot be naively linked the metallicity gradients in the star-forming ISM. Instead, as we differentiate in equation (\ref{eq:Lzt}), the observed radial metallicity gradient in stars in a typical galactic disc is determined by three nearly equipollent factors: i) the averaged radial gradient of the star-forming ISM, ii) the inside-out formation of the disc, and iii) radial mixing in the stellar populations. Typically we expect the inside-out formation of a disc to take place on a comparable timescale to its initial enrichment (by typically $\gtrsim 1 \dex$ within the first $\sim 2 \Gyr$). Consequently, the inside-out formation easily makes a positive contribution of $\gtrsim 0.1 \dex / \kpc$ to the observed stellar metallicity gradient, and can result in a positive/inverted radial metallicity gradient. Thus, findings of inverse metallicity gradients in extra-galactic observations are not a surprise, but a natural consequence of inside-out formation.
As seen in eq.(\ref{eq:Lzt}), the radial mixing of stars to first order averages the metallicity gradients from different radii. Strong radial mixing contributes to a flattening of gradients, and the larger velocity dispersions of older and more metal-poor stellar populations contribute a moderately negative term to $\d \feh / \d R$.

The observed signatures (stellar metallicity gradients, velocity-metallicity correlations) critically depend on the exact relationship between star formation rates and chemical enrichment. In particular, in a simple chemical evolution model without delayed enrichment, without hot gas phase and without radial migration and flows, inside-out formation has nearly no effect on today's stellar metallicity gradients, since the faster enrichment compensates the larger number of stars formed in the inner galaxy early on. As seen in Figure 10, the storage of yields in a hot non-star forming phase is primarily responsible for the strong impact of inside-out formation on the observed metallicity gradient, especially in the inner regions. This process (and the redistribution of the yields by lateral exchange, which has a smaller but non-negligible effect) decouples enrichment from the star-formation rates on short timescales. This also implies that e.g. N-body models painted with a sophisticated chemical evolution should not be used to infer radial abundance gradients and chemo-kinematic relationships, as long as they cannot perfectly match the star formation rates and radial flows between their N-body and chemical evolution parts.

We have further discussed the finding in the Milky Way of an inverse relationship of mean azimuthal speed vs. metallicity for (thick) disc stars \citep[][]{Spagna10, Lee11}. Just as in the question of inverse gradients, these inverse relationships are by no means an indication of inverted radial metallicity gradients of the past star forming ISM, but instead are fully explained by the inside-out formation of the Milky Way disc. Turning this argument around, they provide strong evidence for an early inside-out formation of the Milky Way disc that took place on the same timescale, $\sim 2 \Gyr$, as thick disc formation and SNIa enrichment. Our simple model also reproduces and explains the general structure of the observations: negative $\d \Vphi / \d \feh$ at relatively large $\Vphi$ for the low $\afe$ populations is caused by the metallicity gradient of the disc, altered by radial migration/churning. The latter is responsible for the relatively small value of $|\d \Vphi / \d \feh|$, while positive $\d \Vphi / \d \feh$ in the thick disc is a consequence of inside-out formation.

Gradient inversions in $\d \feh / \d R$ and $\d \Vphi / \d \feh$, while caused by inside-out formation, are not conditional to each other. In particular the inverse relationship,  $\d \Vphi / \d \feh > 0$, in the thick disc, just implies that for a specified metallicity range, more metal-poor stars formed at smaller radii, which is a characteristic of an inside-out forming disc. However, inside-out formation does not have to invert $\d \feh / \d R$ in the stellar populations. We have shown that higher velocity dispersions for older populations act in opposite directions on the two statistics: While increased asymmetric drift, i.e. lower $\Vphi$, gives a positive contribution to $\d \Vphi / \d \feh$, the associated radial expansion gives a negative contribution to $\d \feh / \d R$, helping to differentiate between the different factors at hand. 

In higher precision data, the $\Vphi(\feh)$ for high $\afe$ stars will be an important diagnostic for the formation of the Milky Way's thick disc. As we showed in \figref{fig:velgrad} the original radial metallicity gradient of the high $\afe$/thick disc directly imprints on $\Vphi(\feh)$: if $\d \feh / \d R$ is negative in the star-forming disc at that time, the knee in the $\feh - \afe$ plane lies at lower $\feh$ for outer galactic radii, and hence $\Vphi(\feh)$ would decline towards larger metallicities after reaching a maximum. This is, however, a tricky measurement to make, since contamination of this sequence with the faster rotating lower $\afe$ stars biases the metal-rich end towards larger $\Vphi$, while decreasing disc contamination and increasing halo contamination exacerbate the decline of $\Vphi$ towards lower metallicities. Uncertain abundance errors, made worse by the common strategy to adopt a sample selection sloping in $\afe$ versus $\feh$, make it difficult to judge the current survey results \citep[this contamination problem is mentioned by][]{Bekki11}, but a notable decrease of $\Vphi$ towards larger $\feh$ in the high $\afe$ subset of \cite{Wojno16}, which we tested to be similar to our models when folded with $0.2 \dex$ errors, gives an interesting indication for a negative gradient in the past MW disc.

We also examined the consequences of inside-out formation for $\d \feh / \d z$, in the disc. At the solar galactocentric radius, inside-out formation contributes of order $\sim -0.2 \dex / \kpc$ to $\d \feh / \d z$, bringing the model in line with observed vertical metallicity age gradients, while fitting constraints for the vertical change in mean stellar age. We also note that the models predict flatter/more negative radial $\d \feh / \d R$ metallicity gradients at larger altitudes $z$. Testing against a model without radial migration reveals that the main driver for the change of radial gradients with altitudes is radial migration with the outwards increase of scale-heights under action conservation, with the contributing factor that the more metal-poor old inner disc populations become more dominant at high altitudes in the inner disc.

If a disc has a star-formation cut-off (steepening of the Kennicutt-Schmidt law below a threshold surface density), the shifts of the cut-off radius with time will affect the local metallicity gradients even more strongly than the classic inside-out formation. E.g. an outwards shift of the cut-off radius at relatively early times (in particular if the disc has some radial mixing of the gas) will contribute very positively to $\d \feh / \d R$. Also, near the cut-off, the blurring term of radial mixing gains particular importance due to the strong radial density contrasts. We see this in our models both in the metallicity gradients around the cut-off, and in the large effects on the vertical structure of the disc near that region.

While this paper helps to link and also differentiate these different observational findings, a quantitative study of these processes has to be left to specific modelling of each survey. We used a model that is largely consistent with the Milky Way, but has to be fitted in detail to the upcoming data. At the current stage we cannot reliably derive tighter constraints on the inside-out formation timescales. We have neither a good constraint on the enrichment timescale of the system, as not even the \cite{Bensby14} sample offers the necessary precision at ages greater than $\sim 10 \Gyr$, nor do we know e.g. the SNIa timescales to sufficient precision to provide more than a rough estimate, that the inside-out formation timescale of the Milky Way should have been $\sim 2 \Gyr$, comparable to the SNIa timescale and the timescale of initial metallicity enrichment. With all its quantitative uncertainties, this qualitative result can be considered robust.

Our models also underline the need for a better understanding of gas flows in galaxies. There are several processes that can produce inverse radial metallicity gradients in the star forming ISM. The traditional explanations range from the piercing of cold accretion into the central galaxy (an explanation in the literature, which we consider less plausible) to intrinsic gradient inversions in classical chemical evolution models in the absence of flows, which we identify, however, with manual changes of the Schmidt-Kennicutt law of star formation. Instead, we suggest other natural explanations: i) higher losses of processed material from the central regions, ii) re-accretion of some of the enriched material in galactic fountaining, aided by inside-out formation, which favours stronger relative accretion rates on outer radii, or iii) a general outwards push of the enriched material, which would be natural, because high angular momentum stellar yields push into the partially pressure and turbulence supported hot medium. Again, these processes will have to be differentiated and quantified in future models and comparisons with observations. All we can achieve here, is pointing out the possibility and feasibility, and leave the quantification to future publications.

\section{Acknowledgements}

It is a pleasure to thank J. Binney for interesting discussions and helpful comments on the draft. We would like to thank M. Aumer, M. Bergemann, D. Kawata, G. Pezzulli and J. Stott for helpful comments.
This work was partly supported by the European Union FP7 programme through ERC grant number $320360$. 
PJM gratefully acknowledges financial support from the Swedish National Space Board and the Royal Physiographic Society in Lund.

\section{Appendix: Examinating the effects of a changing circular velocity curve}

\begin{figure}
\epsfig{file=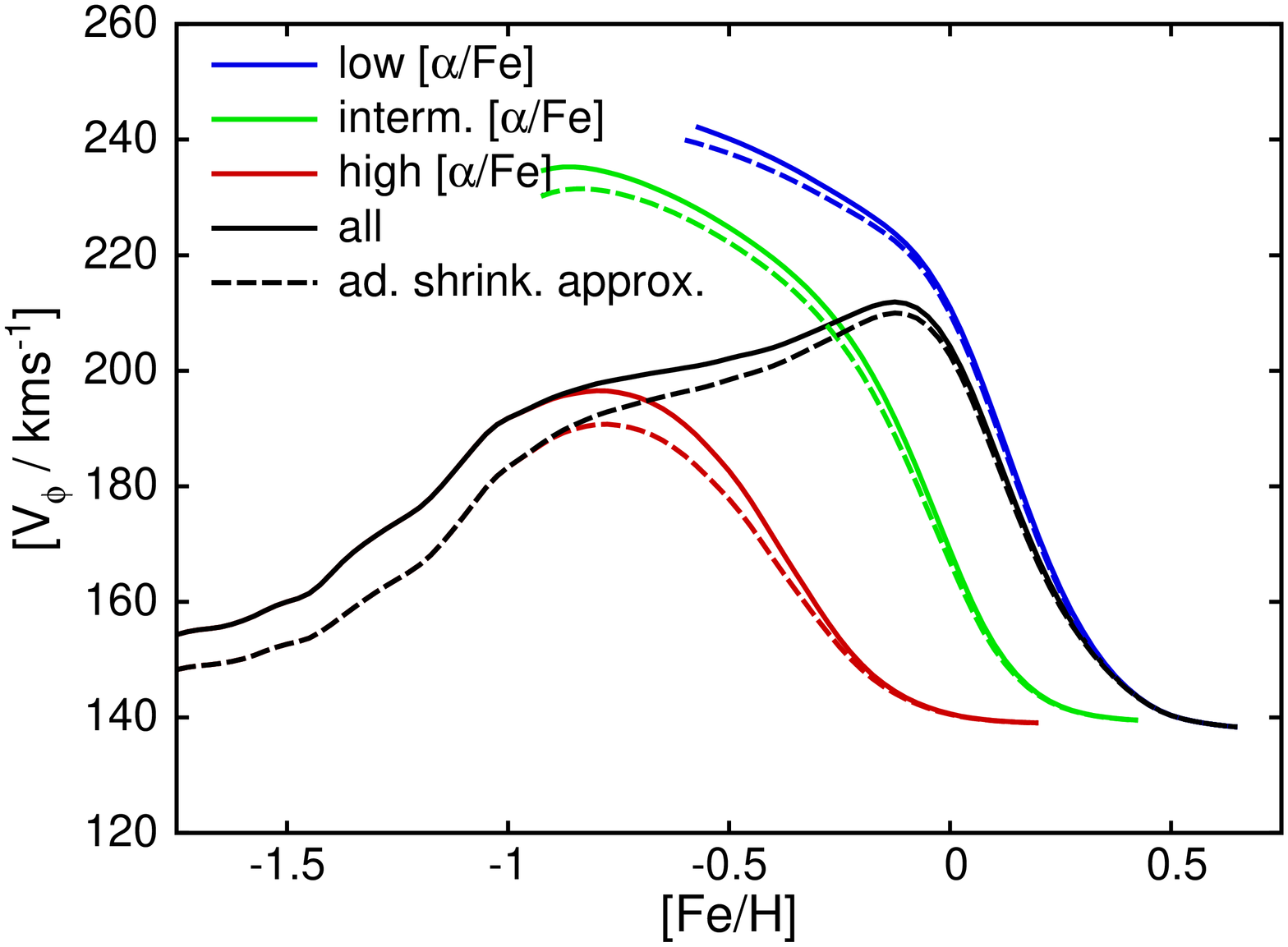,angle=-90,width=\hsize}
\epsfig{file=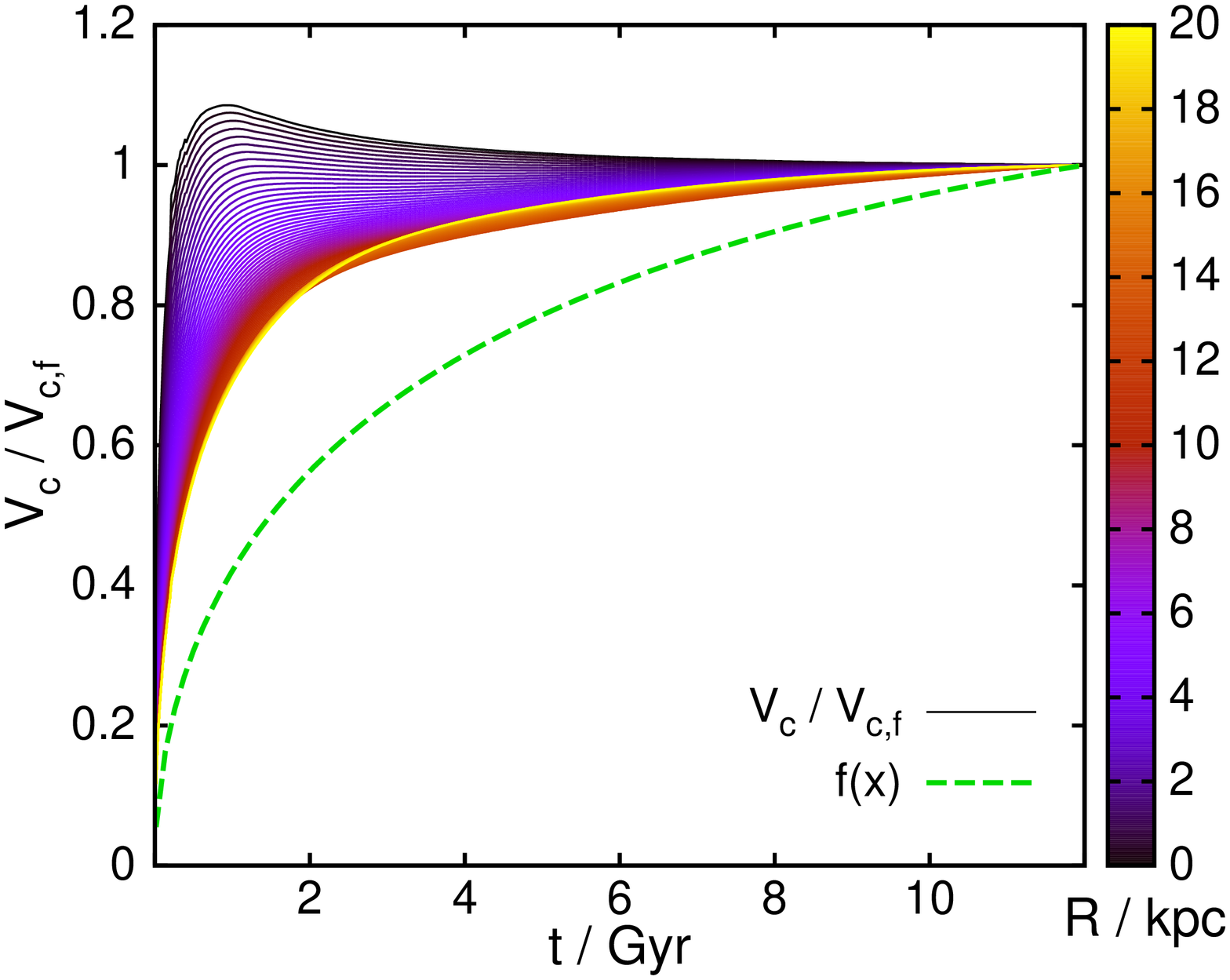,angle=-90,width=\hsize}
\caption{Effects of adiabatic contraction of the disc as the galaxy grows. The lower panel shows the value of the disc’s contribution to $\Vc$ as a fraction of its final contribution, as a function of time. The different colours correspond to different radii. The green dashed line is the naive expectation from the gas accretion rate. The upper panel shows the minor effect of this growth on the $\feh$-$\Vphi$ trends for different components in the inside-out formation model, see \figref{fig:velgrad}.} \label{fig:adcon}
\end{figure}

The purpose of this Appendix is to examine the effects of the changing potential associated with the growing galaxy, which causes adiabatic contraction. While the disc is accreting more mass, one would expect some increase in the circular speed, which leads to an adiabatic contraction of older disc populations.

There are two possible ways to think about this problem:
\begin{itemize}
 \item A straight-forward way is to use the fact that our problem is from the beginning formulated in the angular momentum $\Lzo$. The spatial dimension only enters in the chemical evolution model for assessing star formation and radial flows. However, the radial flows in the gas are at least an order of magnitude faster than the adiabatic contraction. The uncertainties in the time-dependent star formation efficiencies from the Schmidt-Kennicutt-law are far larger than the associated corrections to the surface density - the adiabatic contraction looks from this perspective similar to a very minor addition to radial flows and a very slightly increased coefficient in the star formation efficiency. With this argumentation, we can argue that treatment of adiabatic contraction is not relevant.
 \item We can keep a fixed spatial interpretation of the chemical evolution model and impose a zeroeth order correction for changes of $\Vc$ by inserting an additional matrix for the adiabatic correction $A(R, \Lzo, t)$ that maps the original radius to the angular momentum $\Lzo$ in front of the churning matrix $M$. We will quickly move along this line of argumentation to give a rough estimate of the effect.
\end{itemize}

From \cite{Wang11} and other cosmological simulations, we know that most of the inner halo is already in place before a disc can form. Also, the turbulent behaviour in the first period of mass accretion likely prevents the formation of a disc. Thus, most of the halo contribution should already be in place. For the sake of simplicity, we will assume for the halo/dark matter contribution to the circular speed:
\begin{equation}
\Vch^2(t) = \Vch^2(12 \Gyr) \cdot \left(\frac{3}{4} + \frac{1}{4}N\int_{0}^{t} I(t')dt'\right) ,
\end{equation}
where $I$ is the gas infall over time, and $N = (\int_{0}^{t'}Idt')^{-1}$. The square-root of the infall term is depicted with a green line in the bottom panel of \figref{fig:adcon}. It would be tempting to use just the same term for the contribution for the galactic disc, but three factors lead to a fast, early increase of the circular speed in the disc. The square-root of the disc contribution as a function of radius and time is depicted with coloured lines in the bottom panel of \figref{fig:adcon}. To get the approximation plotted here (which excludes effects of the changing scale-height of the components), we calculated the circular speed contribution of the disc using the surface density profile of our disc and a fixed exponential vertical density profile with $300 \pc$ scale-height.
We identify three reasons for this behaviour:
\begin{itemize}
 \item The loss of material with star formation means that a somewhat higher fraction of the final mass than one would expect accumulates in the early history of the galaxy. Intuitively this is also rooted in the first build-up of the gas disc and it's subsequent decline in mass.
 \item The inside-out formation leads to a far quicker accumulation of mass in the inner radii, moving the increase in circular speed particularly in the inner parts, but also further out, to far earlier times
 \item The disc mass distribution is far from spherical, and hence the radial surface density gradient near each point contributes significantly to the radial force. I.e. a steeper radial surface density gradient will deliver a significant positive contribution to the circular velocity. The increasing scale-length of the disc hence gives a decreasing contribution to $\Vc(R,t)$ over time and can even dominate in the inner disc, as we see from \figref{fig:adcon}.
\end{itemize}
We note that an increase in scale-height over time would yield a minor effect in the same direction, and a bulge would limit adiabatic contraction further. We now add our two approximate terms for disc and halo with weights $0.6$ and $0.4$ together, to obtain
\begin{equation}
\Vc^2 (t) = \Vch^2 (t) + \Vcd^2 (t) ,
\end{equation}
and calculate from this term the transition matrix $A$. The effect of this change on the [Fe/H]-$v_\phi$ relationships for different components of our model with inside-out formation is shown in \figref{fig:adcon} (compare to Fig. 11, top panel). As expected the effect of the adiabatic contraction (compare the dashed lines to the solid lines) is very minor, primarily causing a small decrease of the average azimuthal velocity for lower metallicity stars.

\label{lastpage}

\begin{thebibliography}{}

\bibitem[Aguirre et al.(2001)]{Aguirre01}
Aguirre A., Hernquist L., Schaye J., Katz N., Weinberg D.H., Gardner J., 2001, ApJ, 561, 521

\bibitem[Andrews et al.(2016)]{Andrews16}
Andrews B.H., Weinberg D.H., Sch{\"o}nrich R., Johnson J., 2016, arXiv:1604.08613

\bibitem[Aumer \& Binney(2009)]{AB09}
Aumer M., Binney J., 2009, MNRAS, 397, 1286

\bibitem[Aumer et al.(2016)]{Aumer16}
Aumer M., Binney J., Sch\"onrich R., 2016, MNRAS, 459, 3326

\bibitem[Bekki \& Tsujimoto(2011)]{Bekki11}
Bekki K., Tsujimoto T., ApJ, 738, 4

\bibitem[Bensby et al.(2014)]{Bensby14}
Bensby T., Feltzing S., Oey M.S., 2014, A\&A, 562, 71

\bibitem[Bergemann et al.(2014)]{Bergemann14}
Bergemann M. et al., 2014, A\&A, 565, 89

\bibitem[Bird et al.(2013)]{Bird13}
Bird J., Kazantzidis S., Weinberg D., Guedes J., Callegari S., Mayer L., Madau P., 2013, ApJ, 773, 43

\bibitem[Bilitewski \& Sch\"onrich(2012)]{BS12}
Bilitewski T., Sch\"onrich R., 2012, MNRAS, 426, 2266

\bibitem[Binney \& McMillan(2011)]{BM11}
Binney J, McMillan P.J., 2011, 413, 1889

\bibitem[Binney \& McMillan(2016)]{BM16}
Binney J, McMillan P.J., 2016, 456, 1982

\bibitem[Bland-Hawthorn \& Cohen(2003)]{BH03}
Bland-Hawthorn J., Cohen M., 2003, ApJ, 582, 246

\bibitem[Boeche et al.(2014)]{Boeche14}
 Boeche C. et al., 2014, A\&A, 568, 71

\bibitem[Brook et al.(2012)]{Brook12}
Brook C.B., Stinson G., Gibson B.K., Ro{\u{s}}kar R., Wadsley J., Quinn T., 2012, MNRAS, 419, 771

\bibitem[Casagrande et al.(2011)]{Casagrande11}
Casagrande L., Sch\"onrich R., Asplund M., Cassisi S., Ram{\'i}rez I., Mel{\'e}ndez J., Bensby T., Feltzing S., 2011, A\&A, 530, 138

\bibitem[Casagrande et al.(2016)]{Casagrande16}
Casagrande L. et al., 2016, MNRAS, 455, 987

\bibitem[Chabrier(2003)]{Chabrier03}
Chabrier G., 2003, PASP, 115, 763

\bibitem[Chiappini et al.(2001)]{Chiappini01}
Chiappini C., Matteucci F., Romano D., 2001, ApJ, 554, 1044

\bibitem[Cheng et al.(2012)]{Cheng12}
Cheng J. et al., 2012, ApJ, 746, 149

\bibitem[Collins et al.(2007)]{Collins07}
Collins J.A., Shull J.M., Giroux M.L., 2009, ApJ, 657, 271

\bibitem[Cresci et al.(2010)]{Cresci10}
Cresci G., Mannucci F., Maiolino R., Marconi A., Gnerucci A., Magrini L., 2010, Nature, 467, 811

\bibitem[Curir et al.(2012)]{Curir12}
Curir A., Lattanzi M.G., Spagna A., Matteucci F., Murante G., Re Fiorentin P., Spitoni E., 2012, A\&A, 545, 133

\bibitem[Dav{\'e} et al.(2001)]{Dave01}
Dav{\'e} R. et al., 2001, ApJ, 552, 473

\bibitem[Dehnen \& Binney(1998)]{DB98}
Dehnen W., Binney J., 1998, MNRAS, 298, 387

\bibitem[Fox et al.(2016)]{Fox16}
Fox A.J. et al., 2016, ApJ, 816, 11 	

\bibitem[Flynn et al.(2006)]{Flynn06}
Flynn C., Holmberg J., Portinari L., Fuchs B., Jahreiß H., 2006, MNRAS, 372, 1149

\bibitem[Friedli et al.(1994)]{Friedli94}
Friedli D., Benz W., Kennicutt R., 1994, ApJ, 430, 105

\bibitem[Genovali et al.(2015)]{Genovali15}
Genovali K. et al., 2015, A\&A, 580, 17

\bibitem[Gibson et al.(2013)]{Gibson13}
Gibson B.K., Pilkington K., Brook C.B., Stinson G.S., Bailin J., 2013, A\&A, 554, 47

\bibitem[Haywood(2008)]{Haywood08}
Haywood M., 2008, MNRAS, 388, 1175

\bibitem[Holmberg \& Flynn(2004)]{Holmberg04}
Holmberg J., Flynn C., 2004, MNRAS, 352, 440

\bibitem[Holmberg et al.(2009)]{Holmberg09}
Holmberg J., Nordstr\"om B. Andersen J., 2009, A\&A, 501, 941

\bibitem[Ivezi{\'c} et al.(2008)]{Ivezic08}
Ivezi{\'c} Z. et al., 2008, ApJ, 684, 287

\bibitem[Jones et al.(2010)]{Jones10}
Jones T.A., Ellis R.S., Jullo E., Richard J., 2010, ApJL, 725, 176

\bibitem[Jones et al.(2013)]{Jones13}
Jones T.A., Ellis R.S., Richard J., Jullo E., 2013, ApJ, 765, 48 

\bibitem[Juri{\'c} et al.(2008)]{Juric08}
Juri{\'c} et al., 2008, ApJ, 673, 864

\bibitem[Kawata et al.(2017)]{Kawata16}
Kawata D., Grand R., Gibson B., Casagrande L., Hunt J., Brook A., 2017, MNRAS, 464, 702

\bibitem[Kordopatis et al.(2011)]{Kordopatis11}
Kordopatis G. et al., 2011, A\&A, 535, 107

\bibitem[Kordopatis et al.(2013)]{Kordopatis13}
Kordopatis G. et al., 2013, MNRAS, 436, 3231 

\bibitem[Lacey \& Fall(1985)]{Lacey85}
Lacey C.G., Fall S.M., 1985, ApJ, 290, 154

\bibitem[Lee et al.(2011)]{Lee11}
Lee Y.S. et al., 2011, ApJ, 738, 187

\bibitem[Loebman et al.(2011)]{Loebman11}
Loebman S., Ro{\v{s}}kar R., Debattista V., Ivezi{\'c} {\v{Z}}., Quinn T., Wadsley J., 2011, ApJ, 737, 8

\bibitem[Luck \& Lambert(2011)]{Luck11}
Luck R.E., Lambert D.L., 2011, AJ, 142, 136

\bibitem[Maeder(1992)]{Maeder92}
Maeder A., 1992, A\&A, 264, 105

\bibitem[Marasco et al.(2012)]{Marasco11}
Marasco A., Fraternali F., Binney J., 2012, MNRAS, 419, 1107

\bibitem[Marinacci et al.(2011)]{Marinacci11}
Marinacci F., Fraternali F., Nipoti C., Binney J., Ciotti L., Londrillo P., 2011, MNRAS, 415, 1534

\bibitem[McKee et al.(2015)]{McKee15}
McKee C.F., Parravano A., Hollenbach D.J., 2015, ApJ, 814, 13

\bibitem[McWilliam et al.(2008)]{McWilliam08}
McWilliam A., Matteucci F., Ballero S., Rich R., Fulbright J., Cescutti G., 2008, AJ, 136, 367

\bibitem[Meyer \& York(1987)]{Meyer87}
Meyer D.M., York D.G., 1987, ApJ, 315, 5

\bibitem[Minchev et al.(2012)]{Minchev12}
Minchev I., Famaey B., Quillen A., Dehnen W., Martig M., Siebert A., 2012, A\&A, 548, 127

\bibitem[Minchev et al.(2014)]{Minchev14}
Minchev I., Chiappini C., Martig M., 2014, A\&A, 572, 92

\bibitem[Miranda et al.(2015)]{Miranda15}
Miranda M.S. et al., 2015, arXiv:1512.04559

\bibitem[Nath \& Trentham(1997)]{Nath97}
Nath B.B., Trentham N., 1997, MNRAS, 291, 505

\bibitem[Pezzulli et al.(2015)]{Pezzulli15}
Pezzulli G., Fraternali F., Boissier S., Muñoz-Mateos J.C., 2015, MNRAS, 451, 2324

\bibitem[Pezzulli \& Fraternali(2016)]{Pezzulli16}
Pezzulli G., Fraternali F., 2016, MNRAS, 455, 2308

\bibitem[Piffl et al.(2014)]{Piffl14}
Piffl T. et al., 2014, MNRAS, 445, 3133

\bibitem[Prantzos \& Boissier(2000)]{Prantzos00}
Prantzos N., Boissier S., 2000, MNRAS, 313, 338

\bibitem[Przybilla et al.(2008)]{Przybilla08}
Przybilla N., Nieva M.-F., Butler K., 2008, ApJL, 688, 103

\bibitem[Queyrel et al.(2012)]{Queyrel12}
Queyrel J. et al., 2012, A\&A, 539, 93

\bibitem[Rahimi et al.(2011)]{Rahimi11}
Rahimi A., Kawata D., Allende-Prieto C., Brook C., Gibson B., Kiessling A., 2011, MNRAS, 415, 1469

\bibitem[Rahimi et al.(2014)]{Rahimi14}
Rahimi A., Carrell K., Kawata D., 2014, RAA, 14, 1406

\bibitem[Read(2014)]{Read14}
Read J.I., 2014, Journal of Physics G Nuclear Physics, 41, 063101

\bibitem[Ro{\v{s}}kar et al.(2013)]{Roskar13}
Ro{\v{s}}kar R., Debattista V., Loebman S., 2013, MNRAS, 433, 976

\bibitem[Rupke et al.(2010)]{Rupke10}
Rupke D.S.N., Kewley L.J., Barnes J.E., 2010, ApJ, 710, 156

\bibitem[Salpeter(1955)]{Salpeter55}
Salpeter E., 1955, ApJ, 121, 161

\bibitem[Sanders \& Binney(2015)]{Sanders15}
Sanders J., Binney J., 2015, MNRAS, 449, 3479

\bibitem[Sch\"onrich \& Binney(2009a)]{SB09a}
Sch\"onrich R., Binney J., 2009, MNRAS, 396, 203

\bibitem[Sch\"onrich \& Binney(2009b)]{SB09b}
Sch\"onrich R., Binney J., 2009, MNRAS, 399, 1145

\bibitem[Sch\"onrich, Binney \& Dehnen(2010)]{SBD}
Sch\"onrich R., Binney J., Dehnen W., 2010, MNRAS, 403, 1829

\bibitem[Sch\"onrich \& Binney(2012)]{SB12}
Sch\"onrich R., Binney J., 2012, MNRAS, 419, 1546

\bibitem[Schlesinger et al.(2014)]{Schlesinger14}
Schlesinger K. et al., 2014, ApJ, 791, 112

\bibitem[Sellwood \& Binney(2002)]{Sellwood02}
Sellwood J.A., Binney J.J., 2002, MNRAS, 336, 785

\bibitem[Shopbell \& Bland-Hawthorn(1998)]{Shopbell98}
Shopbell P.L., Bland-Hawthorn J., 1998, ApJ, 493, 129

\bibitem[Shull et al.(2011)]{Shull11}
Shull J.M., Stevans M., Danforth C., Penton S.V., Lockman F.J., Arav N., 2016, ApJ, 739, 105

\bibitem[Spagna et al.(2010)]{Spagna10}
Spagna A., Lattanzi M.G., Re Fiorentin P., Smart R.L., 2010, A\&A, 510, 4

\bibitem[Spitoni et al.(2008)]{Spitoni08}
Spitoni E., Recchi S., Matteucci F., 2008, A\&A, 484, 743

\bibitem[Spitoni \& Matteucci(2011)]{Spitoni11}
Spitoni E., Matteucci F., 2011, A\&A, 531, 72

\bibitem[Solway et al.(2012)]{Solway12}
Solway M., Sellwood J.A., Sch\"onrich R., 2012, MNRAS, 422, 1363

\bibitem[Stott et al.(2014)]{Stott14}
Stott J.P., Sobral D., Swinbank A.M., Smail I. Bower R., Best P.N., Sharples R.M., Geach J.E., Matthee J., 2014, MNRAS, 443, 2695

\bibitem[Tissot et al.(2016)]{Tissot16}
Tissot F., Dauphas N., Grossman L., 2016, Sci. Adv., 2, 3, e1501400

\bibitem[Tumlinson et al.(2011)]{Tumlinson11}
Tumlinson J. et al., 2011, Science, 338, 6058, 948

\bibitem[Vera-Ciro et al.(2014)]{VC14}
Vera-Ciro C., D'Onghia E., Navarro J., Abadi M., 2014, ApJ, 794, 173 

\bibitem[Walch et al.(2015)]{Walch15}
Walch S. et al., 2015, MNRAS, 454, 238

\bibitem[Wang et al.(2011)]{Wang11}
Wang J. et al., 2011, MNRAS 413, 1373

\bibitem[Wojno et al.(2016)]{Wojno16}
Wojno J. et al., 2016, arXiv:1603.09339

\bibitem[Wuyts et al.(2016)]{Wuyts16}
Wuyts E. et al., 2016, arXiv:1603.01139

\bibitem[Yuan et al.(2011)]{Yuan11}
Yuan T.-T., Kewley L.J., Swinbank A.M., Richard J., Livermore R.C., 2011, ApJ, 732, 14

\end{thebibliography}
\end{document}